\documentclass[preprint,12pt]{elsarticle}

\usepackage[left=2.7cm,right=2.7cm]{geometry}
\usepackage{graphicx}
\usepackage{amssymb}
\usepackage{lineno}
\usepackage{amssymb}
\usepackage{amsmath}
\usepackage{graphics}
\usepackage{subcaption}
\usepackage{siunitx}
\usepackage{enumerate}
\usepackage{cases}
\usepackage{textcomp}
\usepackage{xcolor}
\usepackage{xfrac}
\usepackage{booktabs}
\usepackage[normalem]{ulem}
\usepackage{accents}
\usepackage{numprint}
\usepackage{dirtytalk}
\newcommand{\interval}[2]{[\,#1,#2\,]}

\definecolor{darkgreen}{rgb}{0.0, 0.6, 0.0}
\newcommand{\reva}[1]{\textcolor{black}{#1}}
\newcommand{\revb}[1]{\textcolor{black}{#1}}

\journal{International Journal of Approximate Reasoning}

\begin{document}

\begin{frontmatter}

\title{\reva{Should data ever be thrown away?} \\ 
Pooling interval-censored data sets with different  precision\tnoteref{mytitlenote}}
\tnotetext[mytitlenote]{This work was partially funded by the Engineering and Physical Science Research Council (EPSRC) through programme grant “Digital twins for improved dynamic design”, EP/R006768/1.}

\author[Liverpool]{Krasymyr Tretiak\corref{cor}}
\ead{k.tretiak@liverpool.ac.uk}

\author[Liverpool]{Scott Ferson}
\ead{ferson@liverpool.ac.uk}

\address[Liverpool]{University of Liverpool,  Liverpool L69 7ZX, United Kingdom}
\cortext[cor]{Corresponding author}

\begin{abstract}
\normalsize
Data quality is an important consideration in many engineering applications and projects. Data collection procedures do not always involve careful utilization of the most precise instruments and strictest protocols.  
As a consequence, data are invariably affected by imprecision and sometimes sharply varying levels of quality of the data.  
Different mathematical representations of imprecision have been suggested, 
including a classical approach to censored data which is considered optimal when the proposed error model is correct,
and a weaker approach  called interval statistics based on partial identification that makes fewer assumptions. 
Maximizing the quality of statistical results is often crucial to the success of many engineering projects, and a natural question that arises is whether data of differing qualities should be pooled together or we should include only precise measurements and disregard imprecise data.
Some worry that combining precise and imprecise measurements can depreciate the overall quality of the pooled data.
Some fear that excluding data of lesser precision can increase their overall uncertainty about results because lower sample size implies more sampling uncertainty.
This paper explores these concerns and describes simulation results that show when it is advisable to combine fairly precise data with rather imprecise data
by comparing analyses using different mathematical representations of 
imprecision. Pooling data sets is preferred when the low-quality data set does not exceed a certain level of uncertainty.
However, so long as the data are random,
it may be legitimate to reject the low-quality data if its reduction of sampling uncertainty does not counterbalance the effect of its imprecision on the overall uncertainty.  

\end{abstract}


\end{frontmatter}

\section{Introduction}
Measurements are often performed under varying conditions or protocols, or by independent agents, or with different measuring devices. This can lead to data measurements with different precision.  
Interval-censored data, or simply \emph{interval data}, are measured values known only within certain bounds instead of being observed exactly. Interval data can arise in various cases including non-detects and data censoring \cite{Helsel2005}, 
periodic observations, plus-or-minus measurement uncertainties, interval-valued expert opinions \cite{SpeirsBridge2010}, privacy requirements, theoretical constraints, bounding studies, etc. \cite{SAND2007-0939, A_probabilistic_approach2011, Nguyen2012}. Some common examples of interval-censored data come from clinical and health studies \cite{Sun2006}, survival data analysis \cite{Survival_analysis1997},  chemical risk assessment \cite{Nysen2014, Shoari2017}, mechanics \cite{Sun2020}, etc. The breadth of the uncertainty captured by the widths of these intervals is epistemic uncertainty \cite{Oberkampf2001, epistemic_2007, epistemic_2009}. This uncertainty arises due to lack of knowledge, limited or inaccurate measurements, approximation or poor understanding of physical phenomena.

Over the past couple of decades various statistical methods have been developed for handling interval-censored data. The substitution method \cite{SUB_Hornung1990, Gleit1985} is still the most common procedure in handling nondetect data \cite{Shoari2017} which are special cases of left-censored data where a value is known to reside between zero and some quantitative detection limit.
Substitution replaces nondetects with a zero, the detection limit, half the detection limit, or some other fraction of the detection limit. This method ignores the imprecision and can lead to incorrect estimates \cite{Helsel_data_below_DL1990, Helsel2006}. One might presume that using the detection limit as the substituted value would yield the most conservative upper confidence limit for the mean, but this is not true.  The reason is that the upper confidence limit depends positively on both the mean and variance of the data, and clustering values at a single detection limit reduces the variance. An alternative strategy, what might be considered the zeroth approach, neglects the imprecision altogether and omits from the analysis any data with imprecision beyond some acceptable threshold. It is hard to assess how common this approach is, but it seems to widespread despite its obvious statistical shortcomings.

Another basic approach is maximum likelihood estimation (MLE) which gauges the parameters of a probability distribution by maximizing a likelihood function \cite{Cohen1959, Singh2002, MLE_2003}. MLE requires an assumption that the data follow a specific distribution.
An alternative approach is the Turnbull’s method \cite{Turnbull1974,Turnbull1976,Rodrigues2018}, a generalisation of the Kaplan--Meier method which is commonly used in survival analysis for estimating percentiles, means, or other statistics without substitutions \cite{Kaplan_Meier1958, KM_She1997}. These methods are non-parametric, and do not require specification of an assumed distribution. The Kaplan--Meier method is used for right censoring whereas Turnbull’s method takes into account interval-censored data.

Interval data can also be considered as a special case of ``symbolic data'' modelled with uniform distributions over the interval ranges \cite{Uniform_approach2000, Billard_Diday2000, Billard_Diday2006}.  This approach corresponds to Laplace’s principle of indifference with respect to the censored interval. All values within each interval are assumed to be equally likely. This assumption makes it relatively straightforward to calculate sample statistics for data sets containing such intervals. This \emph{uniforms approach} characterises a sample as a precise distribution, but estimators based on it are \revb{not consistent} because there is no guarantee the distribution approaches the true distribution from which the data were generated as sample size increases. Bertrand and Groupil \cite{Uniform_approach2000} acknowledged that, even as the sample size grows to infinity, the \revb{true distribution} of the underlying population is only approximated by this approach.

Another approach represents interval uncertainty purely in the form of bounds \cite{Manski2003, Vansteelandt2006, Xiang_Gang2006, SAND2007-0939, Kreinovich2009, Ferson_Siegrist2011, Nguyen2012, TRETIAK2023}. This approach originates from the theory of imprecise probabilities \cite{Walley1991, Kuznetsov1991, Augustin2014, Schollmeyer_2021}. It models each interval as a set of possible values, and calculations result in a class of distribution functions, corresponding to the different possible values within the respective intervals. In contrast to the uniforms approach, some operations within this \emph{intervals approach} can be computationally expensive. However the obtained results are arguably more reliable, and the resulting bounds on the distribution will asymptotically enclose the true distribution as sample size grows to infinity. The intervals approach is computationally slower, and weaker than the other approaches if their assumptions are satisfied.

\begin{figure}[ht!]
	\centering
	\begin{subfigure}{0.5\textwidth}
		\includegraphics[width=\linewidth]{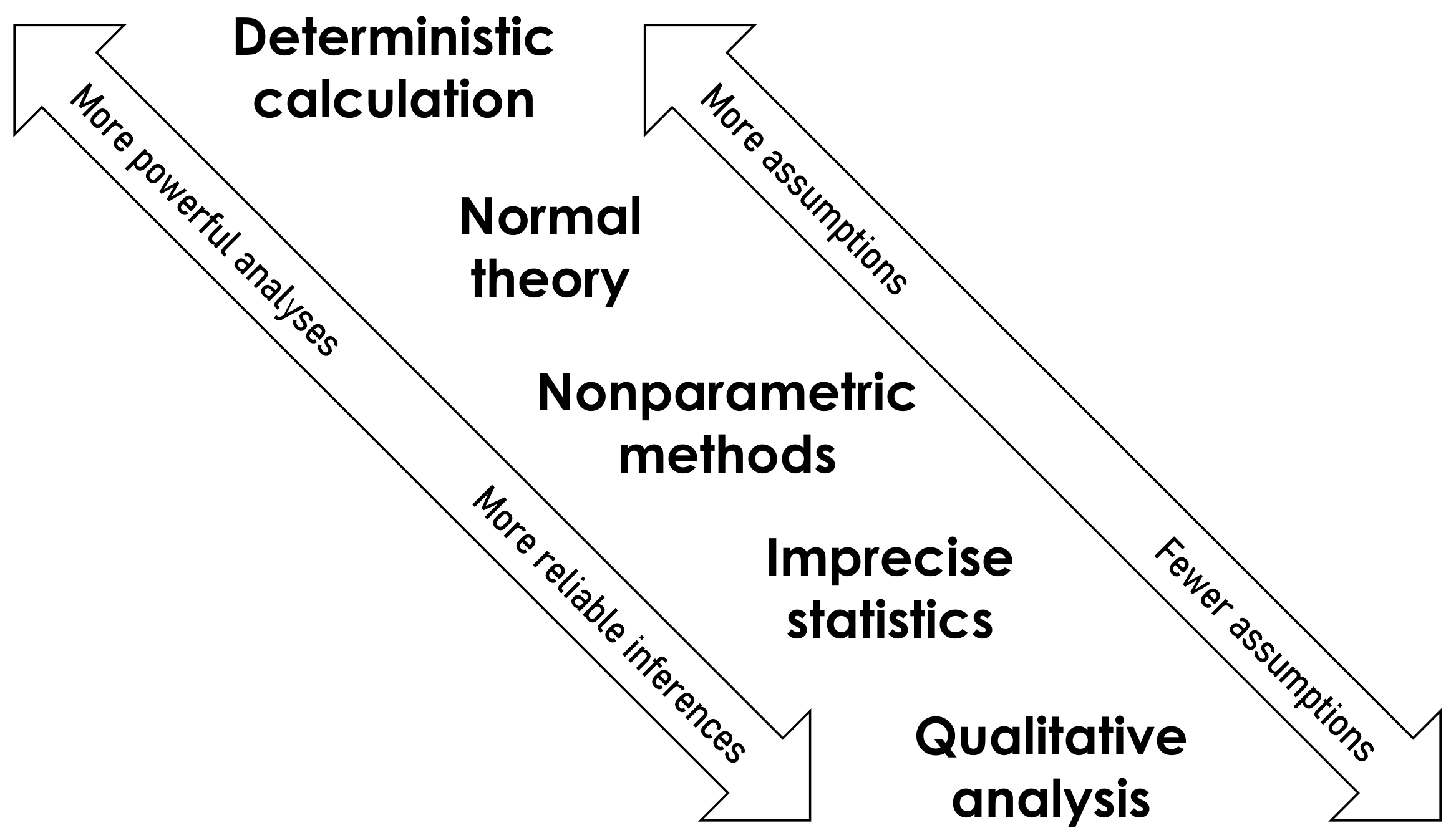}
	\end{subfigure}
	\caption{Schematic representation of assumption-dependent methods that are powerful but possibly unreliable and relatively assumption-free methods that are reliable but may not be as powerful.}
	\label{fig:Continuum}
\end{figure}

As strategies for descriptive statistics, these approaches represent different positions along a continuum between assumption-dependent methods that are powerful but possibly unreliable and relatively assumption-free methods that are reliable but may not be as powerful. Fig.~\ref{fig:Continuum} argues that this continuum connects the poles of deterministic calculations and purely qualitative approaches (such as narrative analysis, word clouds, etc.).
Normal theory, which developed in the first half of the 20th century, is extremely powerful, but it requires assumptions that are often hard to justify (such as normality, independence, homoscedasticity, stationarity, etc.). 
Non-parametric methods that allowed analysts to relax these assumptions flowered in the second half of that century. 
Imprecise statistics, motivated by the recognition of non-negligible epistemic uncertainties such as interval censoring, has developed over the last two decades to provide approaches that further relax assumptions when they are untenable or in doubt.
Modern statistics recognises there is not just one tool for any given statistical problem. Various methods, representing sometimes disparate sets of assumptions can be deployed, and it is often useful to compare the results from such different methods.

A natural question asks about guidelines for making practical decisions whether data of differing qualities should be pooled together. By \textit{pooling} we mean combining the data sets and analysing them as a single sample. Certainly any data sets considered for pooling are assumed to be measuring the same quantity or distribution of quantities. A common belief within the statistics community suggests that all available data should be included in any analysis. Others feel that imprecise data could potentially contaminate precise data and that pooling might need to be avoided lest any bias that hides in the imprecision be introduced.

Sample size is a crucial consideration in each statistical study and, in two-sample tests for instance, it should depend on the minimum expected difference, desired statistical power, and the significance criterion \cite{SampleSizeEstimation}. 
Increasing the sample size improves the characterization of variability, lowering the degree of sampling uncertainty. A lower level of sampling error is advantageous because it permits statistical inferences to be made with more certainty. But is it possible that some data are so imprecise as to warrant their exclusion from an analysis?

Clearly, it would be inappropriate to exclude data because of its imprecision,
even if it is very low quality, if the method of analysis is sensitive to whether and how its imprecision depends on the magnitude of the value being measured.
For instance, when asking people about their salaries, high-earners often tend to give vague answers \revb{\cite{Fukuoka2007}}. Excluding these answers could bias the results dramatically.
However, so long as the imprecision is independent of the underlying magnitudes of the random data, it might be legitimate to reject the low-quality data when
the effect of the measurement imprecision is greater on the overall uncertainty than the reduction in sampling uncertainty from including the low-quality data. 
This can be true even when the imprecision depends on other features of the data including group membership. Data for which measurement imprecision has nothing to do with the data values are said to be \textit{irrelevantly imprecise}. \reva{Mathematically,
this means that breadth of the imprecision of a quantity is independent of its underlying true magnitude.}

This  paper  is  organized  as  follows.
Section~\ref{Interval_statistics} introduces
an approach to analysing statistical data whose measurement uncertainties are intervals. Section~\ref{Generating_data} describes the process of generating synthetic interval data sets for our simulations. Section~\ref{Sec_CI} describes using confidence intervals on the mean to assess uncertainty for precise and imprecise data sets. Section~\ref{KolmogorovSmirnov_bands} presents numerical simulations and comparisons with interval data sets containing different levels of uncertainty using distribution-free Kolmogorov--Smirnov confidence limits to assess uncertainty. Section~\ref{MLE} uses maximum likelihood for fitting a named distribution to precise and imprecise interval data sets 
and contrasts the traditional approach with 
the imprecise probabilities approach to maximum likelihood for which we compute confidence intervals.  \reva{Section~\ref{Discussion} discusses the assumptions used in the preceding analyses, when they might and might not be appropriate and alternative 
assumptions that could be used to characterise imprecision.}
Summaries of  the  application  of several statistical methods for estimating the overall level of uncertainty of interval data, and conclusions on whether it is advisable to pool data with varying quality are presented in Section~\ref{Conclusion}.

\section{Interval statistics}\label{Interval_statistics}
Analysis of data that combine measurements with varying quality can be done with interval descriptive statistics \cite{Uniform_approach2000, Billard_Diday2000, Manski2003, SAND2007-0939, Ferson_Siegrist2011, TRETIAK2023}. Statistical analysis with measurements modeled as intervals provides a reliable method to account for both the sampling and measurement uncertainties. By applying interval statistics we can determine the overall level of uncertainty and decide when pooling of the data sets is preferable or when we should disregard some data. 
Siegrist \cite{Siegrist_2011} used quantitative examples involving combining \emph{precise} and \emph{sloppy} (imprecise) data sets to illustrate cases where the resulting overall uncertainty of the pooled data can be either lower or higher than that of the precise data by itself. In his studies he used two data sets: precise data were sampled from the normal distribution with known mean, variance and blurred with $\pm 0.1$ uncertainty; sloppy data were sampled from the same distribution but blurred with $\pm 0.1 f$ uncertainty, where $f$ can be any positive number greater than one.

This paper extends the idea proposed in \cite{Siegrist_2011} and provides a broader view on questions of pooling data sets with different levels of uncertainty. We assess combinations consisting of only two data sets: precise and imprecise. 
We consider data sets that are measurements, not guesses, opinions, or expert elicitations. These measurements come from the empiricist with clear statements about their precision and measurement protocols.
Because different approaches could result in different decisions, it is essential to compare several statistical methods for experimental data processing, for instance, estimation of the parameters of a named distribution, construction of their confidence intervals, and determination of Kolmogorov--Smirnov~(K--S) limits on the distribution as a whole. The quality of the pooled data set depends on four design parameters: sample size, dispersion, imprecision factor, and balance. The imprecision factor defines the widths of the imprecise interval data. The balance controls the proportions of intervals (precise and imprecise) in the pooled data set.

We assume that our imprecisely measured values are surely enclosed within respective censoring intervals, without necessarily knowing the distribution function characterizing the position of these values inside the intervals. 
It is also assumed that the proffered precision of each datum is always known before the analysis.
In addition, it is assumed that the underlying values of the measurands represented in the precise and imprecise data sets are identically distributed. That is, both measurement protocols are measuring the same thing, and differ only in their abilities to do so precisely.
The level of uncertainty is fixed and does not vary within a given data set. Because we assume that the precise \revb{values} and the imprecise \revb{values will both be} randomly sampled from the same distribution, we exclude situations when these data sets were taken from different parts of a distribution.

\revb{In this paper, we are focusing on epistemic uncertainty (where there is a true value that empiricists just cannot see precisely).  This kind of uncertainty is commonly recognised in science and engineering. In the following section, the simulation strategy used to create imprecise data sets likewise uses the idea that there is a true value that is masked by some cause or mechanism yielding imprecision about that value in the form of statistical censoring.
Exploring the consequences in more general cases where the nature and the degree of censoring varies and might depend on magnitude is deferred to future work.}

\section{Generating interval-valued data sets}\label{Generating_data}
\begin{figure}[ht!]
	\centering
	\begin{subfigure}{0.8\textwidth}
		\includegraphics[width=\linewidth]{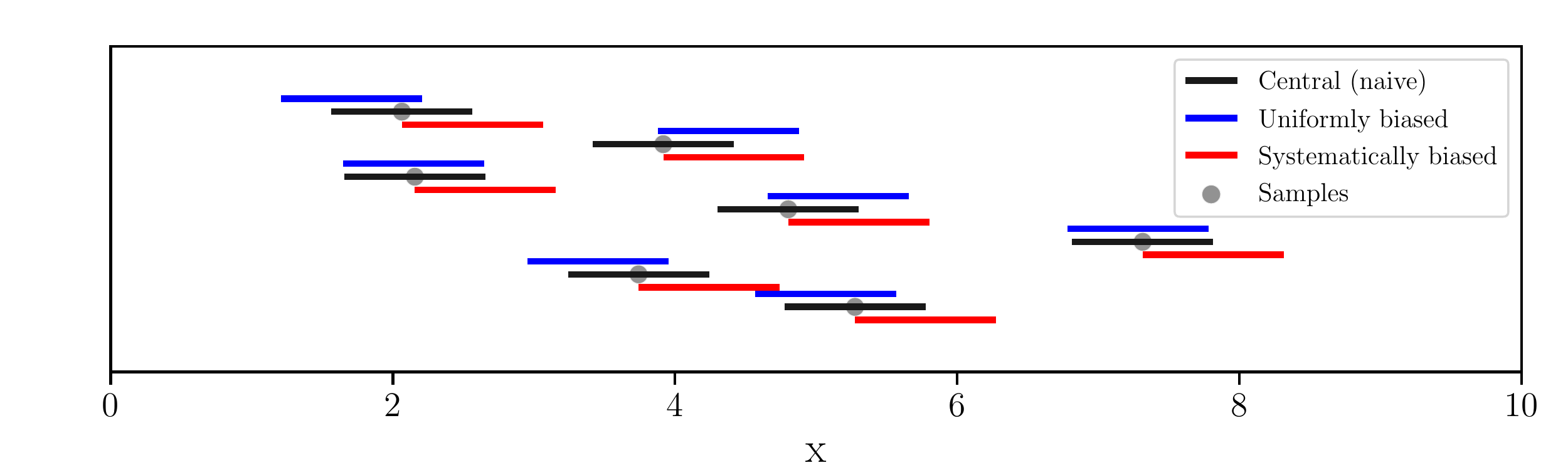}
	\end{subfigure}
	\caption{Example of how single data points can be turned to intervals.}
	\label{fig:intervals}
\end{figure}

The following notation is used  to represent a real interval
\begin{equation*}
    [x] = \interval{\underline{x}}{\overline{x}} = \{ x \in \mathbb{R} \mid \underline{x} \leq x \leq \overline{x} \},
\end{equation*}
where $\underline{x}$ the lower bound and $\overline{x}$ the upper bound. Generation of the underlying sample values and construction of the corresponding interval data sets are carried out in the following order. We use several different distributions, including normal $\mathcal{N}(\mu,\sigma^2)$, uniform $\mathcal{U}(a,b)$, exponential Exp$(\lambda)$, and gamma $\Gamma(k,\theta)$, to populate individual samples in the numerical simulations, where $\mu$ is the mean, $\sigma$ is the standard deviation, $a,b$ are the bounds for the uniform distribution, $\lambda$ is the exponential distribution's rate parameter, and $k$ and $\theta$ are the scale and shape parametrizations for the gamma distribution. We intervalize given samples $x_i$ with one of the following ways:
\begin{itemize}
    \item[] \textit{Central (naive)} method puts the uncertainty characterised by imprecision $\Delta$ symmetrically around the individual value 
    \begin{equation}
        \interval{\underline{x}_i}{\overline{x}_i}  = \interval{x_i-\Delta}{x_i+\Delta},
    \end{equation}
    \item[] \textit{Uniformly biased} method computes intervals as
    \begin{equation}
    \begin{aligned}
            r &\sim \mathcal{U}(-1,1), \\
            m_i &= x_i - \Delta + (1+r)\Delta, \\
            \interval{\underline{x}_i}{\overline{x}_i}  &= \interval{m_i-\Delta}{m_i+\Delta},  
    \end{aligned}
    \end{equation}
    \item[] \textit{Systematically biased} method uses predefined constant bias
    \begin{equation}\label{sys_biased}
    \begin{aligned}
            r &= C, \quad C \in [-1,1] \\
            m_i &= x_i - \Delta + (1+r)\Delta, \\
            \interval{\underline{x}_i}{\overline{x}_i}  &= \interval{m_i-\Delta}{m_i+\Delta}.
    \end{aligned}
    \end{equation}
\end{itemize}
Fig. \ref{fig:intervals} shows the results of applying these three intervalization methods with $\Delta=0.5$ and $C=1$ to random samples generated from the $\mathcal{N}(5,4)$ distribution.
These algorithms produce interval data sets that would be called \say{no-nesting} or ``few-intersections''
under the categorisation scheme described in \cite[][Sec. 3.5]{SAND2007-0939}. The kind of interval data set depends on sample size, width of intervals  and scattering.

\section{Confidence intervals}\label{Sec_CI}
For characterizing data sets we refer to 
statistics which allows us to measure central tendency, dispersion and confidence or prediction intervals. Confidence intervals on the mean are the most common and widely used characterization of the location of a data set. A confidence interval has a confidence level associated with it, which gives the probability that the obtained interval will contain the parameter's true value over many repeated random samples. For normally distributed values $X_1, \dots, X_N$, the confidence interval is
\begin{equation}\label{standart_UCL}
    \interval{L}{U} = \interval{\overline{X} - t_{N-1,1-\alpha/2} \dfrac{s}{\sqrt{N}}}{\overline{X} + t_{N-1,1-\alpha/2} \dfrac{s}{\sqrt{N}}}
\end{equation}
where $\overline{X}$ is the sample mean, $N$ is the number of samples, $s$ is the sample standard deviation, and $t_{N-1,1-\alpha/2}$ is the critical value from Student’s $t$ distribution associated with the desired confidence level and sample size $N$.
The idea of intervalising formula (\ref{standart_UCL}) is 
to find the \emph{union} of all such $\interval{L}{U}$ intervals that arise from any configuration of real-valued points from their respective 
measurements $X_1, \dots, X_N$, some or all of which may be intervals.

Under the rubric of partial identifiability \cite{Manski2003}, several researchers have explored the question of how measurement imprecision should be combined with statistical sampling uncertainty \cite{imbens2004confidence, Vansteelandt2006, Richardson2014}.
Vansteelandt et al.\ \cite{Vansteelandt2006} considered approaches for assessing uncertainty in the case of partial identifiability due to missing data, which is a special case in which many of the $X$-values are points but some are vacuously wide intervals.
They adopt a subtle terminology that distinguishes several notions including \say{ignorance region} which is their phrase for the intervalised sample statistic, and \say{strong uncertainty region} which corresponds to a confidence interval that has been generalised for data with \revb{measurement imprecision}. Estimated ignorance regions reflect the uncertainty due to partial identifiability but not the sampling variability of the data. The uncertainty regions reflect both ignorance from partial identifiability and the sampling variability due to the limitation of sample size. Vansteelandt et al.\ also discuss \say{pointwise uncertainty regions} which are an alternative generalisation of confidence intervals to partially identifiable parameters.

The approach used in this paper, which we might simply call (intervalised) confidence intervals, yields results that we believe are analogous to the strong uncertainty regions of Vansteelandt et al.\ \cite{Vansteelandt2006}. Some suggest \cite{Richardson2014} that these regions are conservative, that is, unnecessarily wide, but we are asking for results that are guaranteed to have the nominal coverage probability on average under the assumptions. Less conservative pointwise uncertainty regions might be good enough in some analyses, but because they \revb{do not simultaneously cover all the parameter values in the true region of ignorance at the nominal coverage probability, they do not offer the statistical guarantee that analysts might be expecting.}

Computing standard deviation and confidence limits for data sets with interval uncertainty may be computationally prohibitive as they are NP-hard problems \cite{NPhard, Kreinovich1998, NP-Hard_Ferson}. This means that computational time grows exponentially with the number of intervals within the given data set. Nevertheless, efficient algorithms have been developed in many common special cases. In order to compute confidence intervals we implemented the algorithm described in \cite{Outlier_detection1, Outlier_detection2} and \cite[][Sec. 4.8]{SAND2007-0939}.
We employ this strategy because of its convenience for no-nesting data for which it has been shown in \cite{SAND2007-0939, Outlier_detection1, Outlier_detection2} that the maximum for $U$ and minimum for $L$ are attained by some extreme configuration which consists of either the right or left endpoint from every interval $X_i=\interval{\underline{x}_i}{\overline{x}_i}$. The algorithm computes the outer bounds on confidence limits and has the following steps:
\begin{enumerate}
    \item Sort the intervals $\interval{\underline{x}_1}{ \overline{x}_1}, \dots,  \interval{\underline{x}_N}{ \overline{x}_N}$ into lexicographic order; this implies that $X_i \leq X_j $ if and only if either $\underline{x}_i \leq \underline{x}_j $, or both $\underline{x}_i = \underline{x}_j $ and $\overline{x}_i \leq \overline{x}_j $.
    \item For each $k=0,1,\dots,N$ compute the configuration $x^{k} = (\underline{x}_1, \dots, \underline{x}_k, \overline{x}_{k+1}, \dots, \overline{x}_N)$ and value $V_k = M_k - E^2_k$, where $E_k = \frac{1}{N}\sum_{j=1}^{N} x_j^k$, $M_k = \frac{1}{N}\sum_{j=1}^{N} (x_j^k)^2$.
    \item Compute $L_k = E_k - t_{\alpha N} \sqrt{V_k/(N-1)}$ and $U_k = E_k + t_{\alpha N} \sqrt{V_k/(N-1)}$.
    \item Finally, to calculate the outer bounds on the confidence interval, return the smallest of the values $\underline{L} = \min (L_0, \dots, L_k)$ and largest for $\overline{U} = \max (U_0, \dots, U_k)$.
\end{enumerate}

Our main question asks whether and when is it advisable to combine precise data with imprecise data. To address this question, we introduce a few distinct interval-valued data sets. A precise data  set, called \textit{skinny}, contains rather narrow intervals which likely 
do not intersect with each other. An imprecise data set, \textit{puffy}, will have wider intervals than skinny which may intersect. Finally, a \textit{pooled} (or mixed) data set is a combination of the skinny and puffy intervals.

\begin{figure}[ht!]
	\centering
	\begin{subfigure}{0.325\textwidth}
		\includegraphics[width=\linewidth]{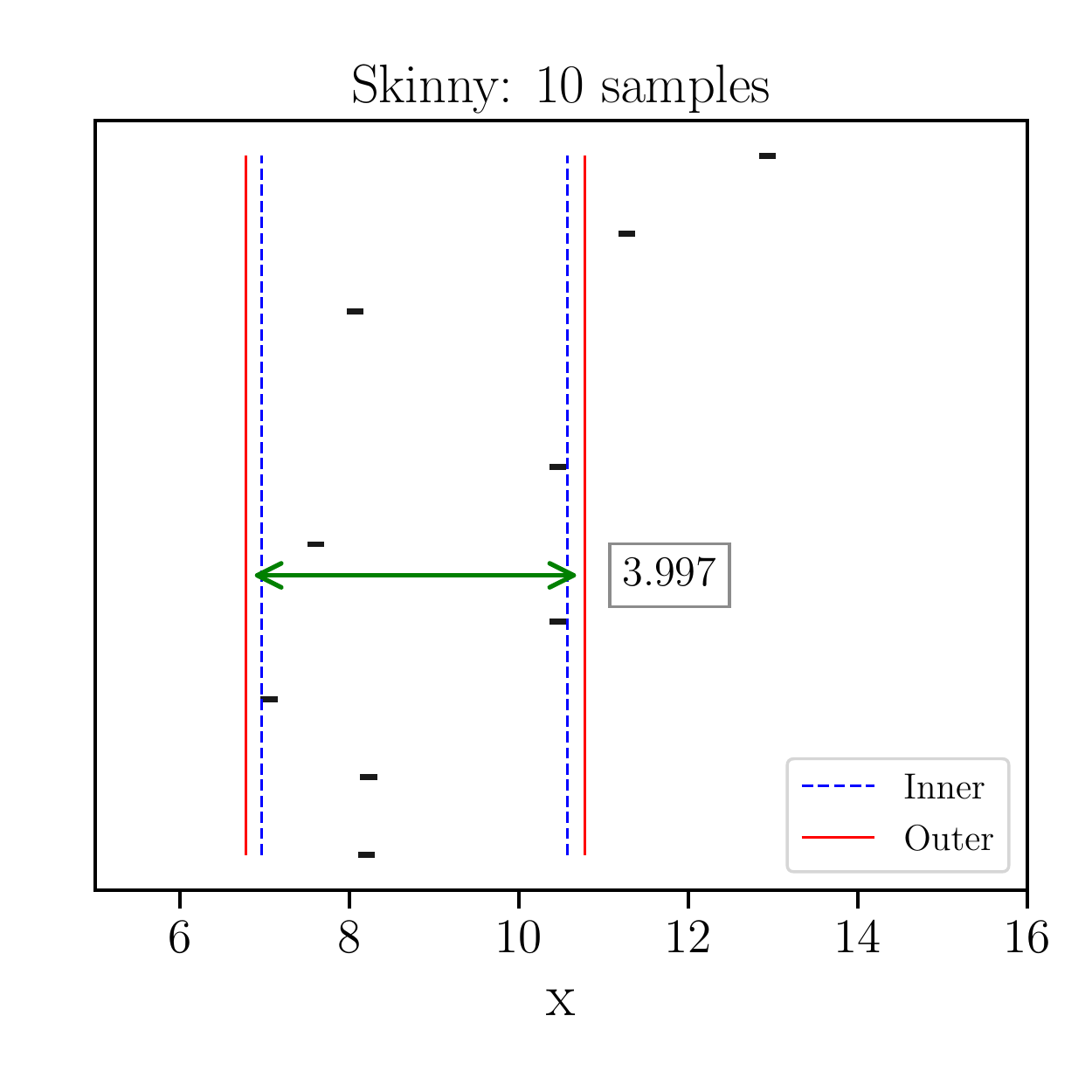}
		\caption{}
	\end{subfigure}
	\begin{subfigure}{0.325\textwidth}
		\includegraphics[width=\linewidth]{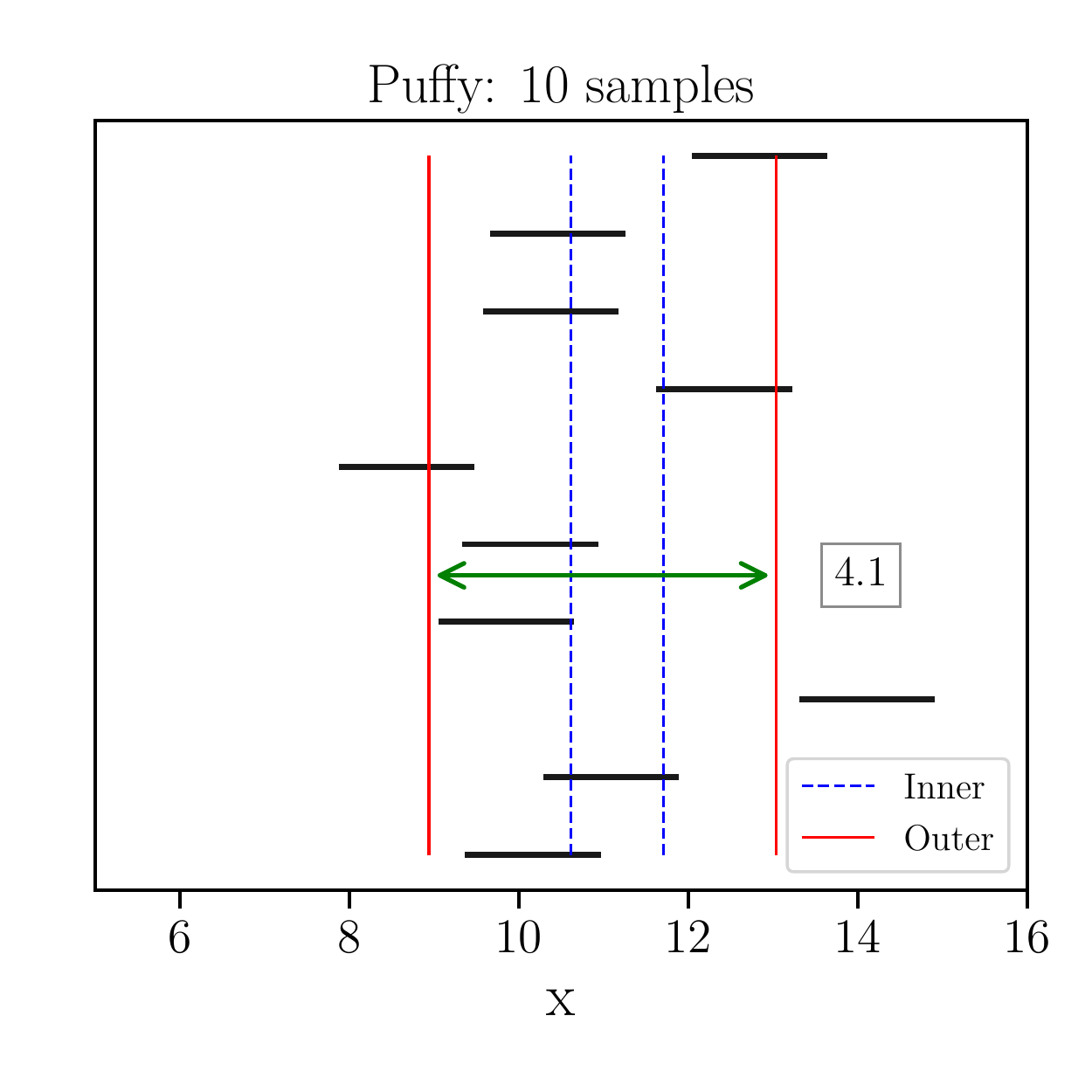}
		\caption{}
	\end{subfigure}
	\begin{subfigure}{0.325\textwidth}
		\includegraphics[width=\linewidth]{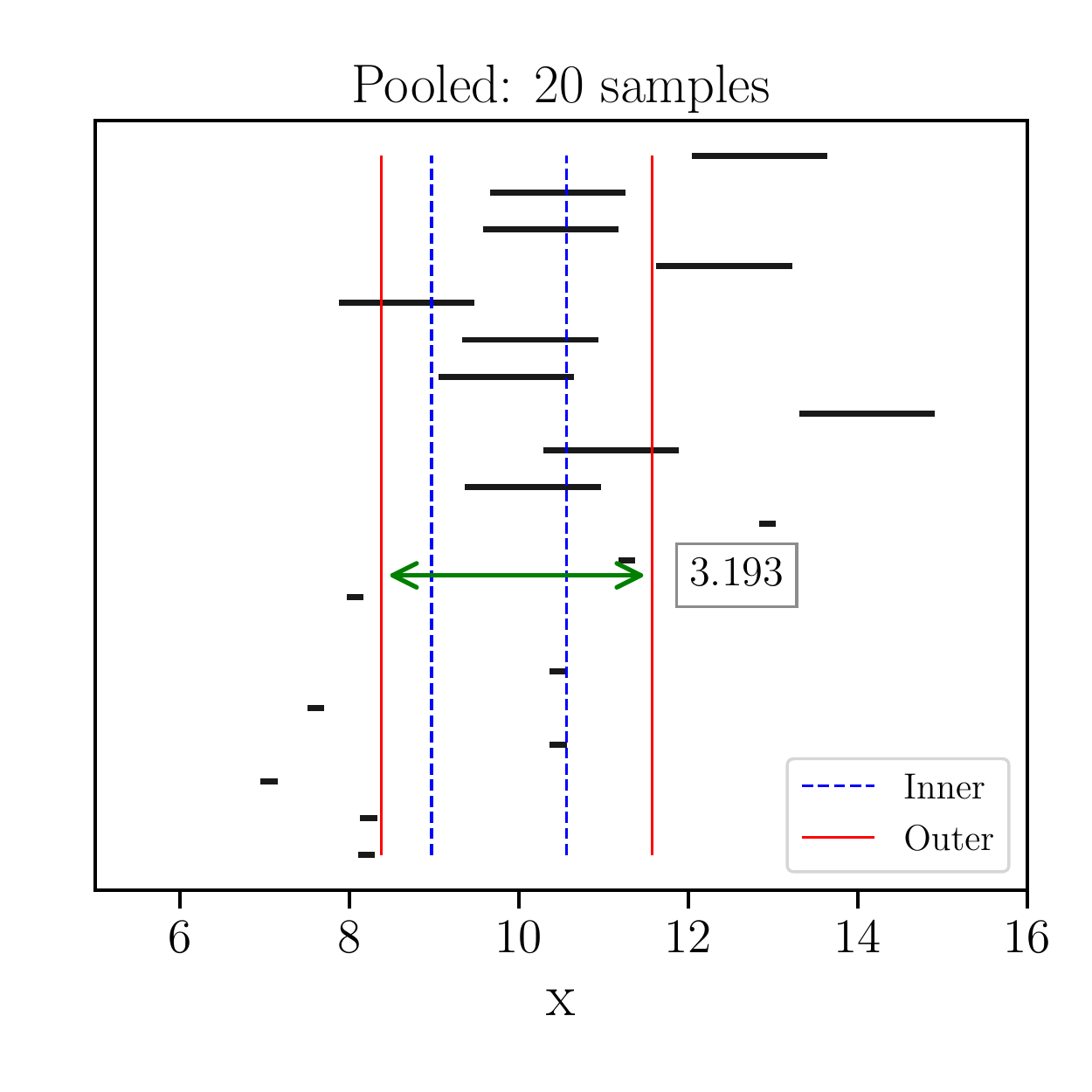}
		\caption{}
	\end{subfigure}
	\begin{subfigure}{0.325\textwidth}
		\includegraphics[width=\linewidth]{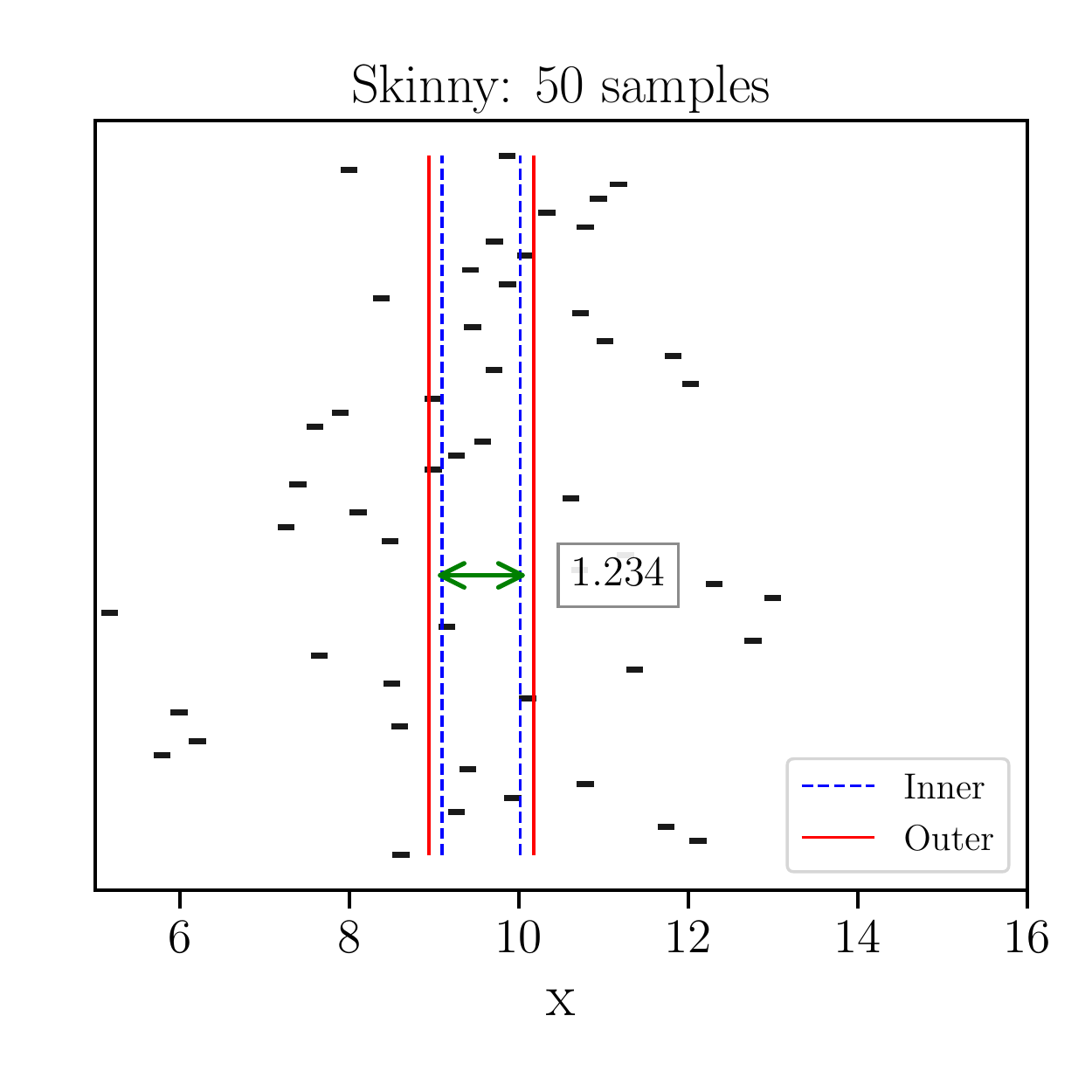}
		\caption{}
	\end{subfigure}
	\begin{subfigure}{0.325\textwidth}
		\includegraphics[width=\linewidth]{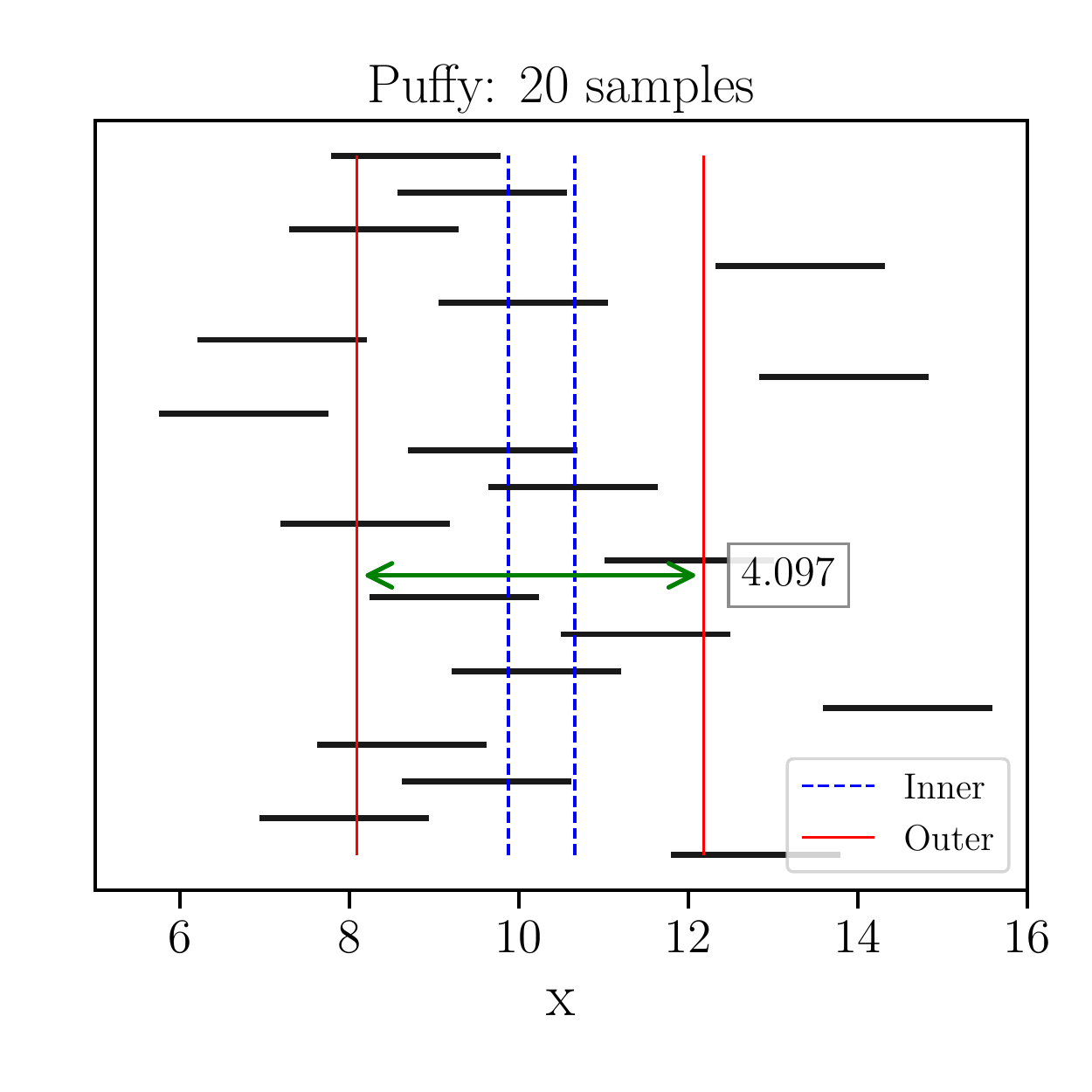}
		\caption{}
	\end{subfigure}
	\begin{subfigure}{0.325\textwidth}
		\includegraphics[width=\linewidth]{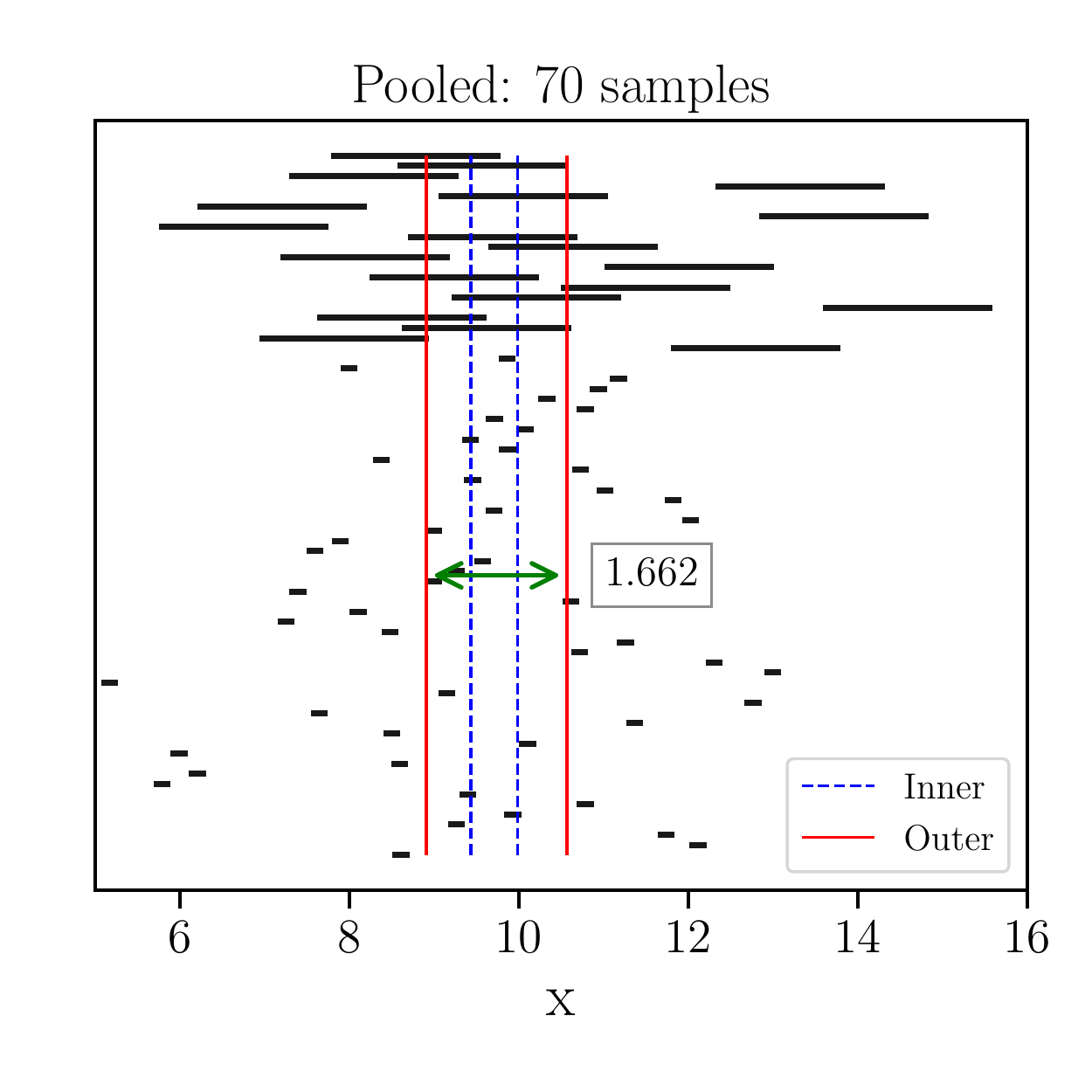}
		\caption{}
	\end{subfigure}
	\caption{Calculated inner and outer confidence intervals on the mean depending on uncertainty level and sample size. The number in the box indicates the width of the outer bounds on the confidence limits. The top example favours pooling but the bottom does not.}
	\label{fig:outerbounds}
\end{figure}

As an introductory example, Fig.~\ref{fig:outerbounds} shows calculated inner and outer bounds 
on the $95\%$ confidence interval on the mean
for different interval data sets. 
We used the $\mathcal{N}(10,4)$ distribution and $\Delta_{\text{skinny}}=0.1$ to define the skinny samples and the same distribution with $\Delta_{\text{puffy}}=f \Delta_{\text{skinny}} $ to define the puffy samples, where $f$ is an imprecision factor which defines the relative size of uncertainty around the puffy data set with respect to the skinny. In the first case with $f=8$, in the top row of Fig.~\ref{fig:outerbounds}a--\ref{fig:outerbounds}c, pooling the data seems to be advantageous because 
the width of the outer bounds on the confidence interval about the mean
with pooling is 3.193, and the width for the precise data without pooling is 3.99.
For the second case with $f=10$, in the bottom row of Fig.~\ref{fig:outerbounds}d--\ref{fig:outerbounds}f, we should avoid pooling because the skinny data set provides better statistics for the central tendency and adding 20 extra-wide intervals increases the overall uncertainty.

We carried out numerical simulations and compared  the width $D$ of the outer bounds for the skinny and pooled data sets depending on sample size, balance, and imprecision factor. For each experiment a fixed number of values were randomly chosen from the $\mathcal{N}(0,4)$ distribution and intervalized using the uniformly biased method with $\Delta_{\text{skinny}} = 0.1$ and $\Delta_{\text{puffy}} = f \Delta_{\text{skinny}}$. The experiment was considered successful if the width of the outer bounds for the pooled data set was less than that for purely precise skinny data set $D_\text{pooled} < D_\text{skinny}$. 

\begin{figure}[ht!]
	\centering
	\begin{subfigure}{0.325\textwidth}
		\includegraphics[width=\linewidth]{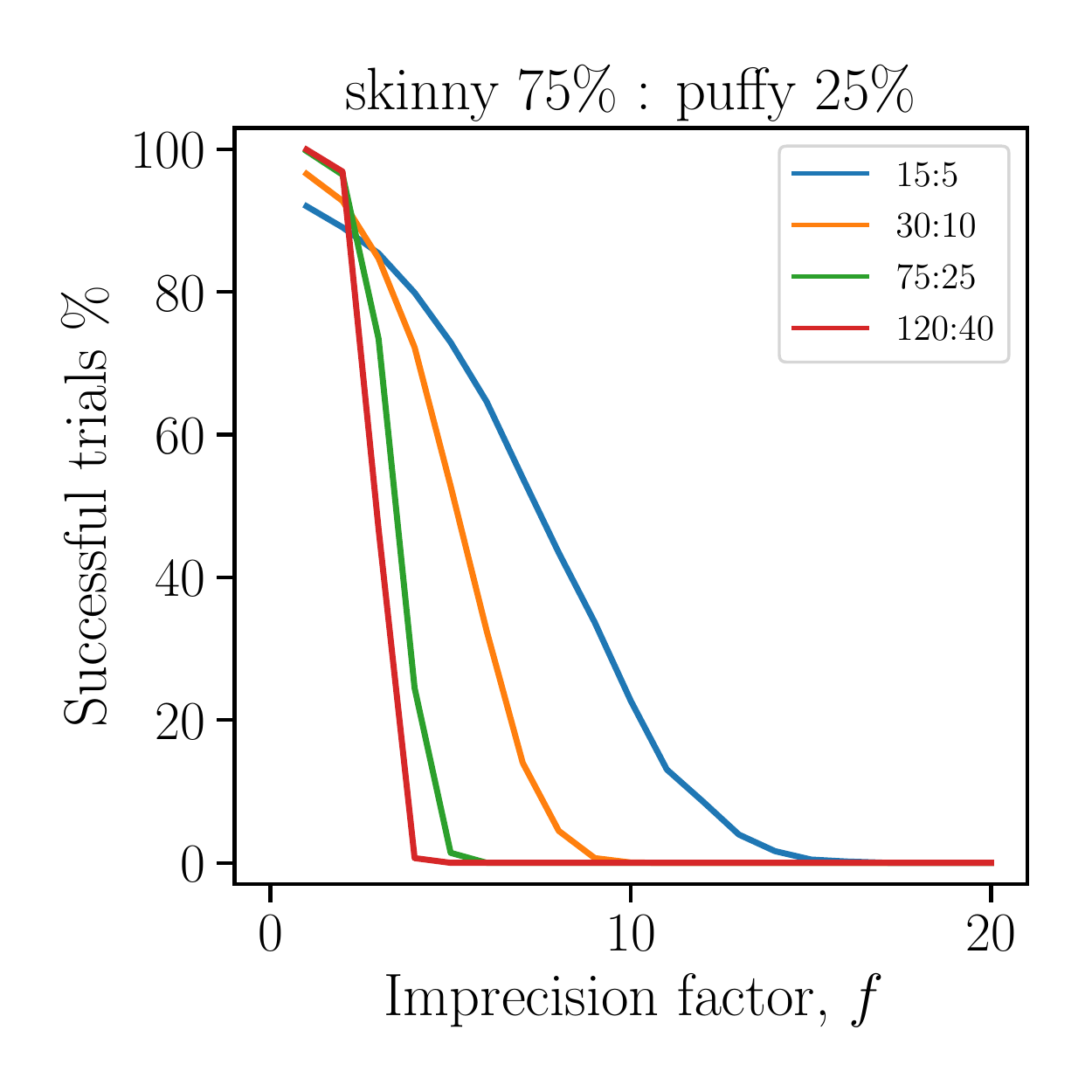}
	\caption{}
	\end{subfigure}
	\begin{subfigure}{0.325\textwidth}
		\includegraphics[width=\linewidth]{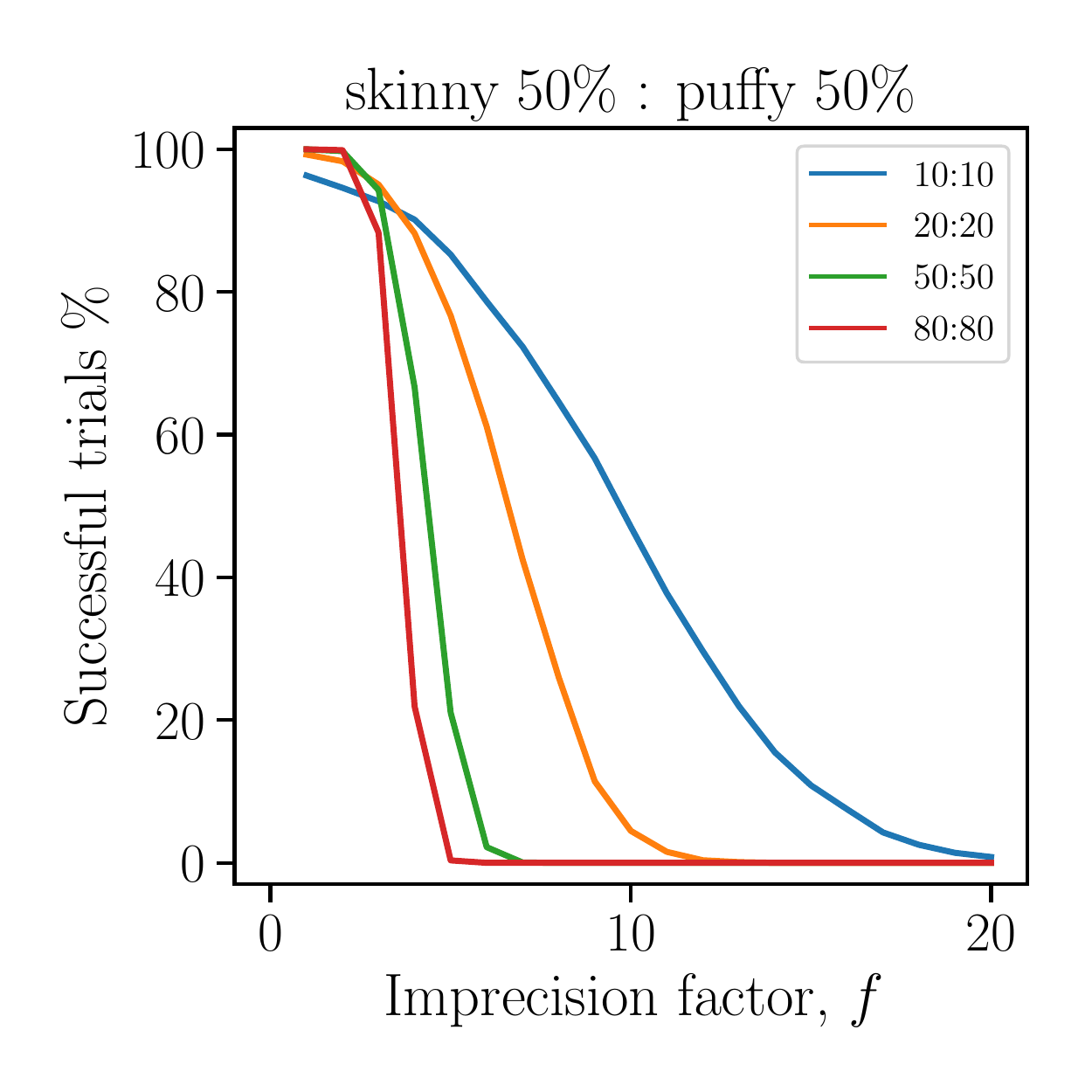}
	\caption{}
	\end{subfigure}
	\begin{subfigure}{0.325\textwidth}
		\includegraphics[width=\linewidth]{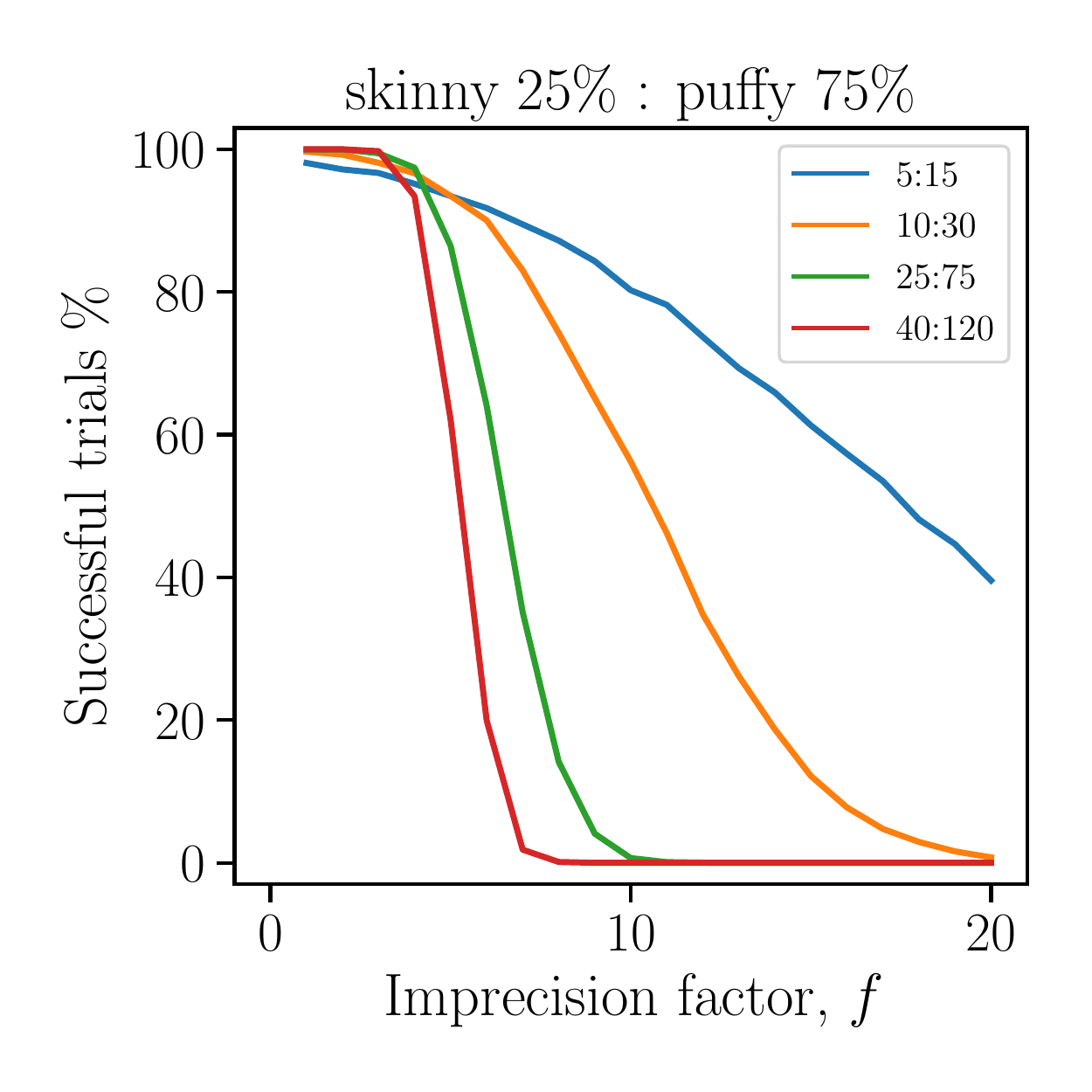}
	\caption{}
	\end{subfigure}
	\caption{Percent of trials in which the pooled data set has less uncertainty when measuring the width of the outer bounds on confidence limits for skinny versus pooled data (that combines skinny and puffy). Insets denote sample sizes for the skinny and puffy data sets.}
	\label{fig:CI_main_simulation}
\end{figure}

Fig. \ref{fig:CI_main_simulation} shows the chance that pooling data together decreases the amount of uncertainty in the output depending on imprecision factor $f$. 
Simulations were repeated $M=5000$ times for each fixed set of parameters and with \numprint{96000000} calls to $\mathcal{N}(0,4)$ in total. The three graphs correspond to different balances between skinny and puffy samples, so the resulting number of samples in the pooled set is $N_\text{pooled} = N_\text{skinny} + N_\text{puffy}$. Different sample sizes are marked with different colors. Thus, the blue curve in Fig. \ref{fig:CI_main_simulation}b indicates that adding 10 imprecise samples (constructed with $f=5$) to 10 relatively precise samples reduces the overall uncertainty $85\%$ of the time when comparing the outer confidence intervals. When the puffy samples are about 10 times ($f=10$) wider than skinny samples, the chance that the pooled data set is better decreases to $47\%$. For 80 precise intervals (red curve) pooling is always a bad idea, unless the imprecision factor of the puffy data is $f \leq 2$ which results in $99\%$ successful trials (i.e., pooling reduces uncertainty). Shifting the balance towards more precise intervals in Fig. \ref{fig:CI_main_simulation}a results in less chance of decreasing the amount of uncertainty when combining data without regard for its quality.

\section{Kolmogorov--Smirnov confidence bands}\label{KolmogorovSmirnov_bands}
\begin{figure}[ht!]
	\centering
	\begin{subfigure}{0.325\textwidth}
		\includegraphics[width=\linewidth]{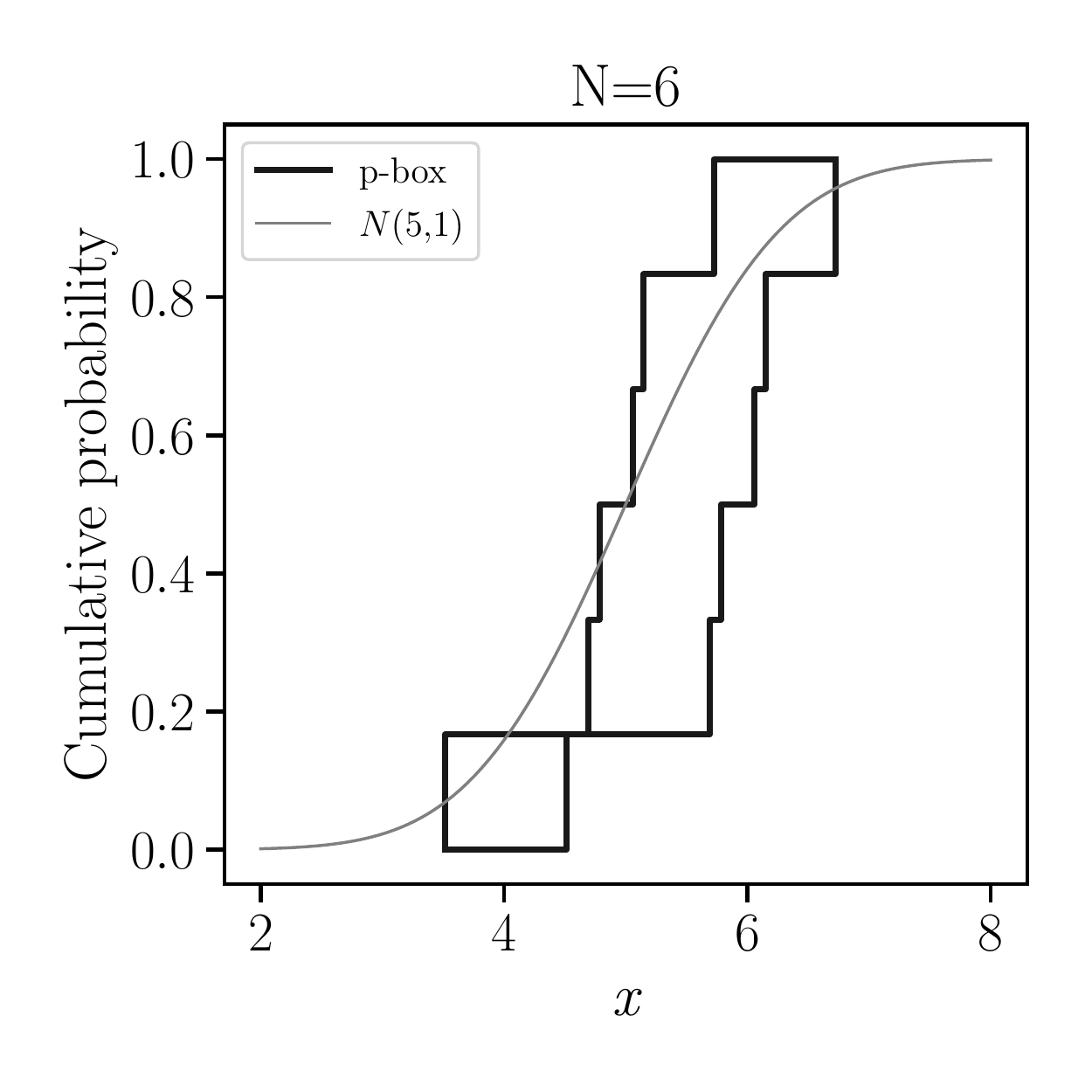}
		\caption{}
	\end{subfigure}
	\begin{subfigure}{0.325\textwidth}
		\includegraphics[width=\linewidth]{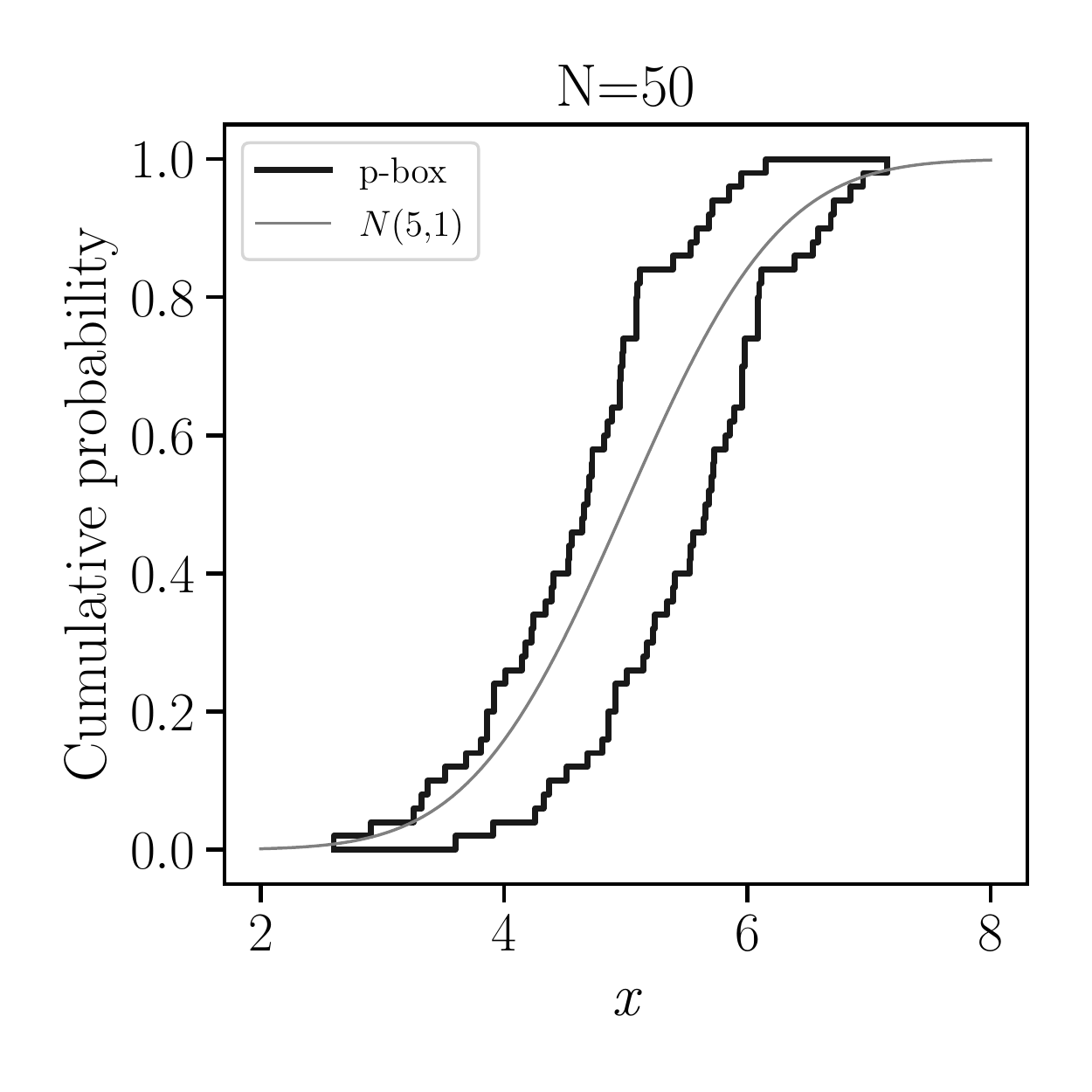}
		\caption{}
	\end{subfigure}
	\caption{Empirical distribution for data sets containing interval imprecision.}
	\label{fig:Pbox-example}
\end{figure}

Independently measured real scalar data  $X = (X_1,X_2,\dots,X_N)$ can be described with an empirical cumulative distribution function (ECDF) which is an estimate of the cumulative distribution function that generated the values in the data set. The ECDF with constant vertical step size $1/N$ is defined as
\begin{equation}
    \hat{F}_{N}(x) = N^{-1} \sum_{i=1}^{N} \text{I}(X_i \leq x),
\end{equation}
where $N$ is the sample size and $\text{I}(X_i \leq x)$ is an indicator function assuming the value 1 if $X_i \leq x$ and 0 otherwise. 
The construction of the empirical distribution function can be generalized for the case when the data set contains intervals. 
The analog of a precise empirical distribution function for a data set containing intervals is a probability box (p-box)~\cite{Ferson_p-box}. 
A p-box is a graphical representation of the uncertainty from imprecision and bounds all possible precise distribution functions that could arise from the data within the respective intervals $\interval{\underline{x}_1}{ \overline{x}_1}, \dots,  \interval{\underline{x}_N}{ \overline{x}_N}$. The bounds for the distribution of an interval data set can be computed by forming empirical distributions for the left and right endpoints respectively
so that 
$\hat{F}_{NL}(x)$ is the ECDF for the values $L = \{\underline{x}_1, \underline{x}_2, \dots, \underline{x}_N\}$
and $\hat{F}_{NR}(x)$ 
is the ECDF for for the values $R = \{\overline{x}_1, \overline{x}_2, \dots, \overline{x}_N\}$. 
The p-box is formed from these empirical distributions as
\begin{equation}
    \interval{\overline{F}(x)}{\underline{F}(x)} = \interval{\hat{F}_{NL}(x)}{\hat{F}_{NR}(x)}.
\end{equation}
For a data set of intervals $\{\interval{5.05}{6.05}$, $\interval{5.72}{6.72}$,  $\interval{5.14}{6.14}$, $\interval{3.51}{4.51}$,  $\interval{4.69}{5.69}$,  $\interval{4.78}{5.78}\}$, the corresponding p-box is presented in Fig.~\ref{fig:Pbox-example}a. These intervals were generated using the $\mathcal{N}(5,1)$ distribution and uniformly biased intervalization with imprecision $\Delta=0.5$. With more sample values, the resulting imprecise ECDF for such data would look something like Fig.~\ref{fig:Pbox-example}b.

Sampling uncertainty arises due to the fact that only some individuals from the population have been measured. 
Kolmogorov--Smirnov (K--S) confidence limits are a traditional way to depict such sampling uncertainty for empirical distributions.
K--S confidence limits, sometimes called confidence bands, are straightforwardly generalised for interval data sets \cite[][Sec.~3.5.4]{Ferson_p-box} by widening the upper and lower bounds of the imprecise ECDF by the Smirnov constant as 
\begin{equation}
        \interval{\overline{B}(x)}{\underline{B}(x)} = [\min(1,\hat{F}_{NL}(x) + D_{\alpha}(N)),
          \max(0,\hat{F}_{NR}(x) - D_{\alpha}(N))]
\end{equation}
where $D_{\alpha}(N)$ is the Smirnov critical value at significance level $\alpha = (1-c)/2$, and $c$ is the confidence level. When the number of samples $N>20$, $D_{\alpha}(N)$ can be approximated as
\begin{equation}\label{KSconstant}
    D_{\alpha}(N) = \sqrt{\dfrac{\ln(1/\alpha)}{2N}} - 0.16693N^{-1} - A(\alpha)N^{-3/2}
\end{equation}
where the first term refers to Smirnov's asymptotic formula and $A(0.1) = 0.00256$, $ A(0.05) = 0.05256$, $A(0.025) = 0.11282$ \cite[p.~113]{Miller1956}. Thus, for the $95\%$ confidence level and $N=30$ measurements, formula (\ref{KSconstant}) gives $D_{0.025}(30) = 0.2417$. For lower sample sizes $N<20$, corresponding values of $D_{\alpha}(N)$ are presented in a statistical table \cite{Miller1956}. K--S confidence intervals are non-parametric constructions, which means that they do not require any information or assumptions about the distribution other than continuity. Increasing sample size results in smaller values of $D_{\alpha}(N)$, and for large numbers of samples the \mbox{K--S} confidence limits will converge to enclose the distribution from which the values were randomly sampled.

\begin{figure}[ht!]
	\centering
	\begin{subfigure}{0.325\textwidth}
		\includegraphics[width=\linewidth]{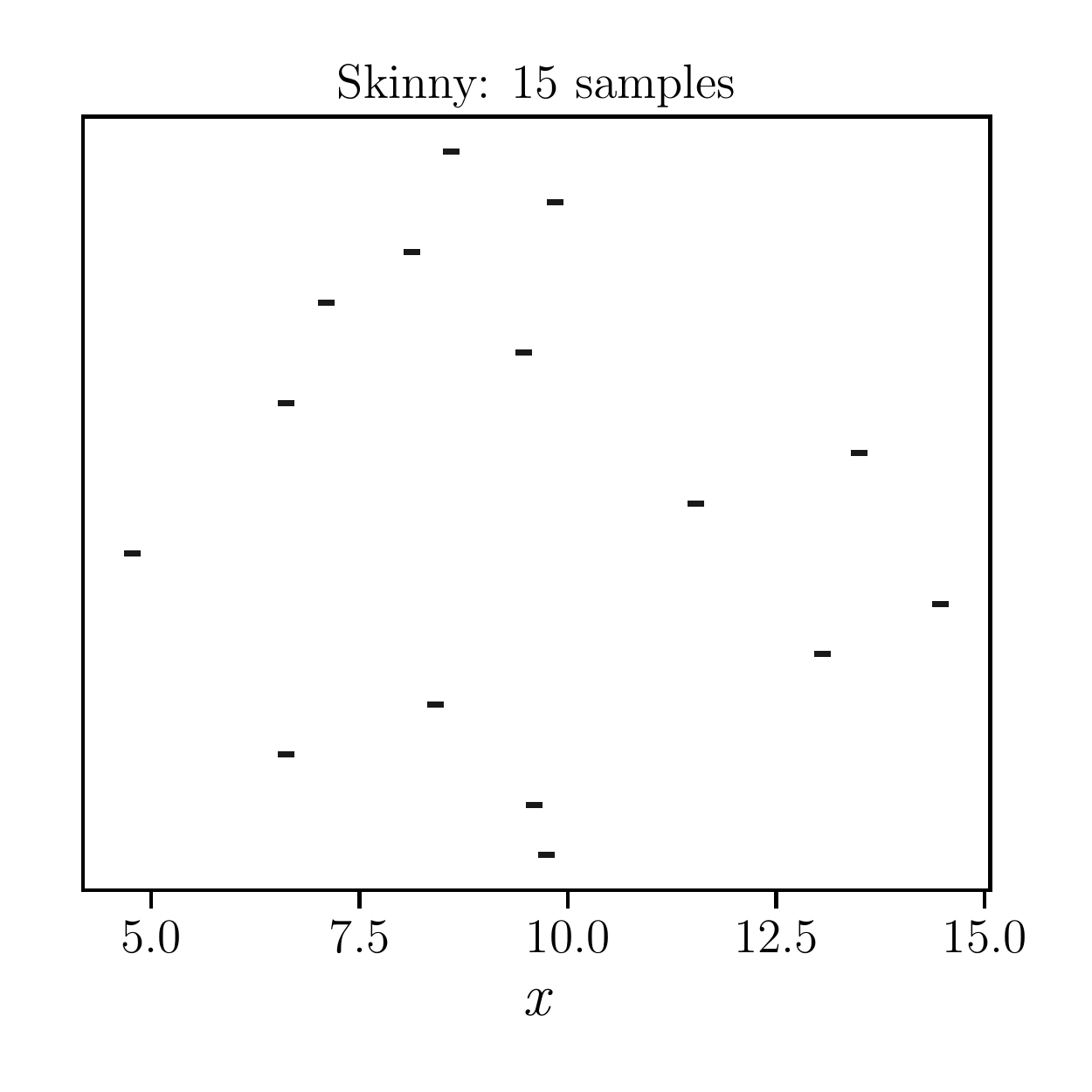}
		\caption{}
		\label{fig:KS-examplea}
	\end{subfigure}
	\begin{subfigure}{0.325\textwidth}
		\includegraphics[width=\linewidth]{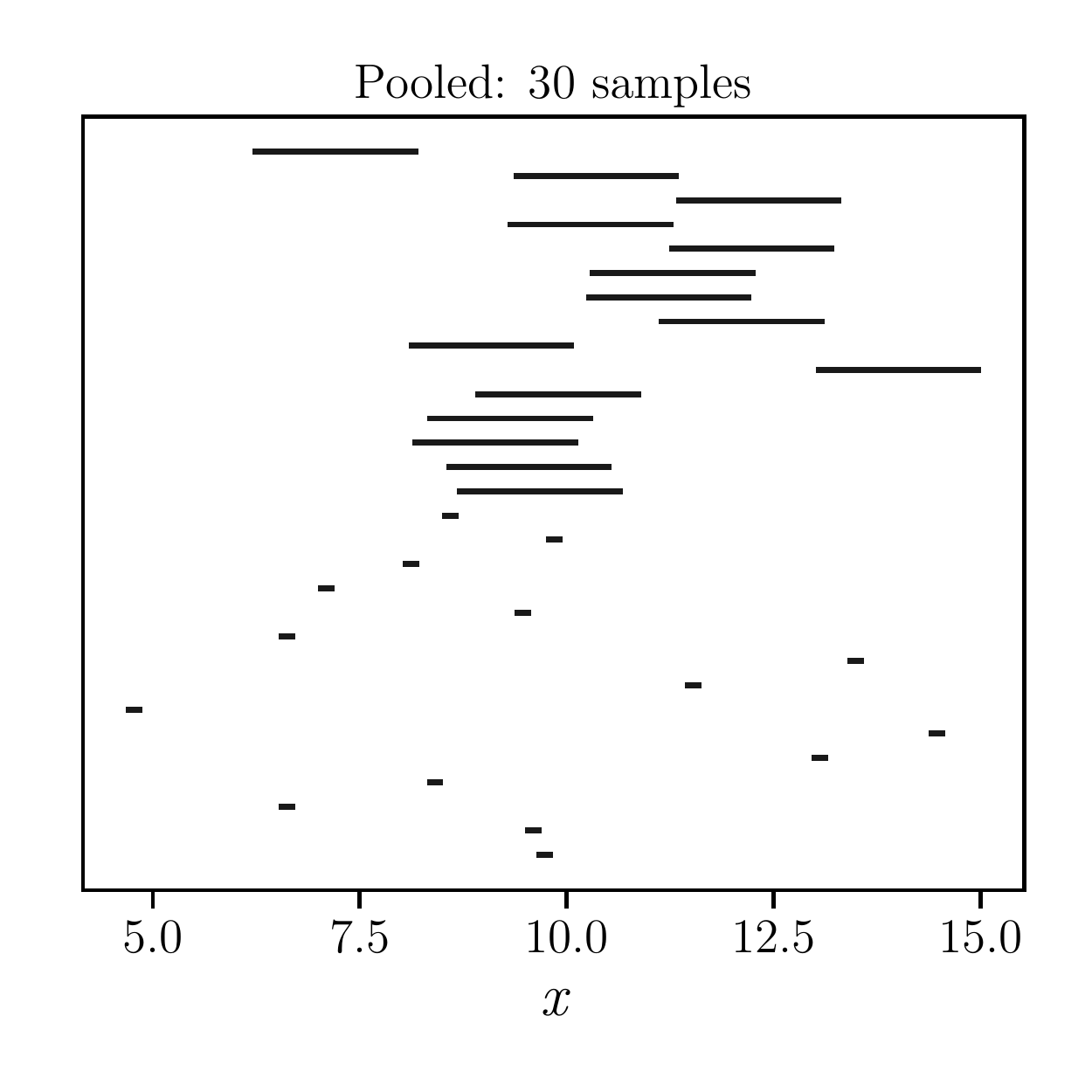}
		\caption{}
		\label{fig:KS-exampleb}
	\end{subfigure}
	\begin{subfigure}{0.325\textwidth}
		\includegraphics[width=\linewidth]{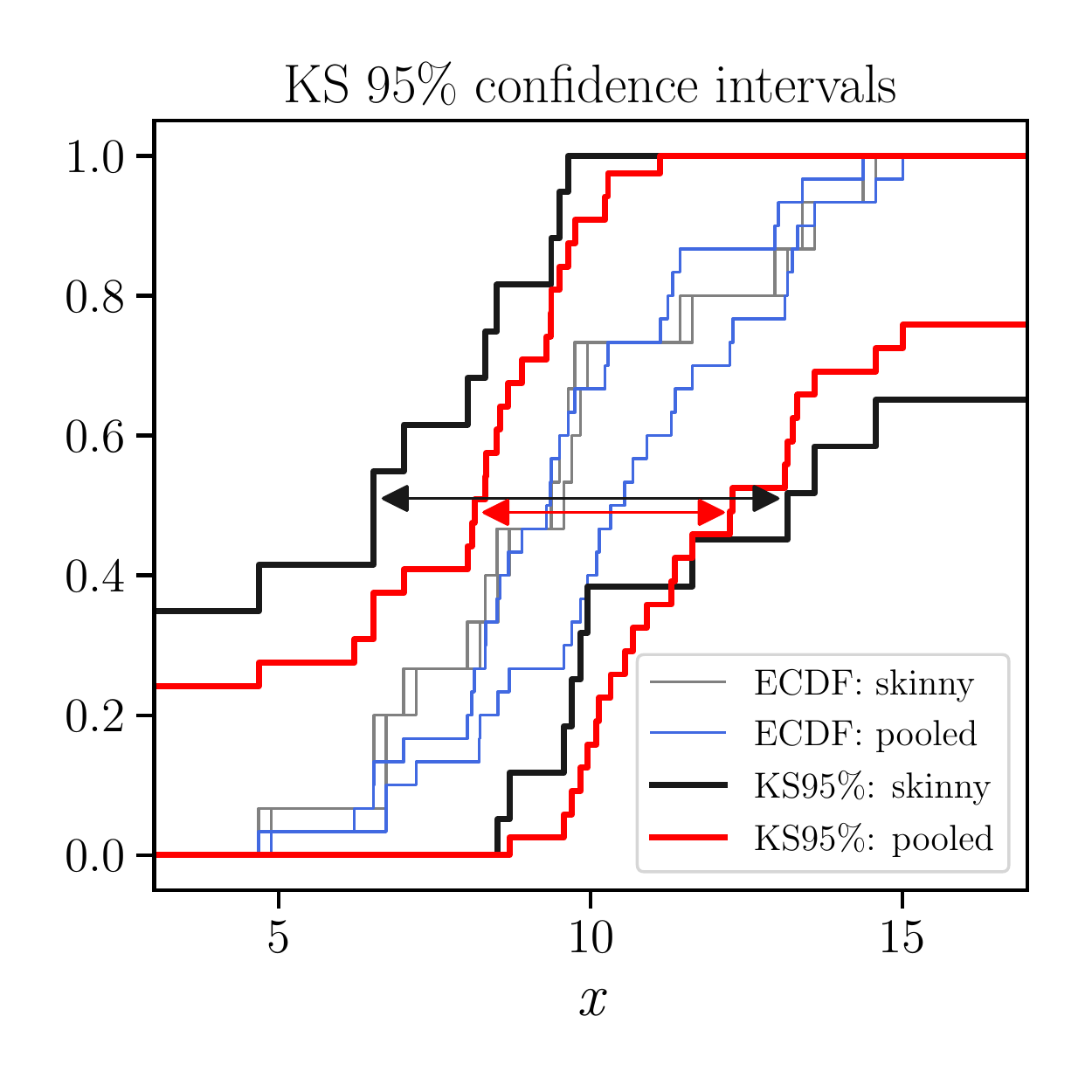}
		\caption{}
		\label{fig:KS-examplec}
	\end{subfigure}
		\begin{subfigure}{0.325\textwidth}
		\includegraphics[width=\linewidth]{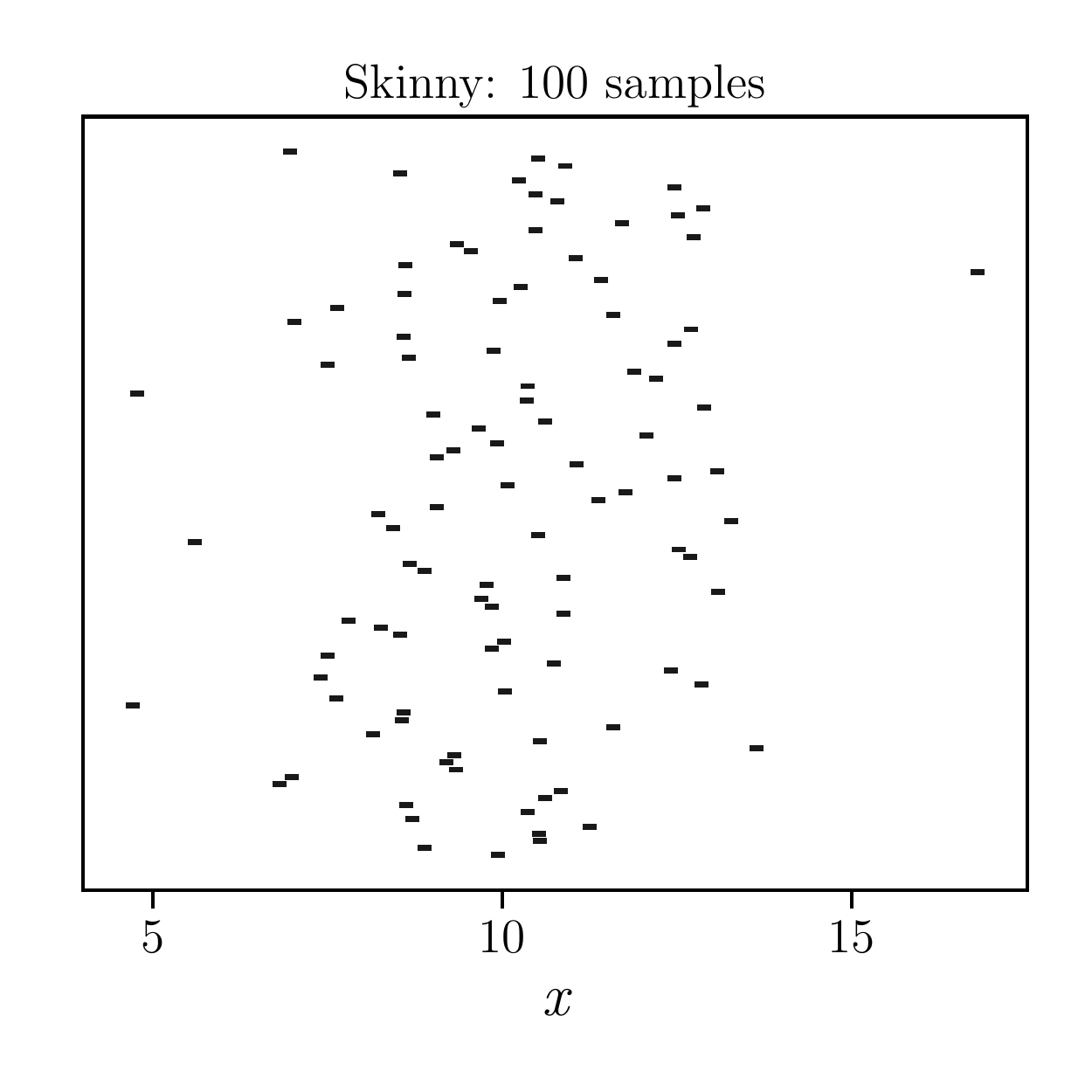}
		\caption{}
		\label{fig:KS-exampled}
	\end{subfigure}
	\begin{subfigure}{0.325\textwidth}
		\includegraphics[width=\linewidth]{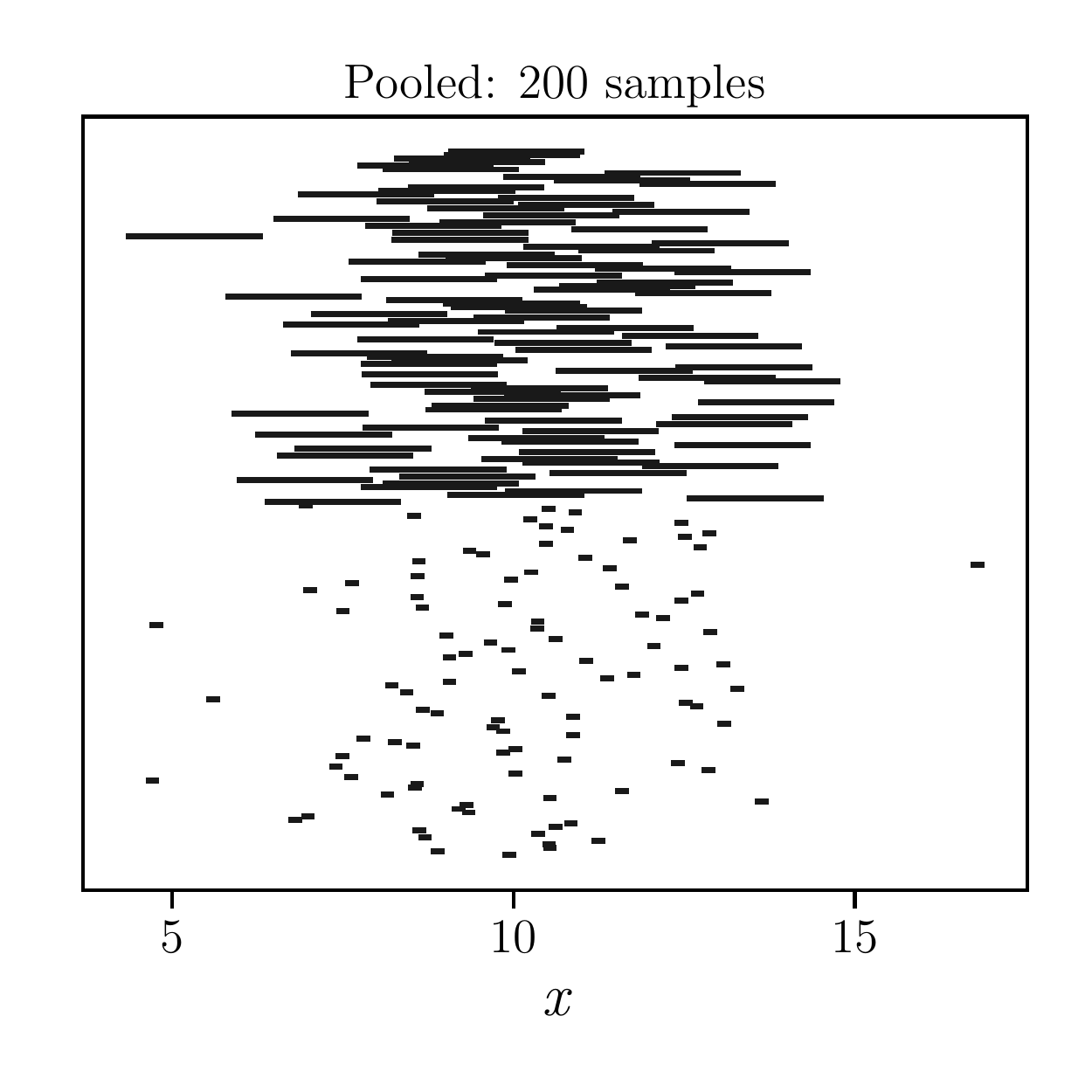}
		\caption{}
		\label{fig:KS-examplee}
	\end{subfigure}
	\begin{subfigure}{0.325\textwidth}
		\includegraphics[width=\linewidth]{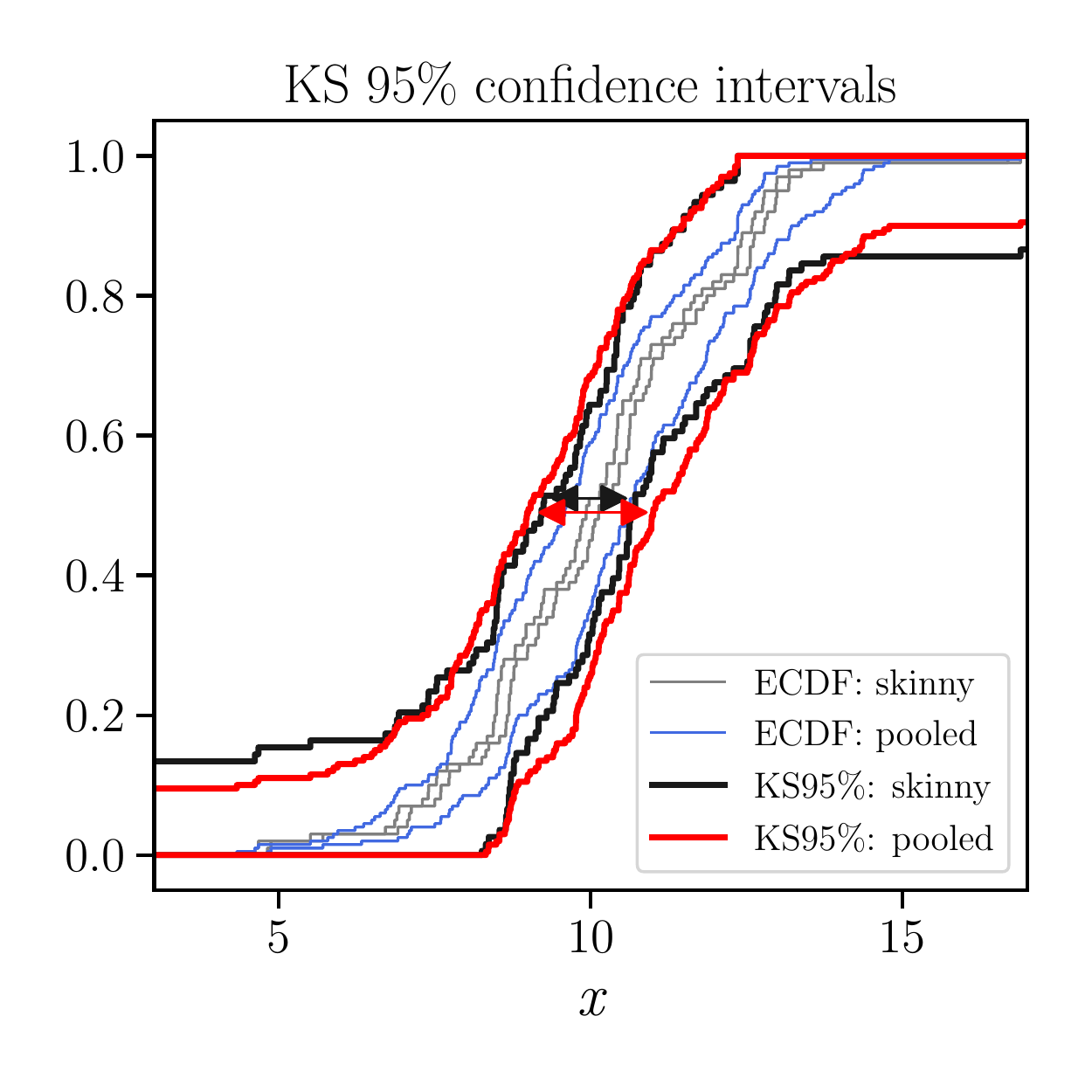}
		\caption{}
		\label{fig:KS-examplef}
	\end{subfigure}
	\caption{Kolmogorov--Smirnov $95\%$ confidence limits on the empirical distributions for two data sets.}
	\label{fig:KS-example}
\end{figure}

The Kolmogorov--Smirnov $95\%$ confidence limits on empirical distributions for skinny and pooled data sets derived from  $\mathcal{N}(10,4)$ are shown in Fig. \ref{fig:KS-example}. 
The pooled data set in Fig.~\ref{fig:KS-exampleb} contains equal numbers of puffy ($\Delta_{\text{puffy}} = 1.0$) and skinny ($\Delta_{\text{skinny}} = 0.1$) intervals for a total of $N=30$. Cumulative distributions and associated $95\%$ K--S limits are shown in Fig.~\ref{fig:KS-examplec}. It can be seen that the K--S limits for the pooled set (red lines) are much narrower despite the imprecision of half the intervals. The opposite case is presented at the bottom row, where the skinny data set has tighter K--S limits for central values.

\begin{figure}[ht!]
	\centering
	\begin{subfigure}{0.5\textwidth}
		\includegraphics[width=\linewidth]{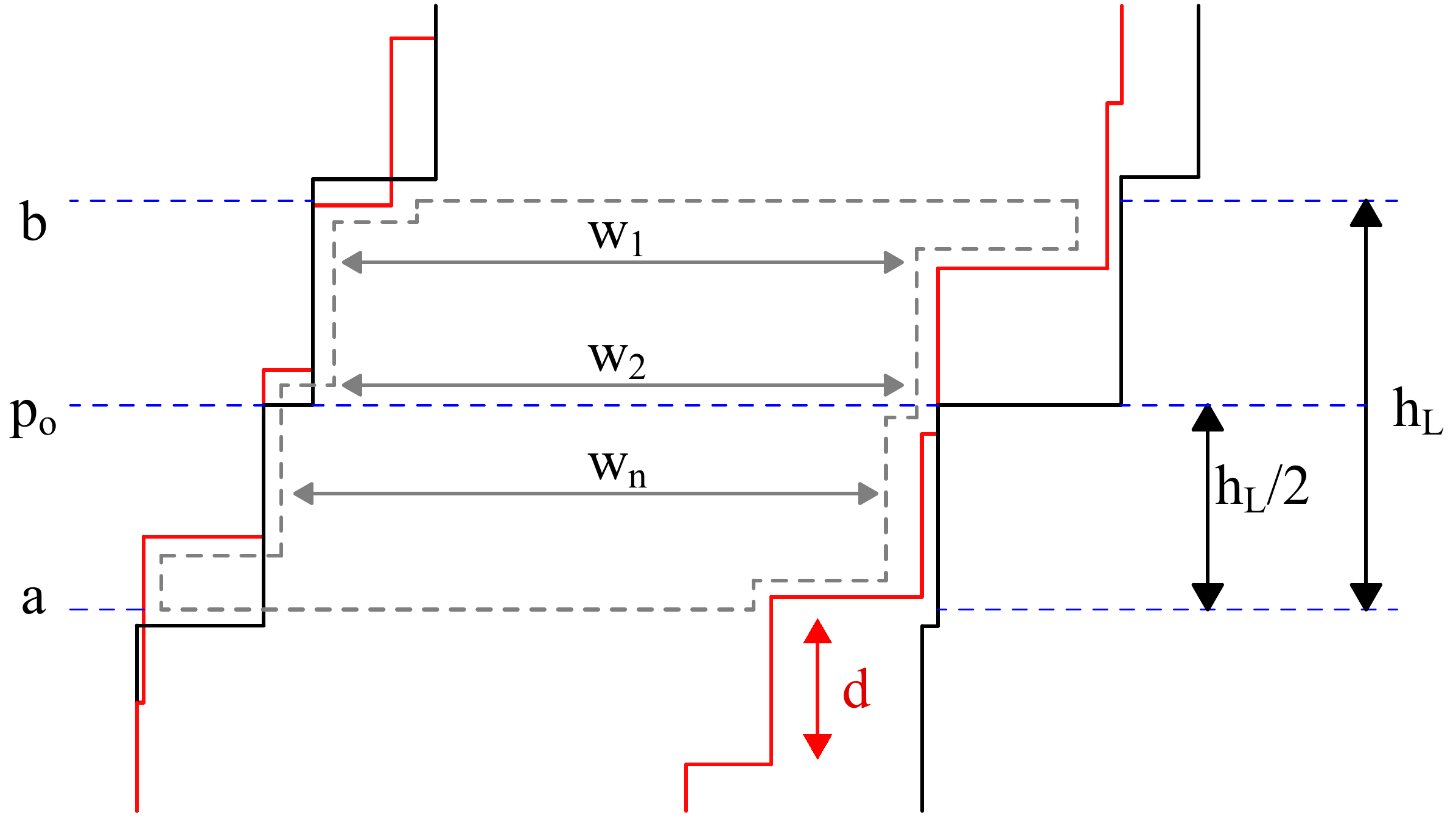}
	\end{subfigure}
	\caption{Schematic representation of a thin horizontal layer between K--S limits.}
	\label{fig:KSthinlayer}
\end{figure}

We performed numerical simulations where the horizontal width of K--S limits around the median was used as the main metric. Measuring the width at $p_0=0.5$ can be ambiguous if it falls on or close to the shoulder of the step function. This situation gives rise to some uncertainty in the measurement because the shoulder has a width itself. To avoid such situations we measure the area of a very thin horizontal layer between K--S limits at level $p_0=0.5$. The vertical height of this layer is defined as $h_{\text{L}} = 2d$, where $d$ is the height of a step of the corresponding ECDF for the pooled data. So, the thin horizontal layer is large enough (in vertical direction) to include several steps of K--S bounds instead of only one point at level $p_0=0.5$, see Fig. \ref{fig:KSthinlayer}. 
The effective horizontal width is defined as 
\begin{equation}
    H = \dfrac{1}{h_L}\sum_{i=1}^{n} w(p_i) \Delta p, \quad \Delta p = \dfrac{b-a}{n}
\end{equation}
where $w(p_i)$ is the single measurement of the width of K--S bounds within the thin layer at level $p_i = a + i \Delta p$, $a = p_0 - h_{\text{L}}/2$ and $b = p_0 + h_{\text{L}}/2$.

\begin{figure}[ht!]
	\centering
	\begin{subfigure}{0.325\textwidth}
		\includegraphics[width=\linewidth]{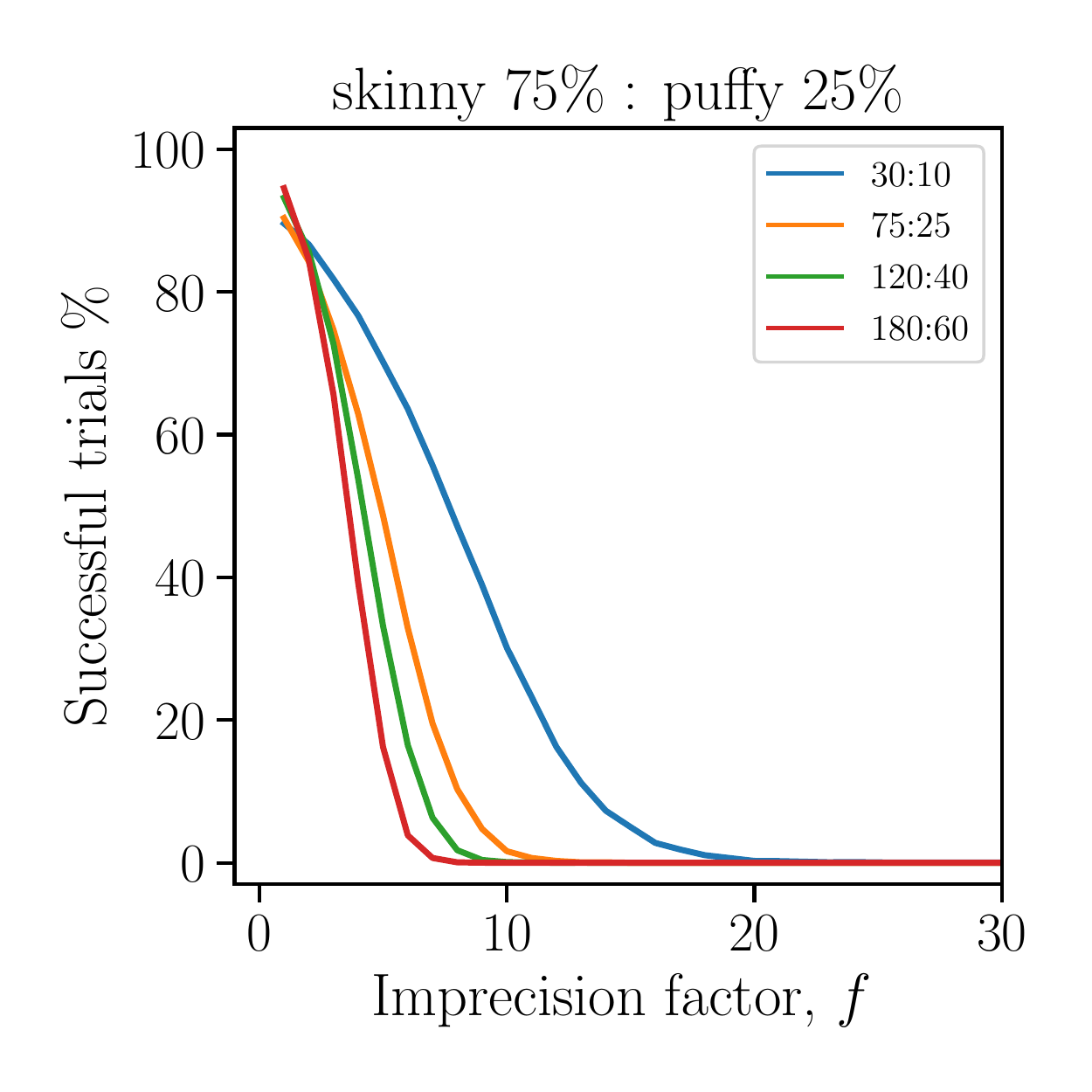}
		\caption{}
	\end{subfigure}
	\begin{subfigure}{0.325\textwidth}
		\includegraphics[width=\linewidth]{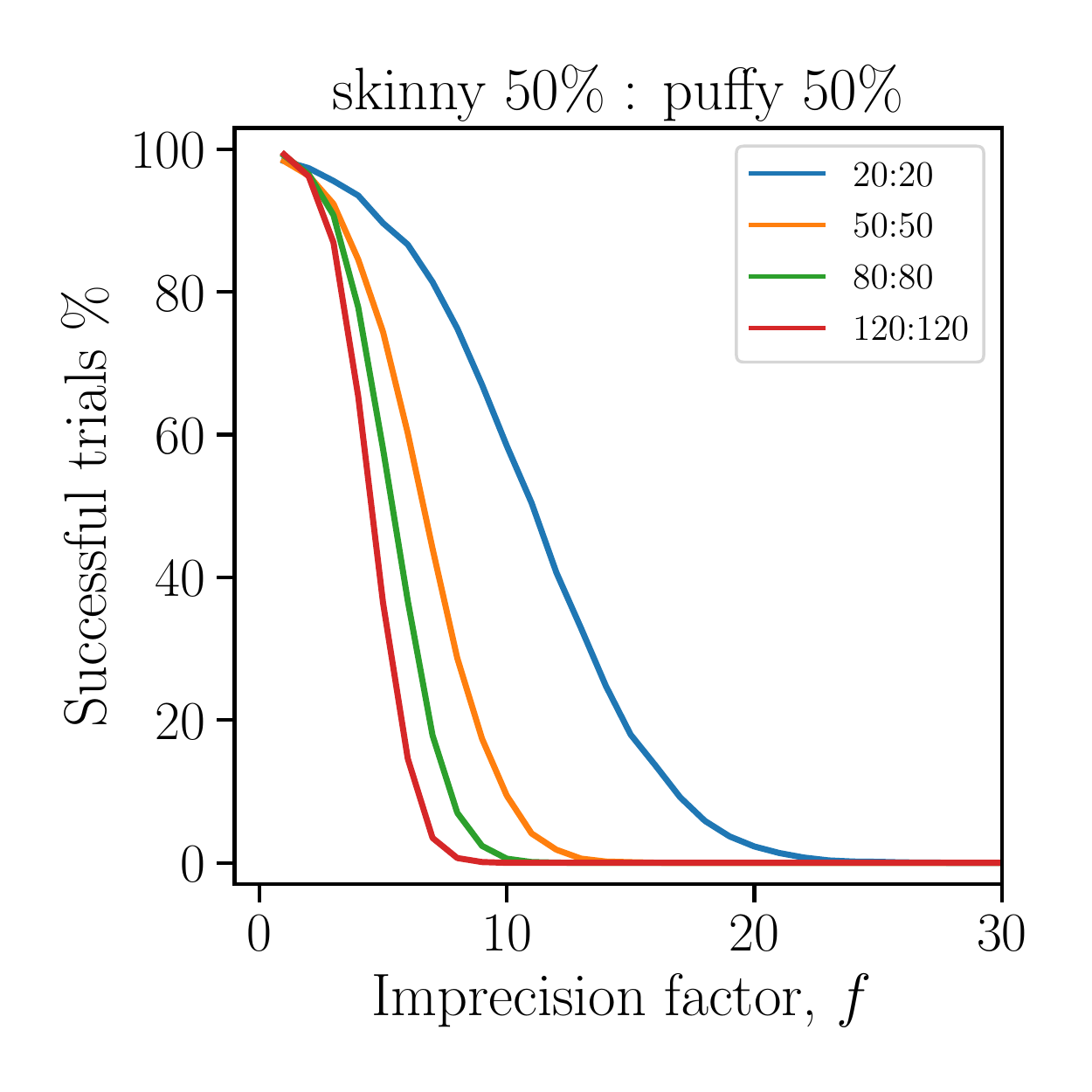}
		\caption{}
	\end{subfigure}
	\begin{subfigure}{0.325\textwidth}
		\includegraphics[width=\linewidth]{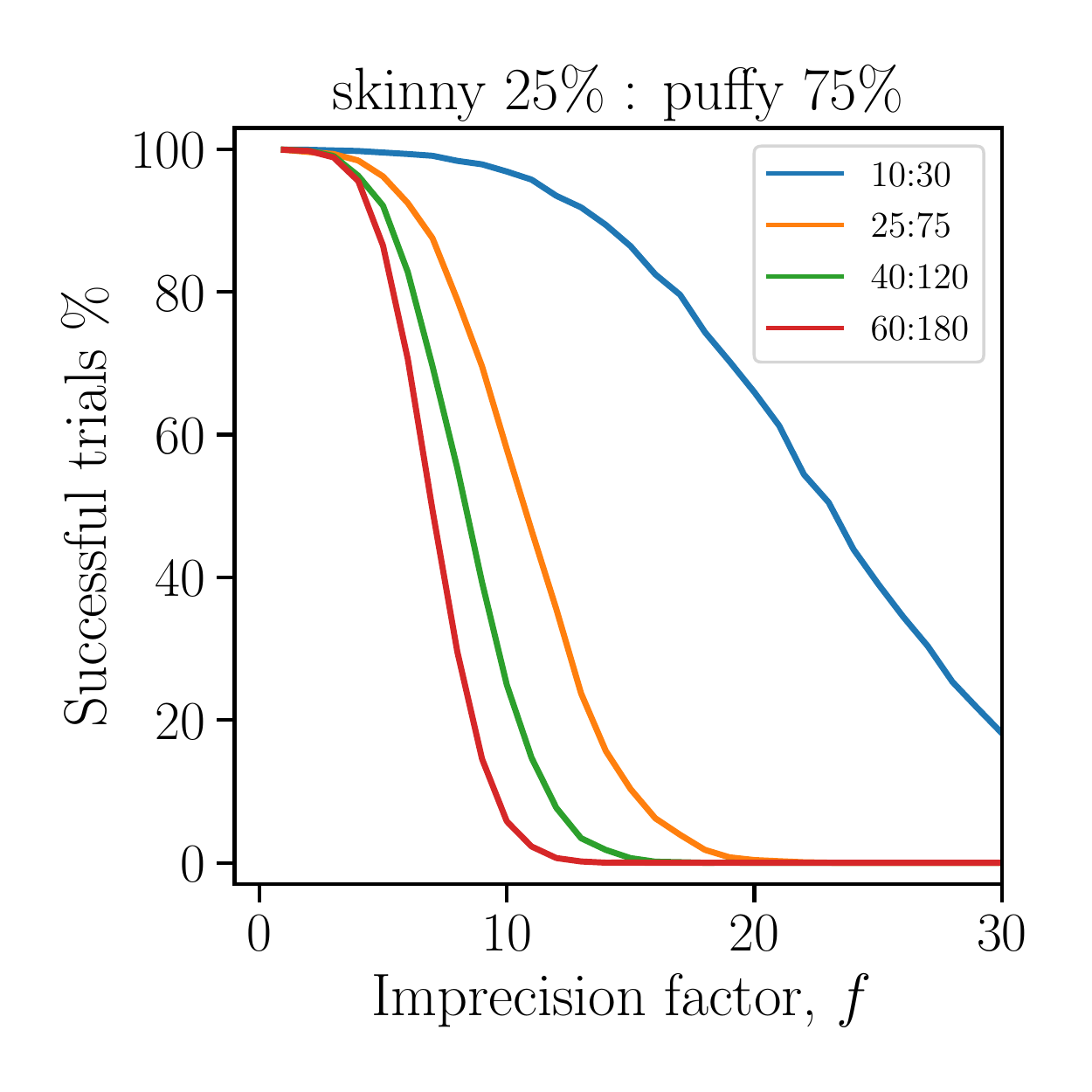}
		\caption{}
	\end{subfigure}
	\caption{Percent of trials in which the pooled data set has less uncertainty around the median for 95\% 
	K--S confidence limits as a function of imprecision factor.}
	\label{fig:KS1Dnormal}
\end{figure}

Calculations are made in the same manner as in Section~\ref{fig:CI_main_simulation} with interval-valued data randomly generated from $\mathcal{N}(0,4)$, where an experiment is considered successful if the width of K--S limits for the pooled data set is less than for the skinny, i.e., $H_{\text{pooled}} < H_{\text{skinny}}$ as it is in Fig.~\ref{fig:KS-examplec}. Fig.~\ref{fig:KS1Dnormal} presents the percent of trials in which the pooled data set has less uncertainty around the $95\%$ K--S bounds for varying imprecision factors. For each of three cases, experiments were replicated $M = \numprint{10000}$ at a fixed imprecision factor and balance.

\begin{figure}[ht!]
	\centering
	\begin{subfigure}{0.325\textwidth}
		\includegraphics[width=\linewidth]{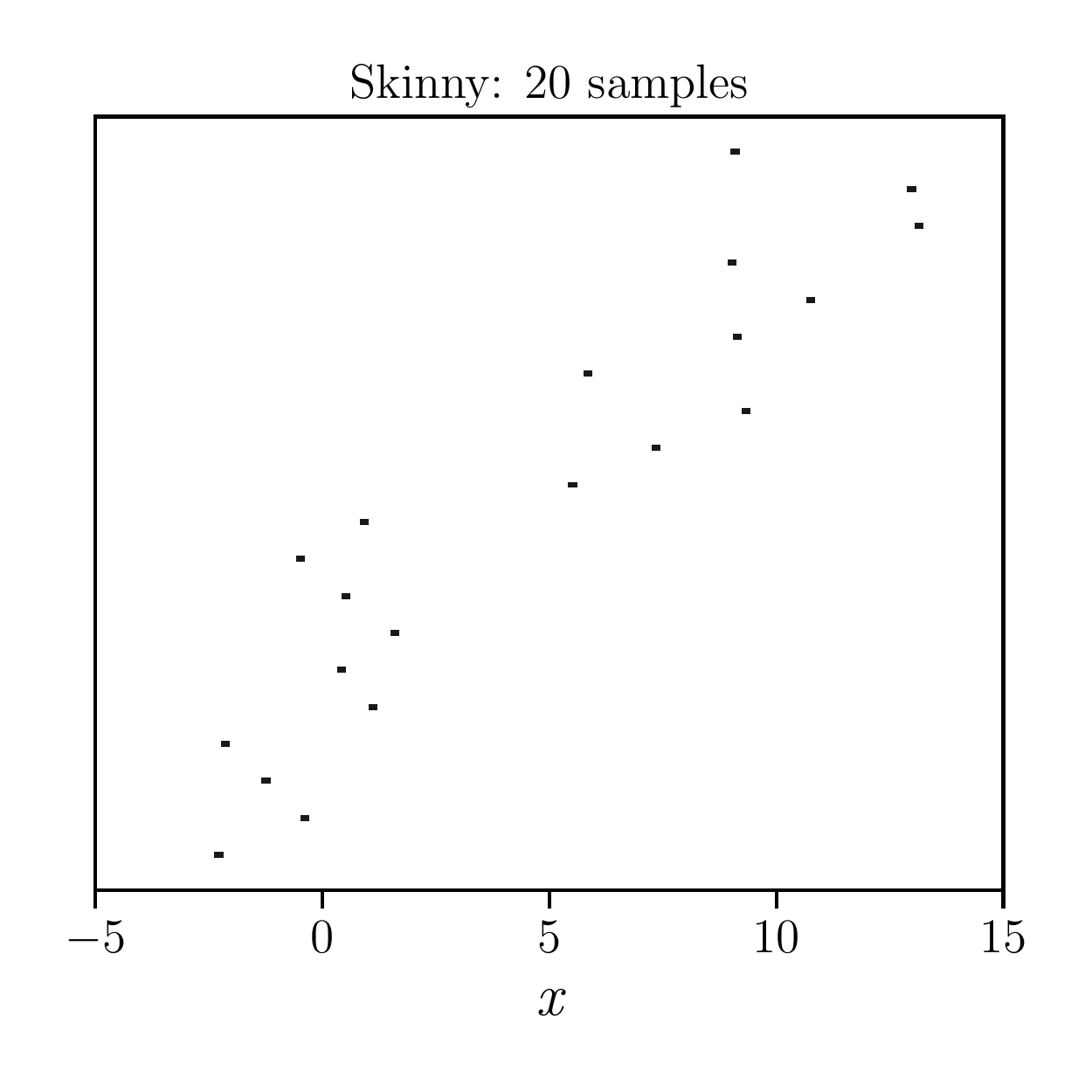}
		\caption{}
		\label{fig:KS-example2a}
	\end{subfigure}
	\begin{subfigure}{0.325\textwidth}
		\includegraphics[width=\linewidth]{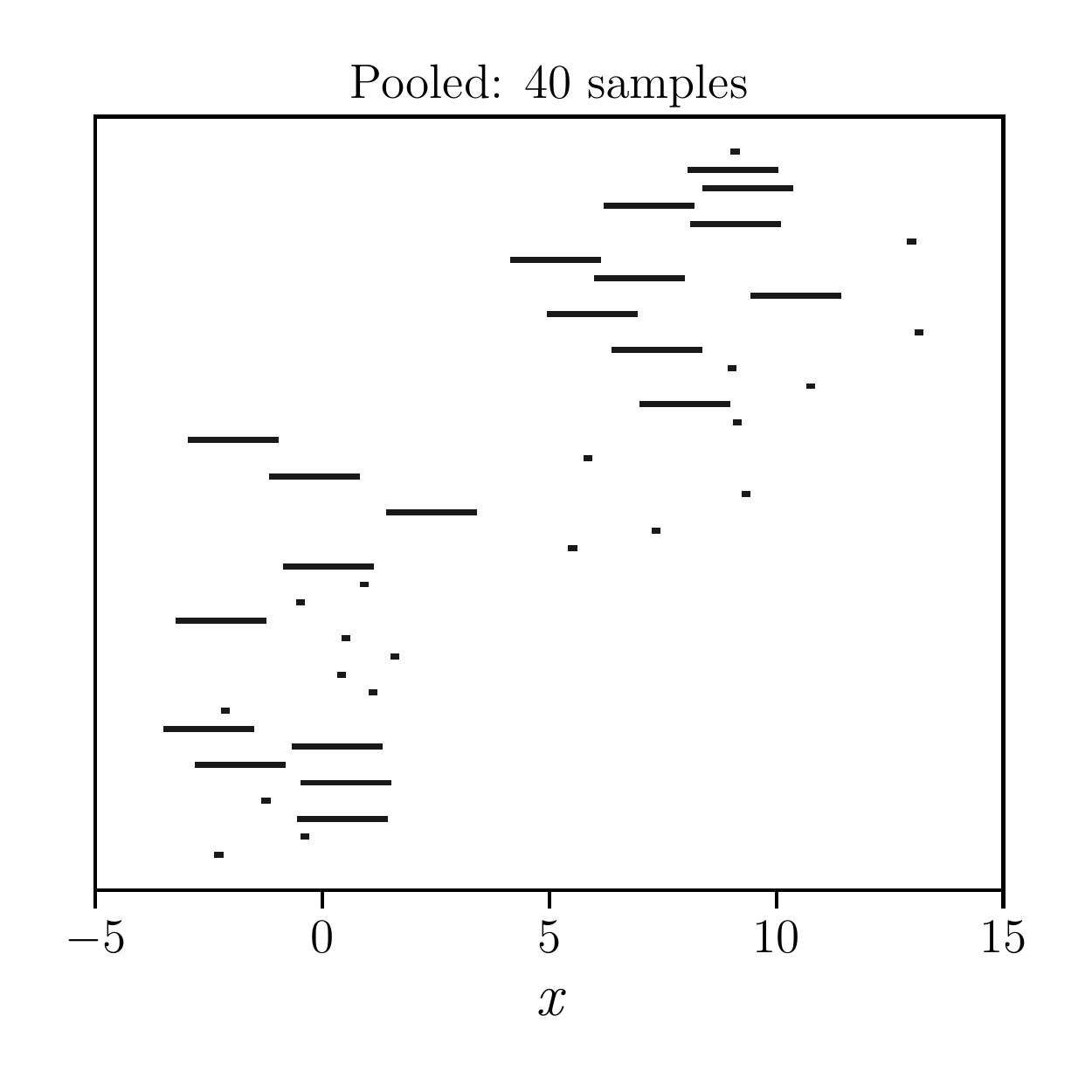}
		\caption{}
		\label{fig:KS-example2b}
	\end{subfigure}
	\begin{subfigure}{0.325\textwidth}
		\includegraphics[width=\linewidth]{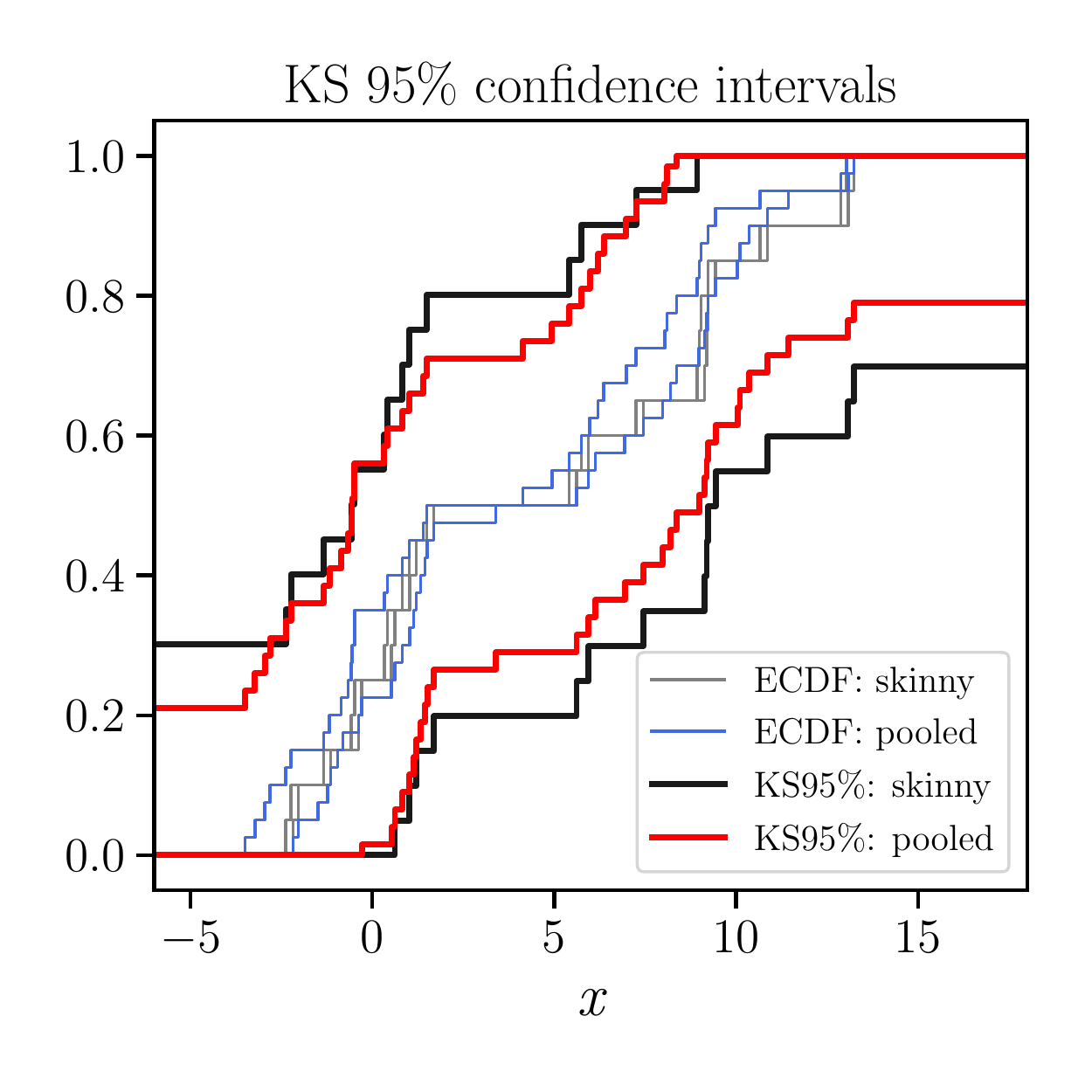}
		\caption{}
		\label{fig:KS-example2c}
	\end{subfigure}
	\caption{Kolmogorov--Smirnov $95\%$ confidence limits on the empirical distributions for two data sets derived from mixture of two normal distributions with $\mu_1 = 0, \sigma_1 = 1$ and $\mu_2 = 8, \sigma_2 = 3$.}
	\label{fig:KS-example2}
\end{figure}

\begin{figure}[ht!]
	\centering
	\begin{subfigure}{0.325\textwidth}
		\includegraphics[width=\linewidth]{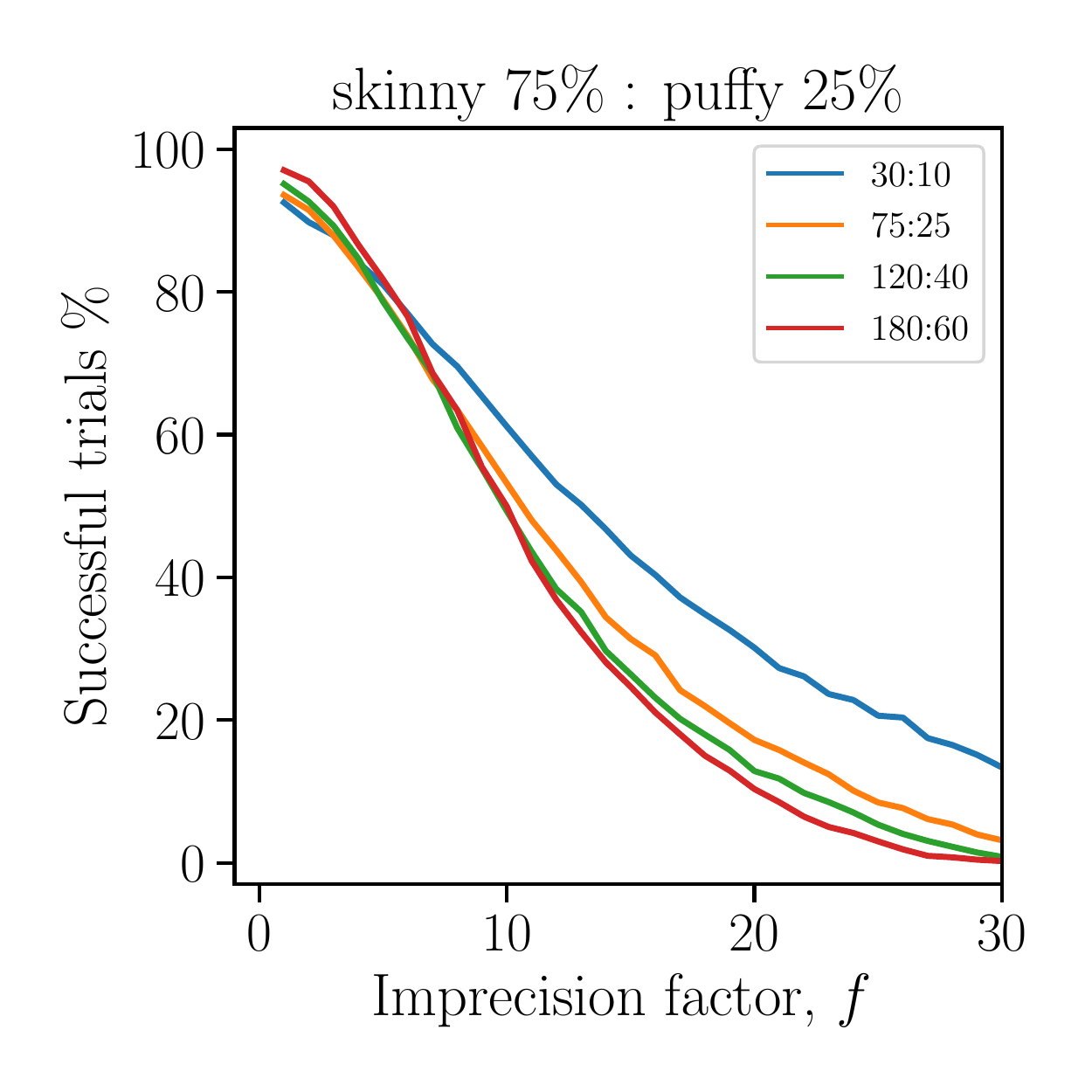}
		\caption{}
	\end{subfigure}
	\begin{subfigure}{0.325\textwidth}
		\includegraphics[width=\linewidth]{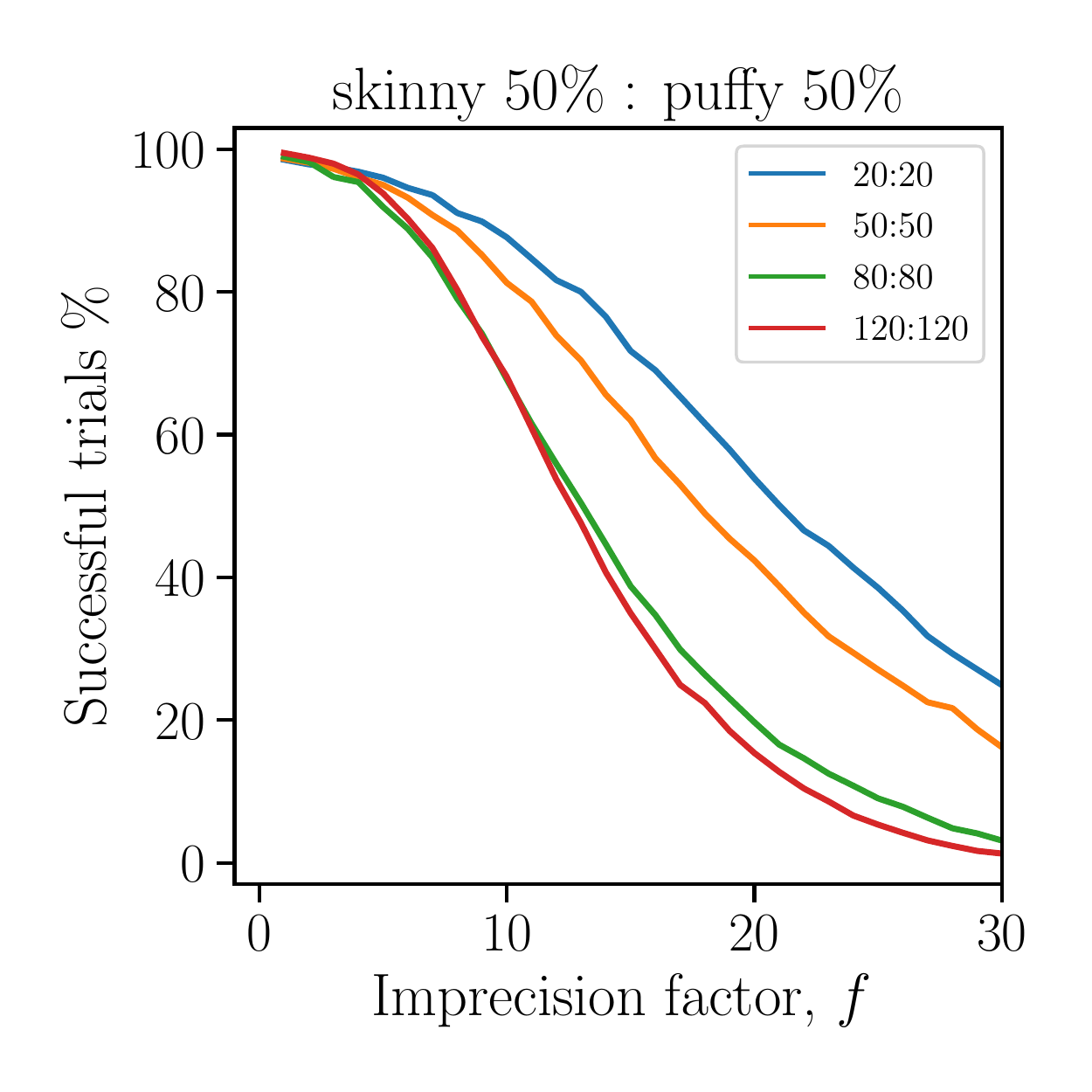}
		\caption{}
	\end{subfigure}
	\begin{subfigure}{0.325\textwidth}
		\includegraphics[width=\linewidth]{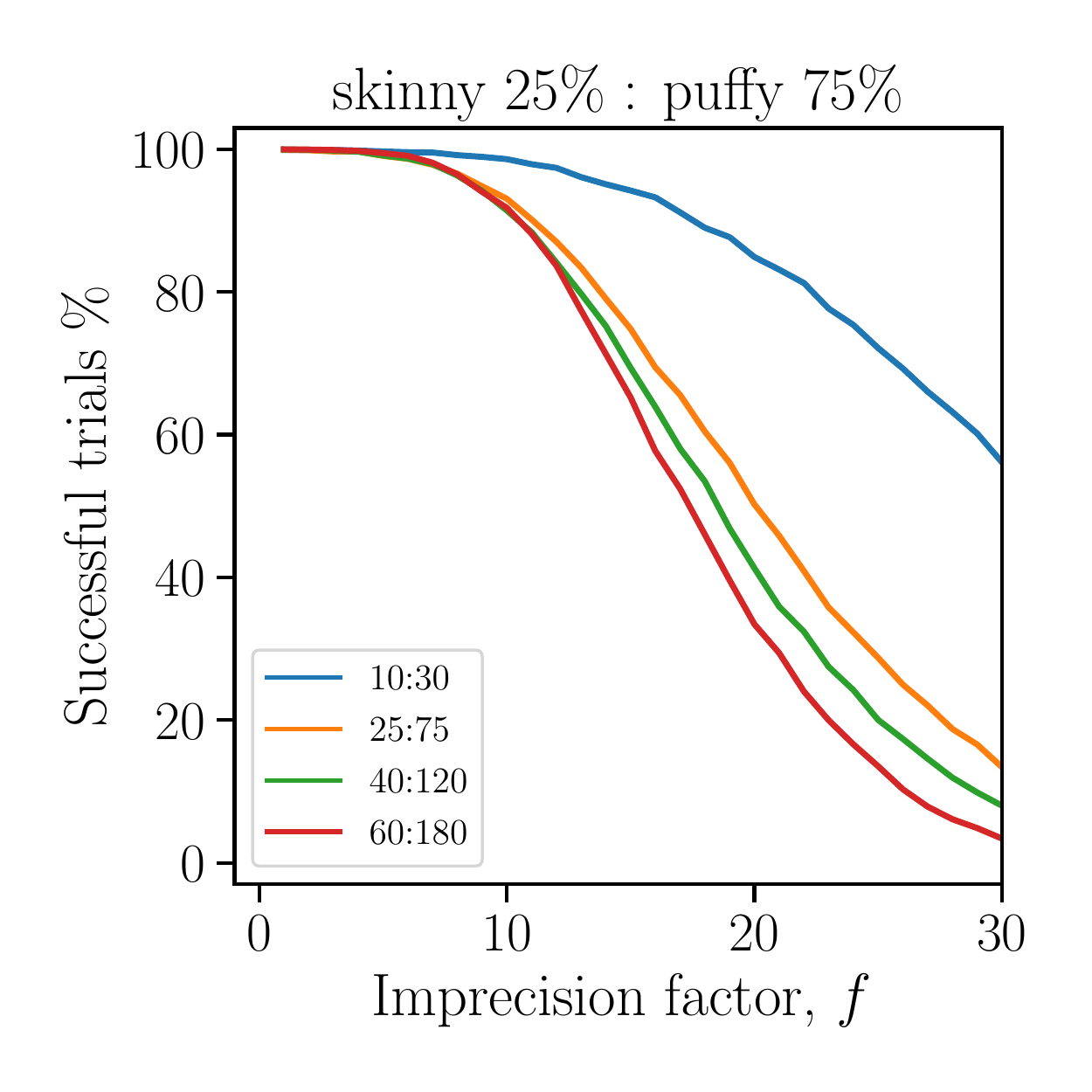}
		\caption{}
	\end{subfigure}
	\caption{Percent of trials in which the pooled data set has less uncertainty around the median for 95\% 
	K--S confidence limits as a function of imprecision factor.}
	\label{fig:KSBi-mod}
\end{figure}

As an additional example, we show the results for a mixture of two weighted normal distributions
\begin{equation}
    \mathcal{N}_1(\mu_1,\sigma_1, n_1), \quad \mathcal{N}_2(\mu_2,\sigma_2, n_2),
\end{equation}
where the weights are the sample sizes  $n_1$ and $n_2 = N - n_1$. Fig.~\ref{fig:KS-example2} shows the Kolmogorov–-Smirnov 95\% confidence limits on the empirical distributions for the
skinny and pooled data sets derived from an equal mixture of two normal distributions. The results from numerical simulations are presented in Fig.~\ref{fig:KSBi-mod}.

\begin{figure}[ht!]
	\centering
	\begin{subfigure}{0.325\textwidth}
		\includegraphics[width=\linewidth]{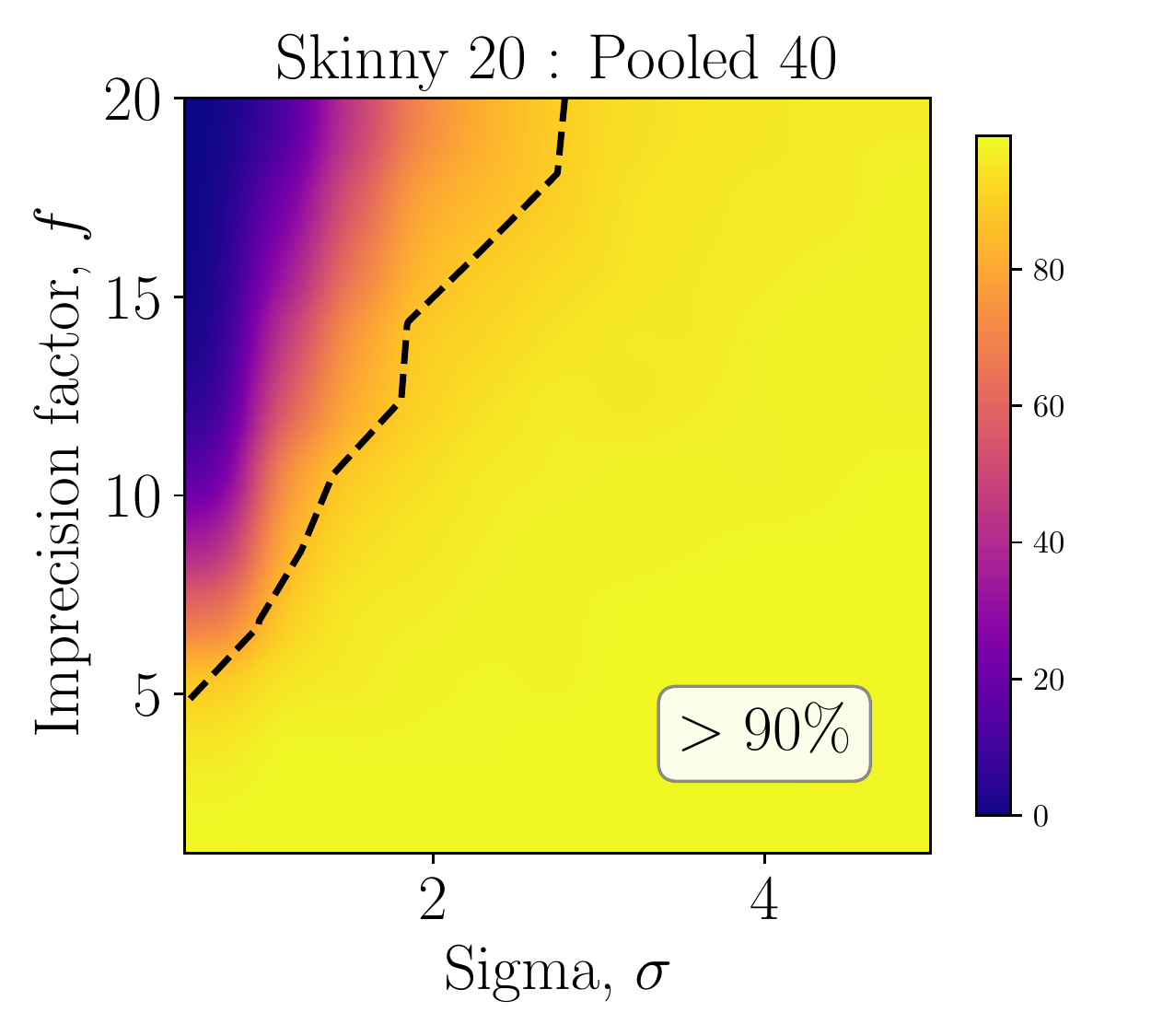}
		\caption{}
	\end{subfigure}
	\begin{subfigure}{0.325\textwidth}
		\includegraphics[width=\linewidth]{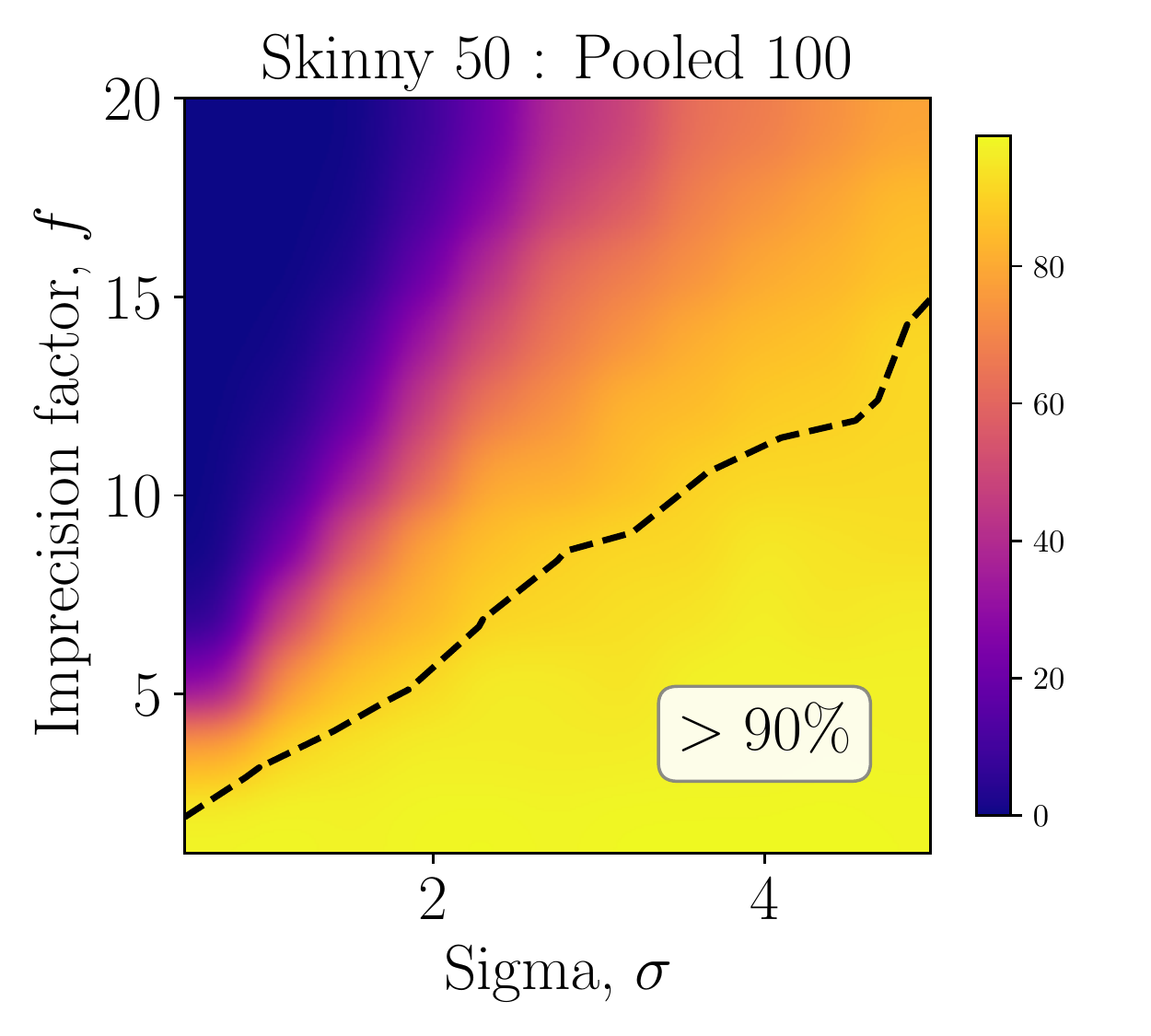}
		\caption{}
	\end{subfigure}
	\begin{subfigure}{0.325\textwidth}
		\includegraphics[width=\linewidth]{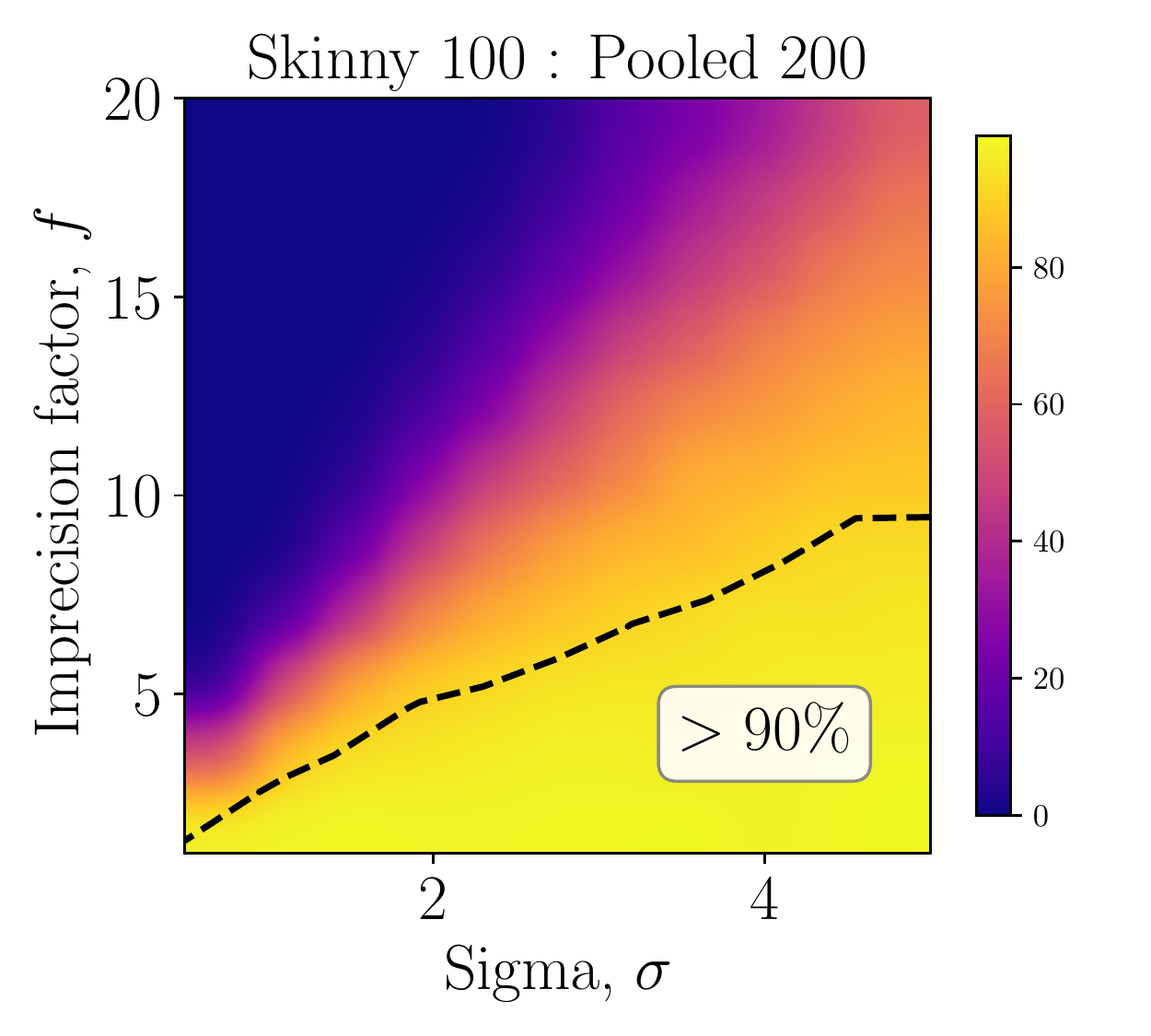}
		\caption{}
	\end{subfigure}
	\caption{Percent of trials in which the pooled data set has less uncertainty around the median for 95\% K--S confidence limits depending on imprecision factor and dispersion. Interval-valued data were randomly generated from a $\mathcal{N}(0,\sigma^2)$ distribution. The dashed line indicates $90\%$ chance of improvement from pooling.}
	\label{fig:KS95_Cnormal}
\end{figure}

\begin{figure}[ht!]
	\centering
	\begin{subfigure}{0.325\textwidth}
		\includegraphics[width=\linewidth]{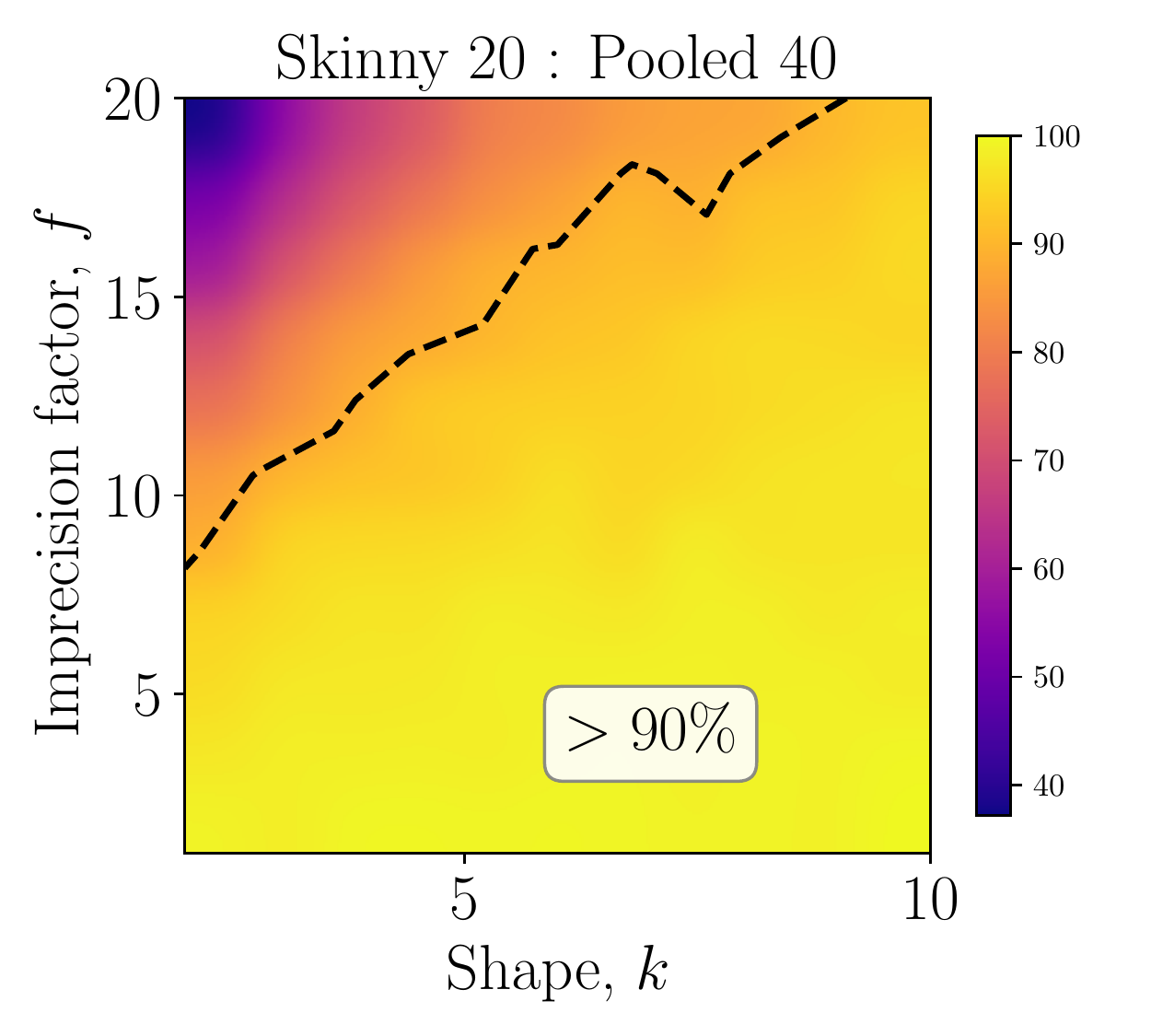}
		\caption{}
	\end{subfigure}
	\begin{subfigure}{0.325\textwidth}
		\includegraphics[width=\linewidth]{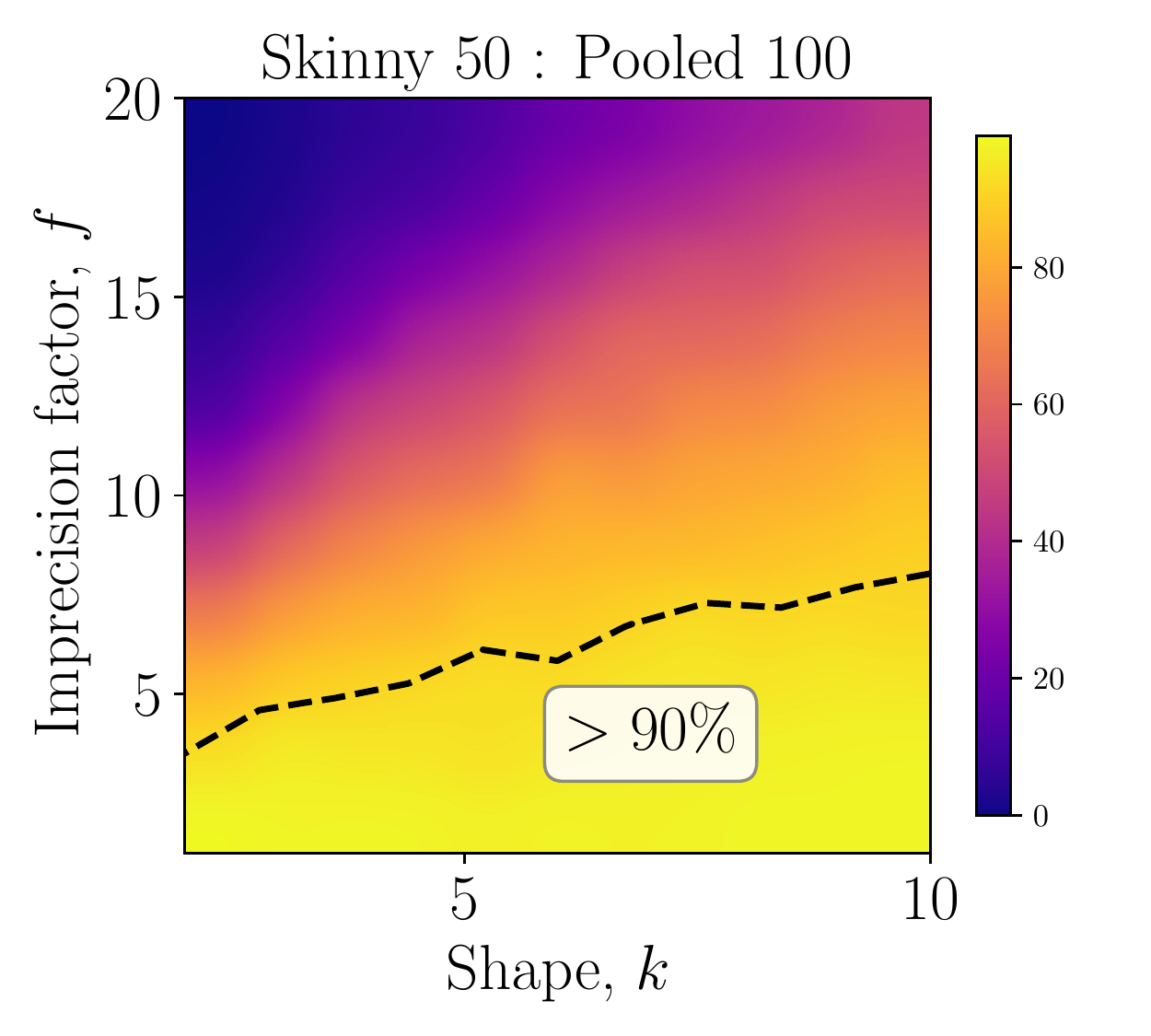}
		\caption{}
	\end{subfigure}
	\begin{subfigure}{0.325\textwidth}
		\includegraphics[width=\linewidth]{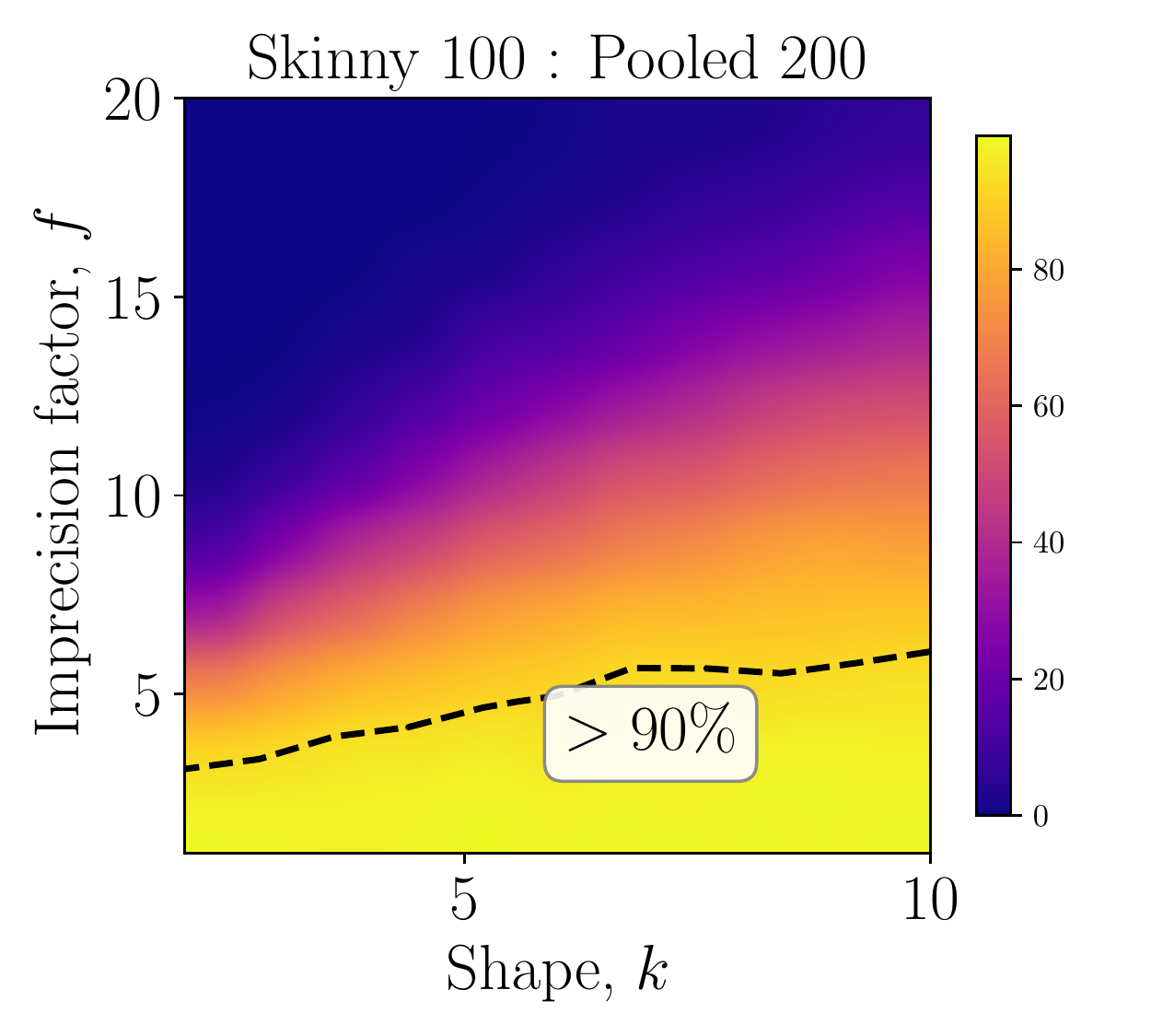}
		\caption{}
	\end{subfigure}
	\caption{Percent of trials in which the pooled data set has less uncertainty around the median for 95\% K--S confidence limits depending on imprecision factor and dispersion. Interval-valued data were randomly generated from a $\Gamma(k,2)$ distribution. Dashed line indicates $90\%$ chance of improvement from pooling.}
	\label{fig:KS95_Cgamma}
\end{figure}

Up to now our simulations kept fixed all the parameters except the imprecision factor $f$. In Fig.~\ref{fig:KS95_Cnormal} and Fig.~\ref{fig:KS95_Cgamma} we vary both the imprecision factor and the dispersion of the underlying distribution. Interval-valued data were randomly
generated from the normal distribution with $\mu=0$ and the gamma distribution with scale parameter $\theta = 2$ for Fig.~\ref{fig:KS95_Cnormal} and Fig.~\ref{fig:KS95_Cgamma} respectively. All the graphs correspond to balanced skinny and puffy sample sizes, so that each form half of the pooled data sets.
These contour plots show the percent of trials in which the pooled data set has less uncertainty around the median for the 95\% K--S bounds.  The yellow regions delimited by the dashed lines correspond to cases in which pooling is almost certainly preferred because it has a very high probability of reducing the uncertainty regardless of differential data quality. 
On the other side of the dashed line, it may be reasonable to check whether pooling would be preferred or should be eschewed for any particular analysis.
In data sets with small sample sizes, or with high dispersion,
it will be almost always be preferable to include additional measurements even with up to 10 times higher imprecision. 
It can be observed that by increasing the number of good quality intervals, we decrease the slope of the $90\%$ level line and minimize situations where pooling can be beneficial.

\section{Fitting parametric distributions: maximum likelihood methods}\label{MLE}

\subsection{Likelihoods and confidence intervals}

Let $X = (X_1, X_2, \dots, X_N)$ are independent and identically distributed random observations coming from some unknown probability density $f(x;\theta)$ which depends on a parameter $\theta$, where  $\theta$ could be a real-valued unknown parameter or a vector of parameters. The assumption of independence implies that
\begin{equation}
    f(X_1, X_2, \dots, X_N; \theta) = f(X_1; \theta) \cdot f(X_1; \theta)\dots f(X_N; \theta)
\end{equation}
which gives the log-likelihood function
\begin{equation}
    l(\theta) = \log L(\theta) = \log \prod_{i=1}^{N} f(X_i; \theta) = \sum_{i=1}^N \log f(X_i; \theta).
\end{equation}
The unknown parameter $\theta$ \revb{is} usually estimated by maximizing the the log-likelihood function
\begin{equation}
    \hat{\theta} = \arg \max_\theta \; l(\theta).
\end{equation}

Usually it is \revb{desirable} to report an interval around this estimate which is likely to contain the actual value. \revb{Such an interval can be approximated} under asymptotic normality \cite{Asymptotic_norm_1970,ibragimov_Asymptotic_norm_1981} of the maximum likelihood method. For a parametric model $f(x;\theta)$, the maximum likelihood estimator $\hat{\theta}$ is asymptotically normal under the regularity conditions~\cite{Hjek1970ACO, bickel1993efficient}
\begin{equation}
    \sqrt{N} (\hat{\theta} - \theta_0) \xrightarrow[]{\mathcal{D}} \mathcal{N}(0, \; \mathcal{I}^{-1}(\theta_0))
\end{equation}
in distribution as $N \rightarrow \infty$, where $\mathcal{I}(\theta)$ is Fisher information defined as
\begin{equation}\label{Fisher}
    \mathcal{I}(\theta_0) = -E_{\theta} \left[ \dfrac{\partial^2 }{\partial \theta^2} \log f(X;\theta)\right],
\end{equation}
assuming that $\log f(X;\theta)$ is twice differentiable with respect to $\theta$. For independent and identically distributed data, the total Fisher information is $N \mathcal{I}(\theta)$ where $\mathcal{I}(\theta)$ is the Fisher information for a single observation. Confidence intervals for the estimated parameter are given by
\begin{equation}\label{CI_main_equation}
    \theta \in \left[\hat{\theta}-t_{N-1,1-\alpha/2}\;\sqrt{\dfrac{1}{N}\;\mathcal{I}^{-1}(\hat{\theta})},\; \hat{\theta}+t_{N-1,1-\alpha/2}\;\sqrt{\dfrac{1}{N}\;\mathcal{I}^{-1}(\hat{\theta})} \right].
\end{equation}

\subsection{Maximum likelihood method: traditional approach}
Traditionally, analysts using maximum likelihood assume the probability for an unknown future value being between $\underline{x}_i$ and $\overline{x}_i$ is
\begin{equation}
    L_i(\theta; x_i) = \int_{\underline{x}_{i}}^{\overline{x}_i} f(x; \theta) dx = F(\overline{x}_{i}; \theta) - F(\underline{x}_{i}, \theta)
\end{equation}
where $\theta$ is the parameter on which data population depends \cite{Meeker_Escobar}.
For a given single interval $\interval{\underline{x}}{\overline{x}}$ chosen randomly from an exponential distribution for example, the likelihood function is
\begin{equation}
    L(\lambda; x) = F( \overline{x}; \lambda) - F( \underline{x}; \lambda),
\end{equation}
where $F = 1 - e^{-\lambda x}, \, x \geq 0$, is the cumulative distribution function. For the case of independent interval observations $X_i = \interval{\underline{x}_i}{\overline{x}_i}$ where $i = 1, \dots, N$ the joint likelihood function is defined as
\begin{equation}
    L(\lambda) = \prod_{i=1}^{N}L_i(\lambda) = \prod_{i=1}^{N} \left[ \exp(-\lambda \; \underline{x}_i) - \exp(-\lambda \; \overline{x}_i) \right].
\end{equation}

\begin{figure}[ht!]
	\centering
	\begin{subfigure}{0.4\textwidth}
		\includegraphics[width=\linewidth]{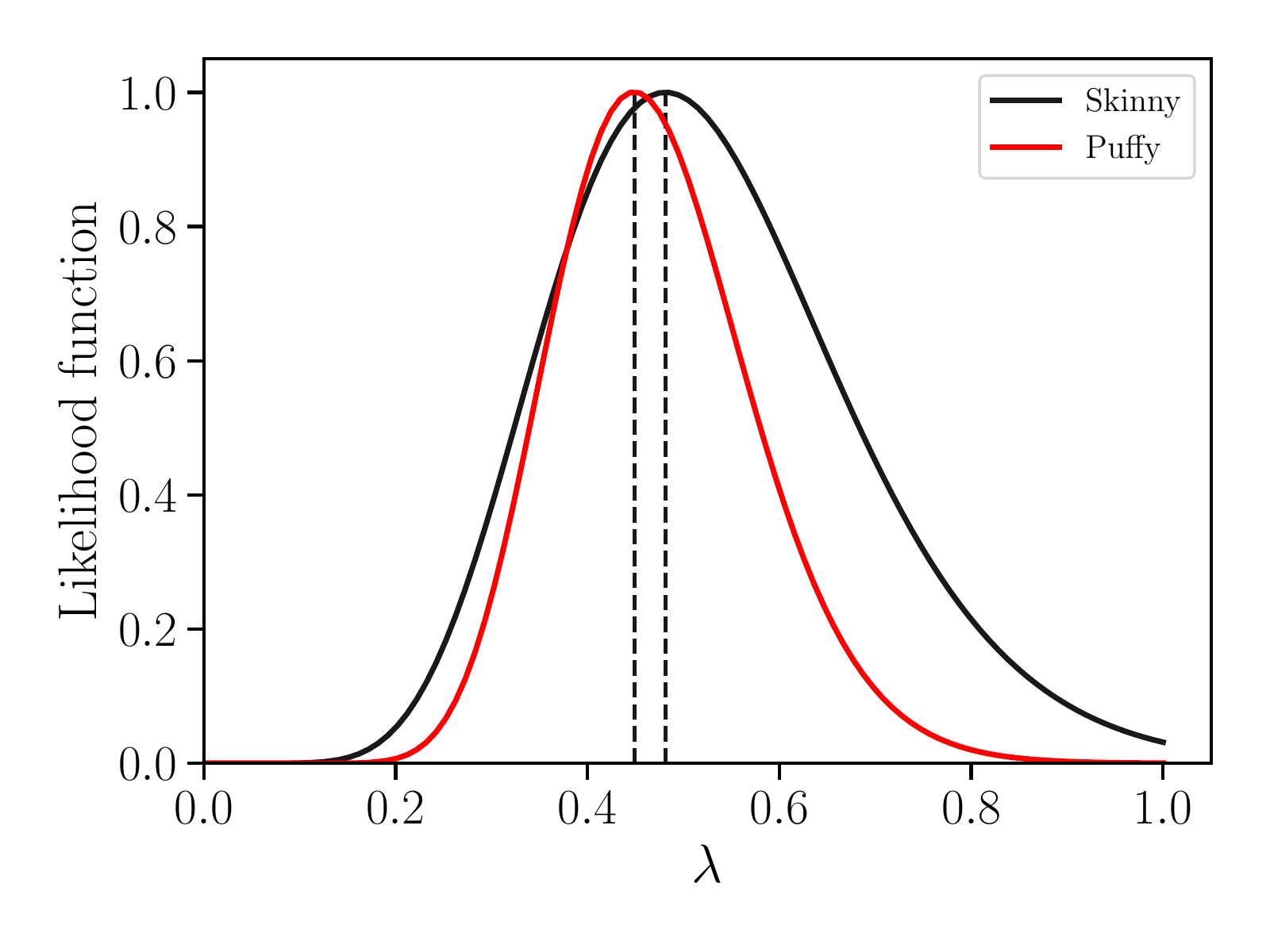}
		\caption{}
	\end{subfigure}
	\begin{subfigure}{0.4\textwidth}
		\includegraphics[width=\linewidth]{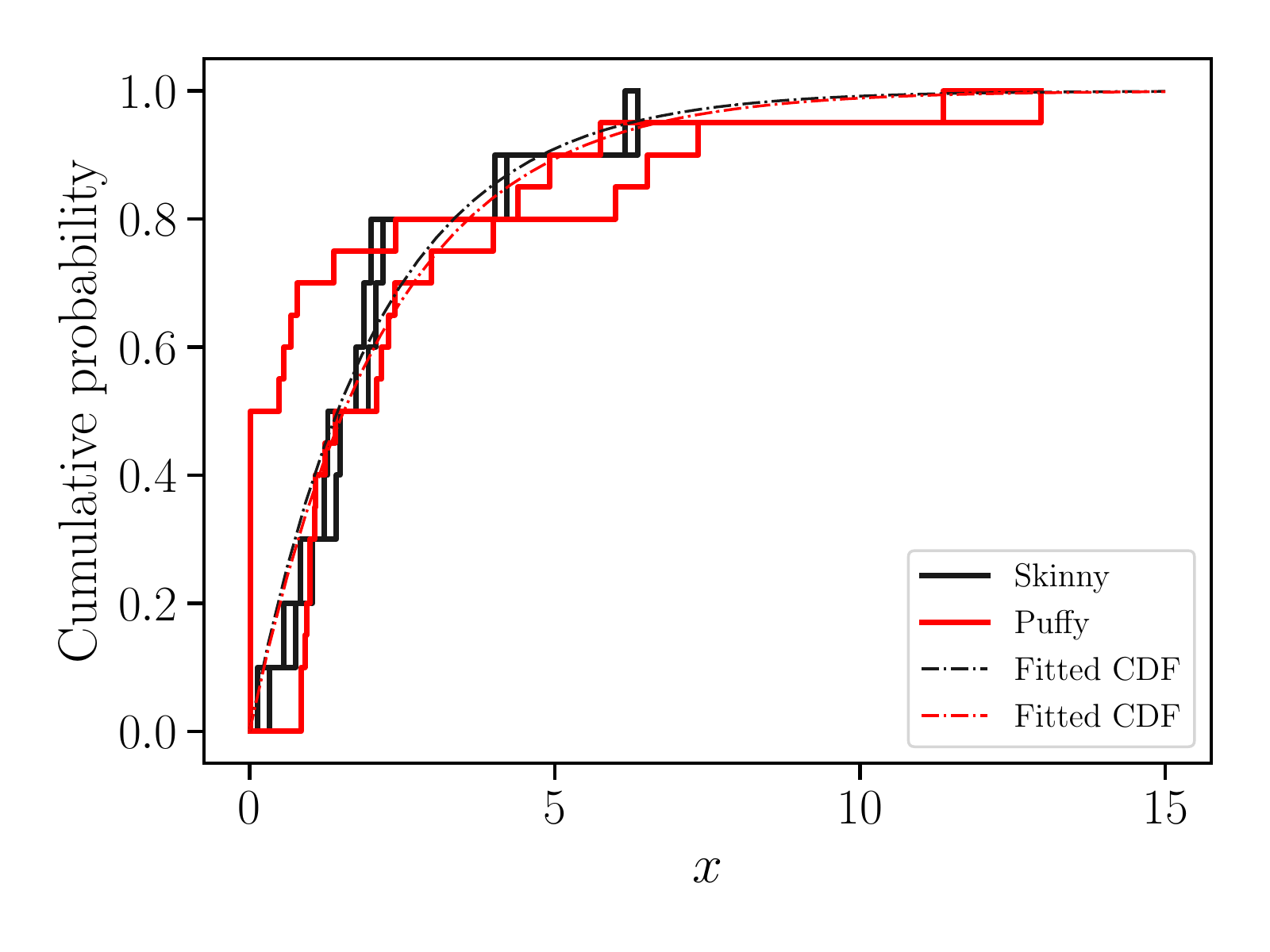}
		\caption{}
	\end{subfigure}
	\caption{Empirical distributions for two data sets with fitted cumulative distributions (on the right) and calculated likelihood functions (on the left).}
	\label{fig:Likelihood_examples}
\end{figure}

Fig.~\ref{fig:Likelihood_examples}a illustrates the likelihood functions and their maxima for two interval data sets: skinny with 10 narrow intervals and puffy with 20 wide intervals. The corresponding empirical cumulative and fitted exponential distributions are shown in Fig.~\ref{fig:Likelihood_examples}b. Both data sets have relatively similar maximum likelihood rate parameters, $\lambda_{\text{skinny}} = 0.481 $ and $\lambda_{\text{pooled}} = 0.448 $, which mostly ignore the large (8 times) difference in their imprecision.

\begin{figure}[ht!]
	\centering
	\begin{subfigure}{0.325\textwidth}
		\includegraphics[width=\linewidth]{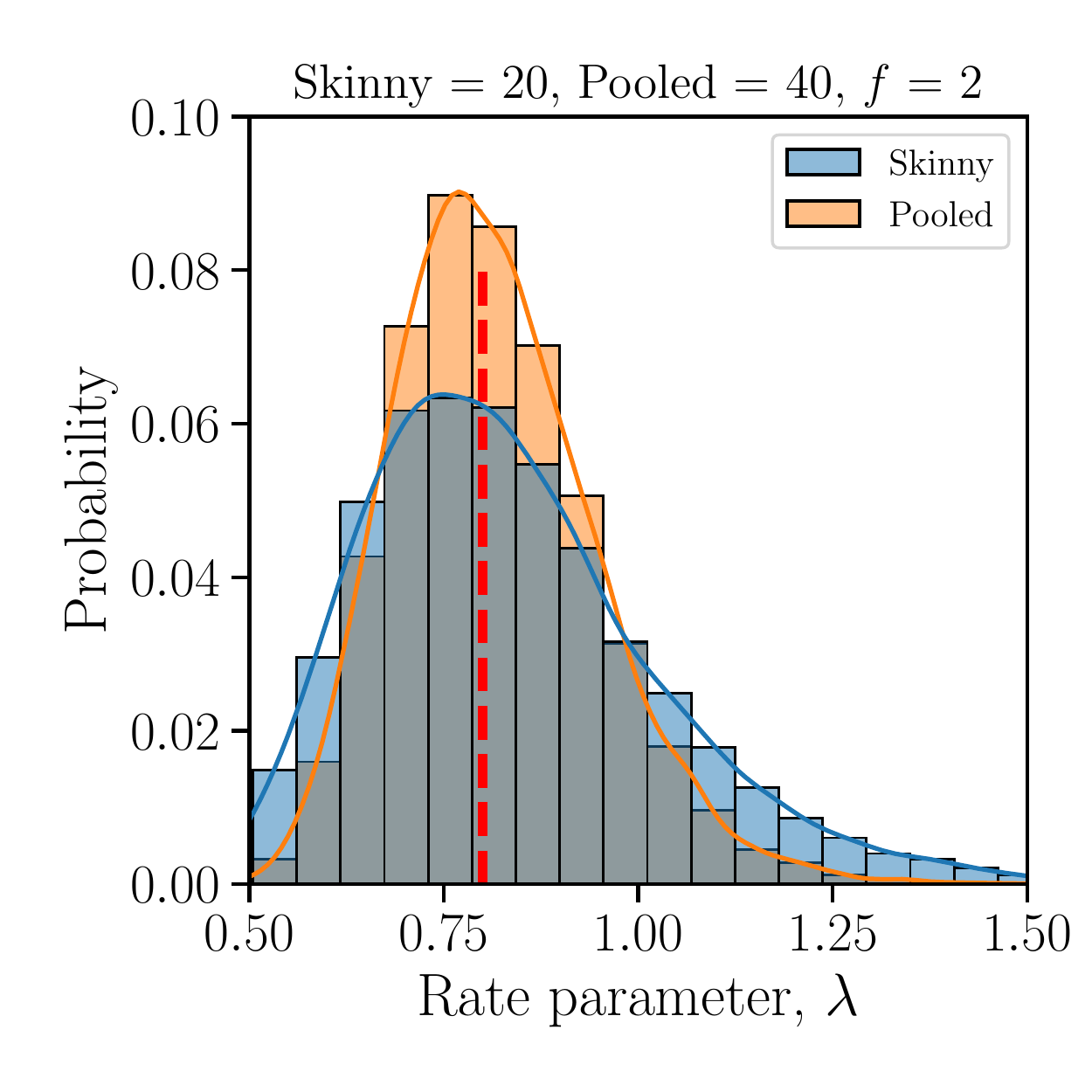}
		\caption{}
	\end{subfigure}
	\begin{subfigure}{0.325\textwidth}
		\includegraphics[width=\linewidth]{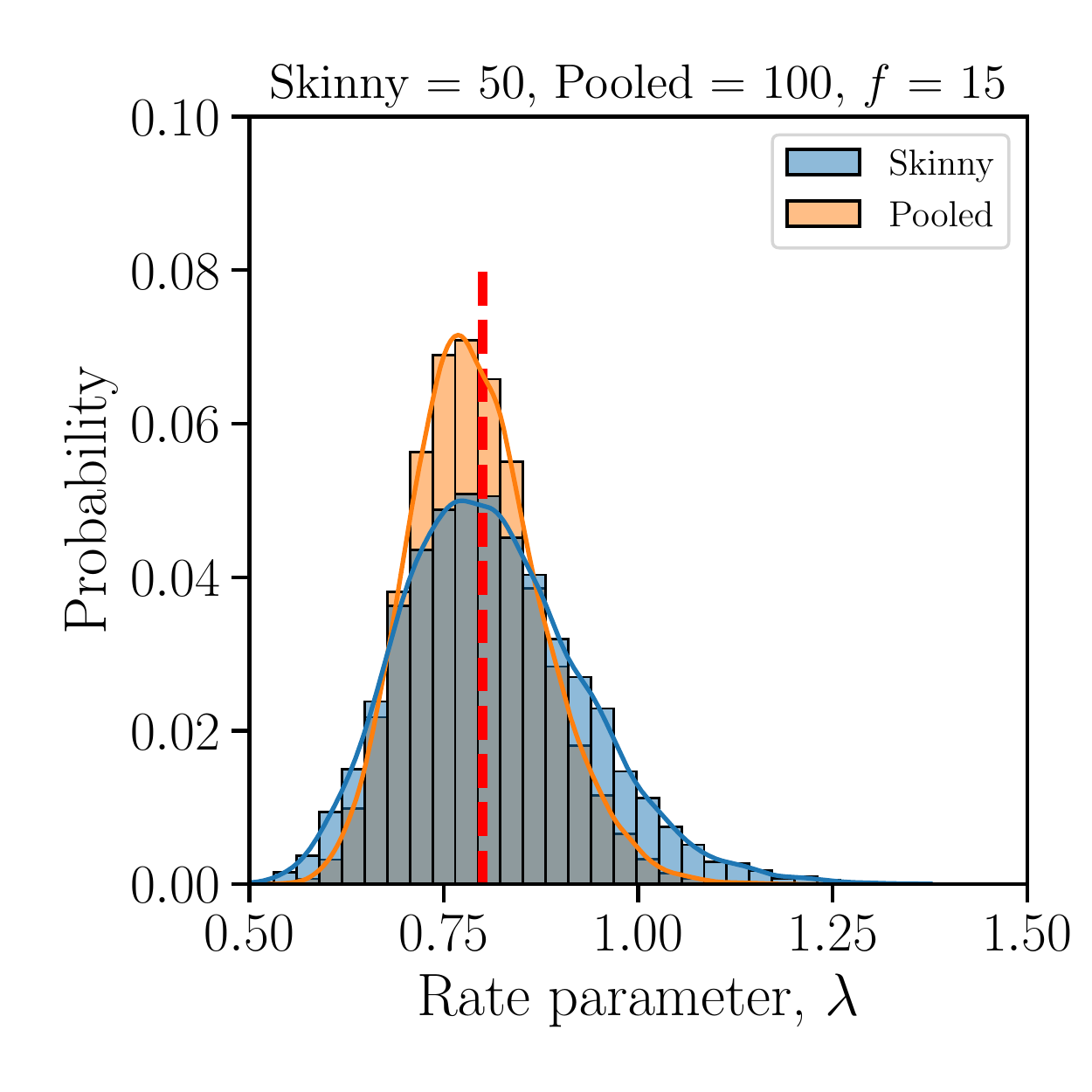}
		\caption{}
	\end{subfigure}
	\begin{subfigure}{0.325\textwidth}
		\includegraphics[width=\linewidth]{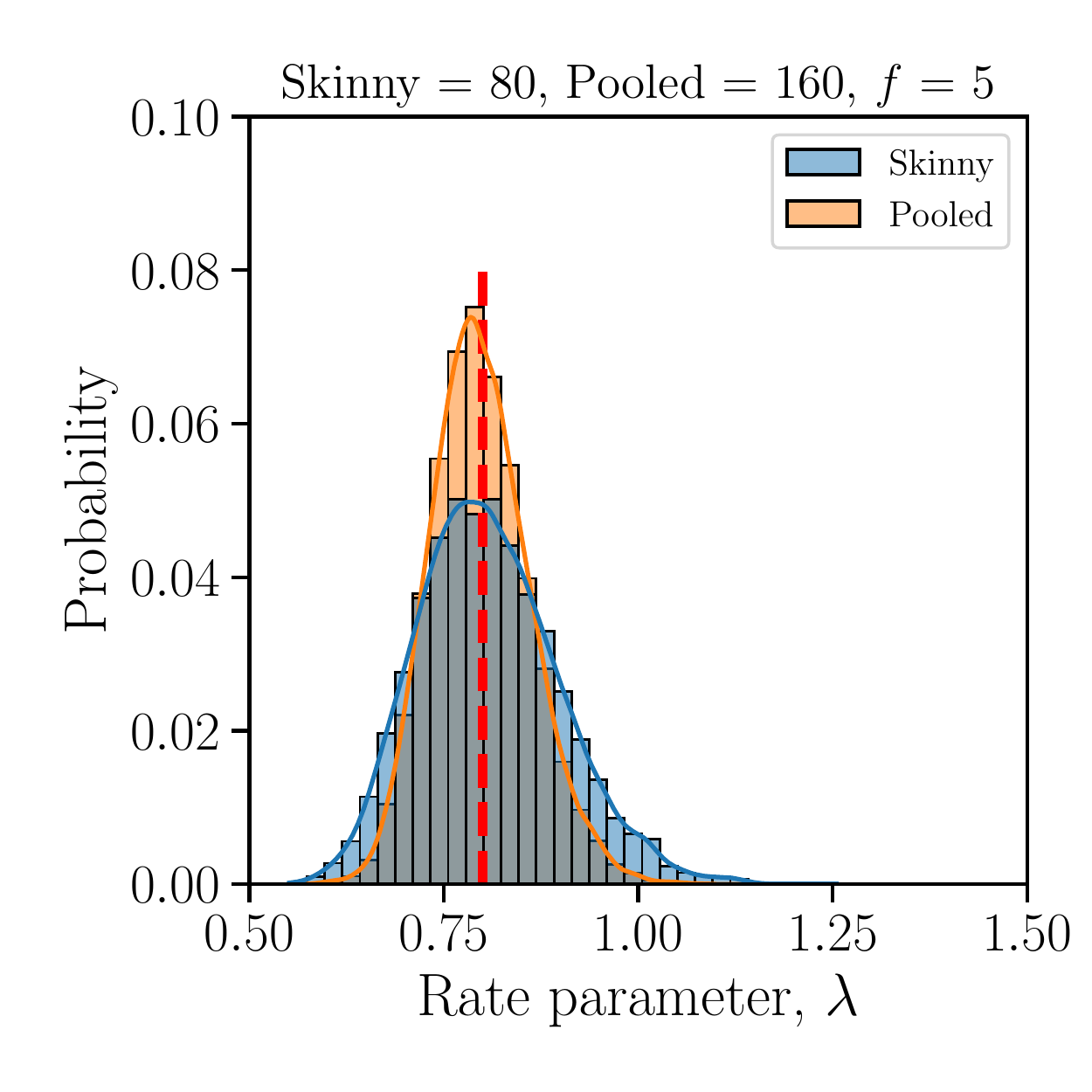}
		\caption{}
	\end{subfigure}
	\caption{Probability distributions for the rate parameter $\lambda$ using traditional maximum likelihood approach: $\interval{\underline{x}_i}{\overline{x}_i} \sim \text{Exp}(0.8)$.}
	\label{fig:MLE1_dist}
\end{figure}

We use the traditional method to estimate the unknown parameter $\lambda$ for the skinny and pooled interval data sets that are independent and identically distributed as Exp$(\lambda_{\text{true}})$ with $\lambda_{\text{true}}=0.8$. Parameters for generating data sets and simulation results are presented in Table~\ref{table:MLE1_inputs}. The first column shows the balance between the skinny and pooled intervals. The imprecision factor $f$ is presented in the second column. For each of three cases experiments were replicated $M = \numprint{10000}$ times. The probability distributions observed for the estimated rate parameter $\lambda$  are shown in Fig.~\ref{fig:MLE1_dist}. 
\begin{table}[ht!]
	\centering
	\begin{tabular}{@{\extracolsep{6pt}}cccccc}
	    \toprule
		\multicolumn{2}{c}{Input parameters} &  \multicolumn{2}{c}{Mean $\Bar{\lambda}$} \\
		\cmidrule{1-2} \cmidrule{3-4} 
		$N$ skinny : pooled    & $f$   & Skinny     & Pooled   \\
        \cmidrule{1-2} \cmidrule{3-4}
		$ 20 : 40 $ & 2 &  0.835 & 0.815\\ 
		$ 50 : 100 $ & 15 &   0.812 &  0.790 \\ 
		$ 80 : 160 $ & 5 &  0.805 & 0.795 \\
		\bottomrule
	\end{tabular}
	\caption{Input parameters for generating interval data sets using Exp$(0.8)$ distribution (on the left) and calculated mean values of rate parameters for skinny and pooled data sets (on the right).}
	\label{table:MLE1_inputs} 
\end{table}

The traditional approach seems to ignore the effect of the censoring which has no discernible effect on the difference between the skinny and pooled mean rates as a function of the uncertainty factor $f$. 
The traditional approach suggests comparable or better accuracy for pooled data sets, no matter how imprecise the wide intervals in them may be.
Consider, for instance, the numerical experiment in the second line of Table~\ref{table:MLE1_inputs} in which 50 narrow intervals are potentially pooled with 50 intervals that are 15 times wider.
The pooled data set gives essentially the same estimate of the $\lambda$ parameter as is given by the 50 narrow intervals by themselves.
Even extremely wide intervals have no practical effect on the estimated means for the rate parameter.

The corresponding confidence interval for the estimate based on the pooled data is actually \emph{narrower} simply because there are twice as many samples, neglecting entirely the fact that they are massively less precise.
Unlike the Kolmogorov--Smirnov approach to interval data discussed in Section~\ref{KolmogorovSmirnov_bands}, the traditional approach to interval censoring in maximum likelihood does not capture this imprecision in a meaningful way.

\subsection{Maximum likelihood method: interval approach}\label{imle}

\begin{figure}[ht!]
	\centering
	\begin{subfigure}{0.4\textwidth}
		\includegraphics[width=\linewidth]{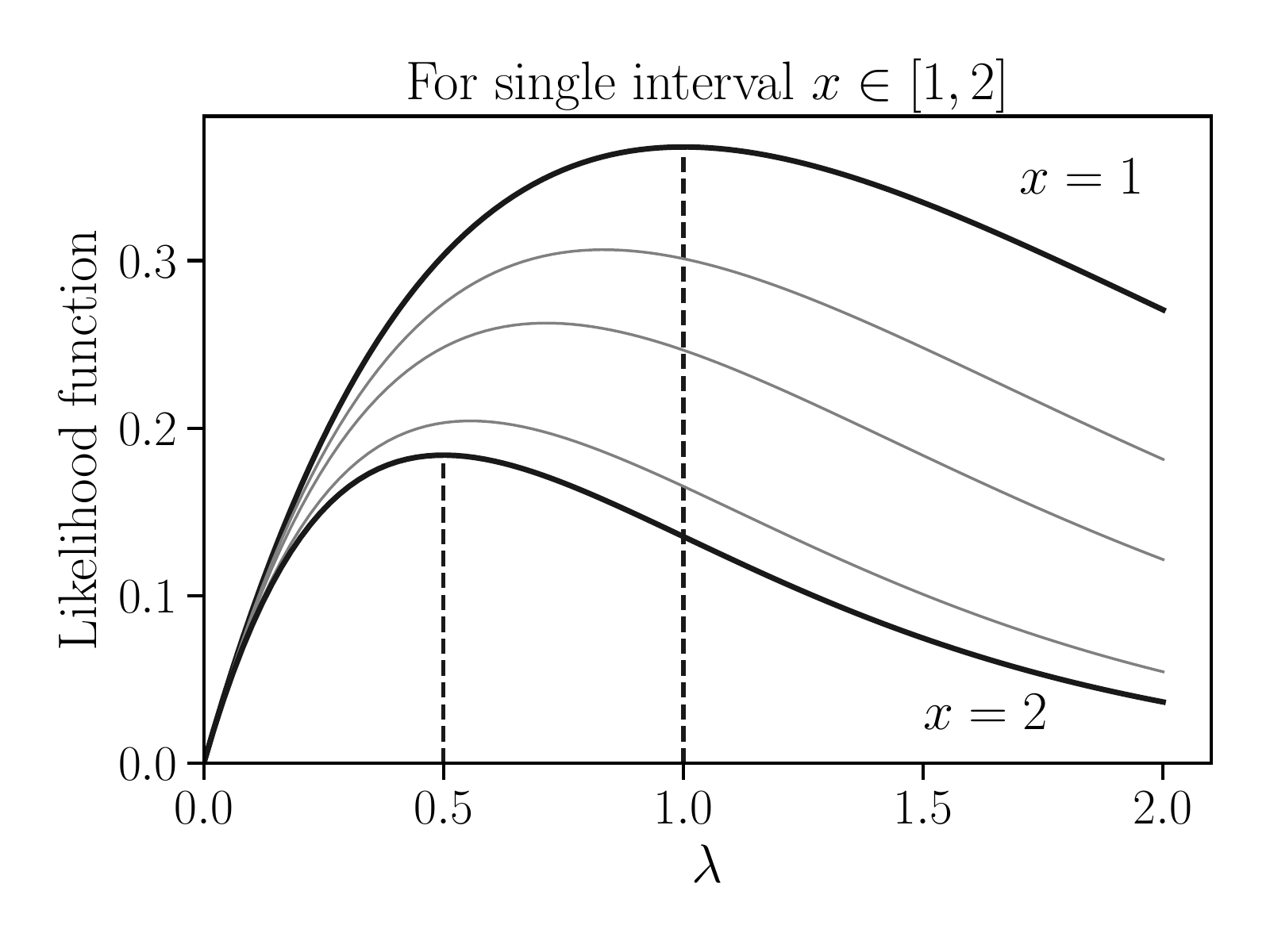}
		\caption{}
		\label{fig:Likelihood_int_approachA}
	\end{subfigure}
	\begin{subfigure}{0.4\textwidth}
		\includegraphics[width=\linewidth]{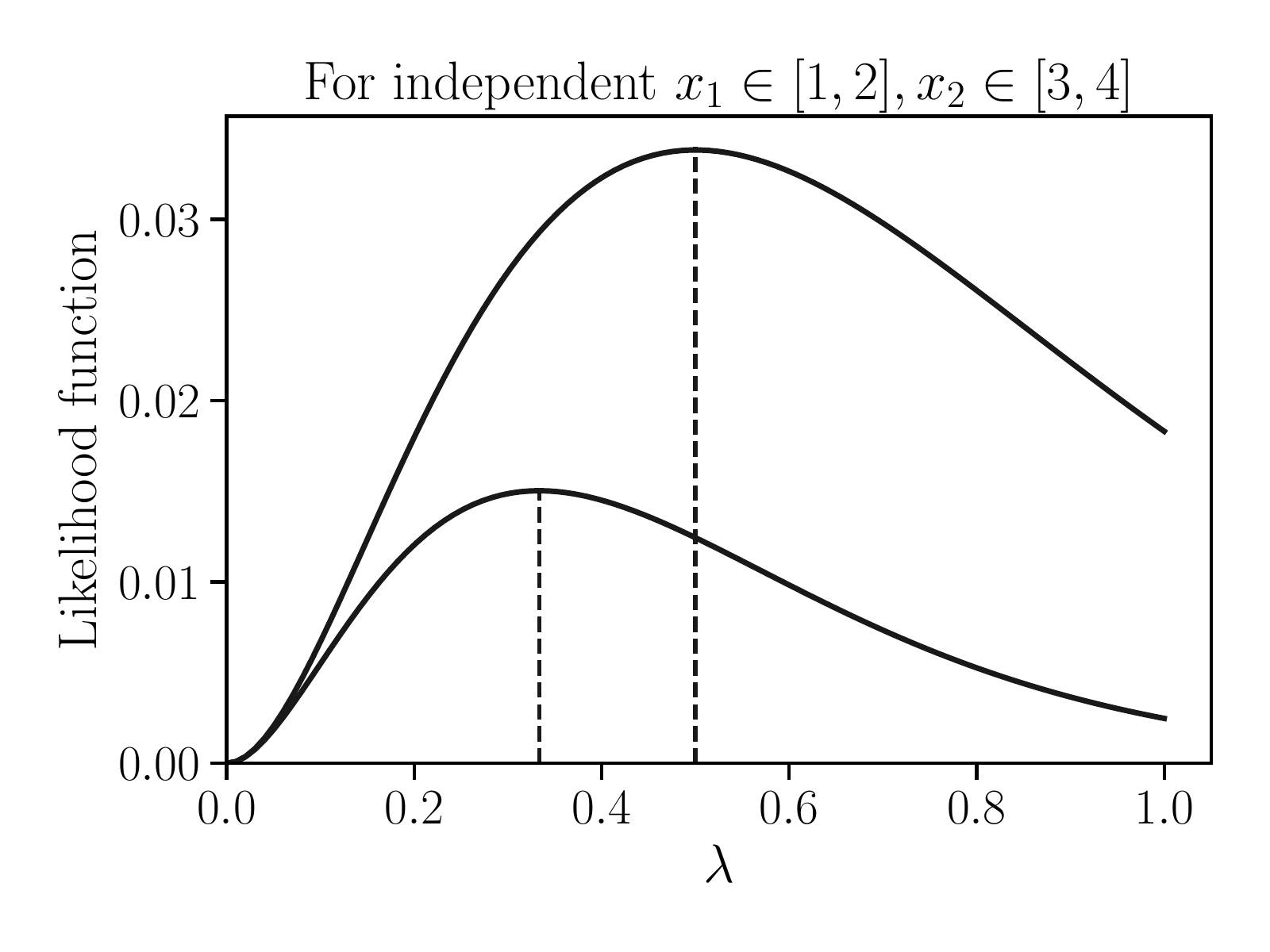}
		\caption{}
		\label{fig:Likelihood_int_approachB}
	\end{subfigure}
	\caption{Examples of the likelihood functions and their peaks: (a) single datum (b) two independent observations.}
	\label{fig:Likelihood_int_approach}
\end{figure}

In contrast to the traditional approach described above,
the interval approach \cite{Walley1991, Manski2003, SAND2007-0939} creates a \emph{class} of maximum likelihood solutions. When the data sets contain intervals rather than only point estimates, a single probability distribution no longer suffices to characterise the epistemic uncertainty.
Fitting an exponential distribution to the interval datum $\interval{\underline{x}}{\overline{x}}$ by a maximum likelihood interval approach produces a class of exponential distributions, see Fig.~\ref{fig:Likelihood_int_approachA}.
That class can be estimated simply by the set of values $\lambda \in \interval{\hat{\lambda}_1}{\hat{\lambda}_2}$ where
\begin{equation}
\begin{aligned}
            \hat{\lambda}_1 = 1 / \overline{x} \quad \text{and} \quad
            \hat{\lambda}_2 = 1 / \underline{x}.
\end{aligned}
\end{equation}

In the case of multiple intervals $\interval{\underline{x}_j}{\overline{x}_j}$, $j=1, \dots, N$, assuming that measurements are independent, the maximum likelihood estimator yields
\begin{equation}
    \hat{\lambda}_1 = \dfrac{N}{\sum_{j=1}^{N} \overline{x}_j} \quad \text{and} \quad \hat{\lambda}_2 = \dfrac{N}{\sum_{j=1}^{N} \underline{x}_j}.
\end{equation}
Fig.~\ref{fig:Likelihood_int_approachB} shows that the bounds on the products of the likelihood functions for multiple imprecise observations under independence.

\begin{figure}[ht!]
	\centering
	\begin{subfigure}{0.4\textwidth}
		\includegraphics[width=\linewidth]{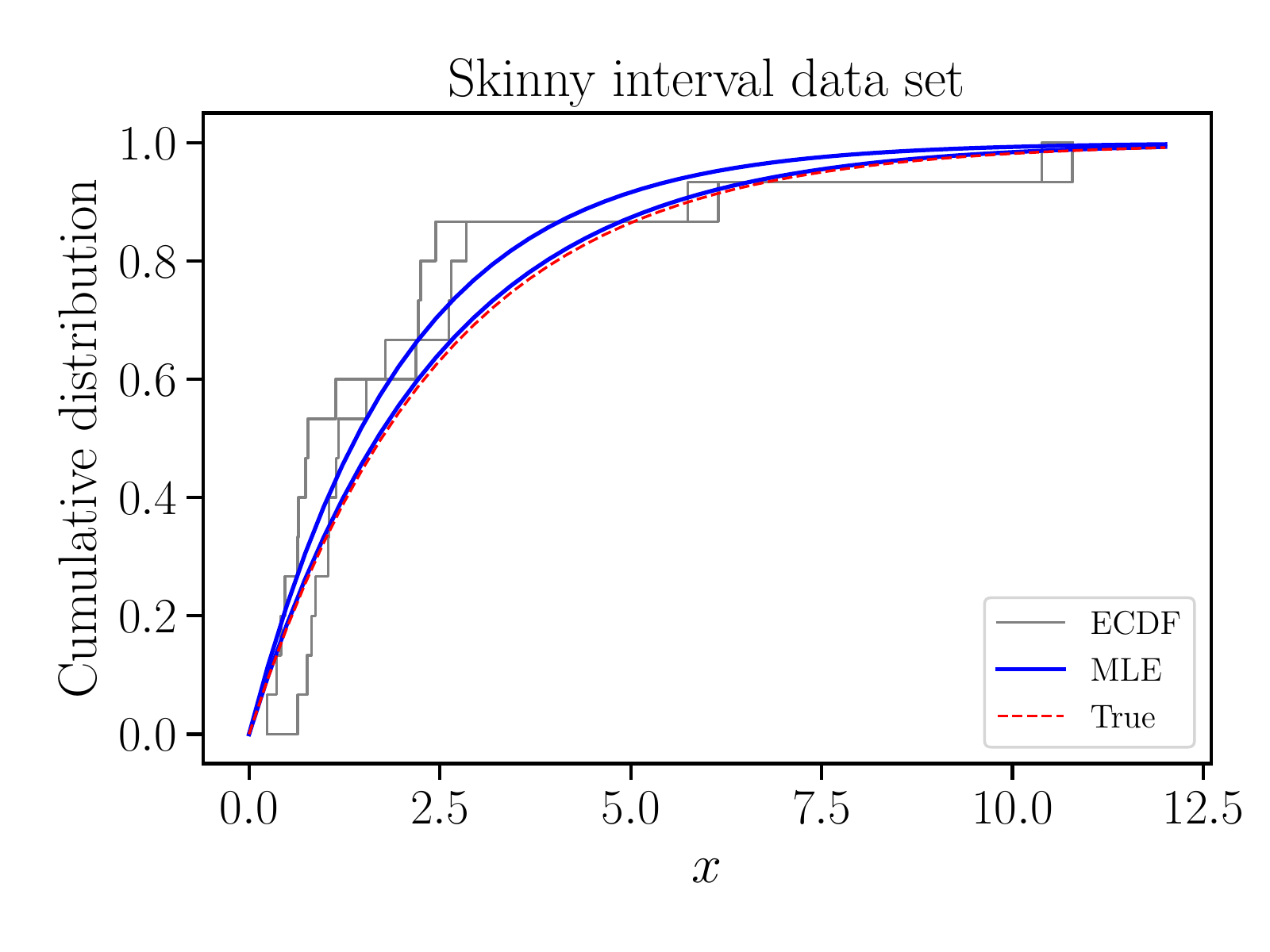}
		\caption{}
	\end{subfigure}
	\begin{subfigure}{0.4\textwidth}
		\includegraphics[width=\linewidth]{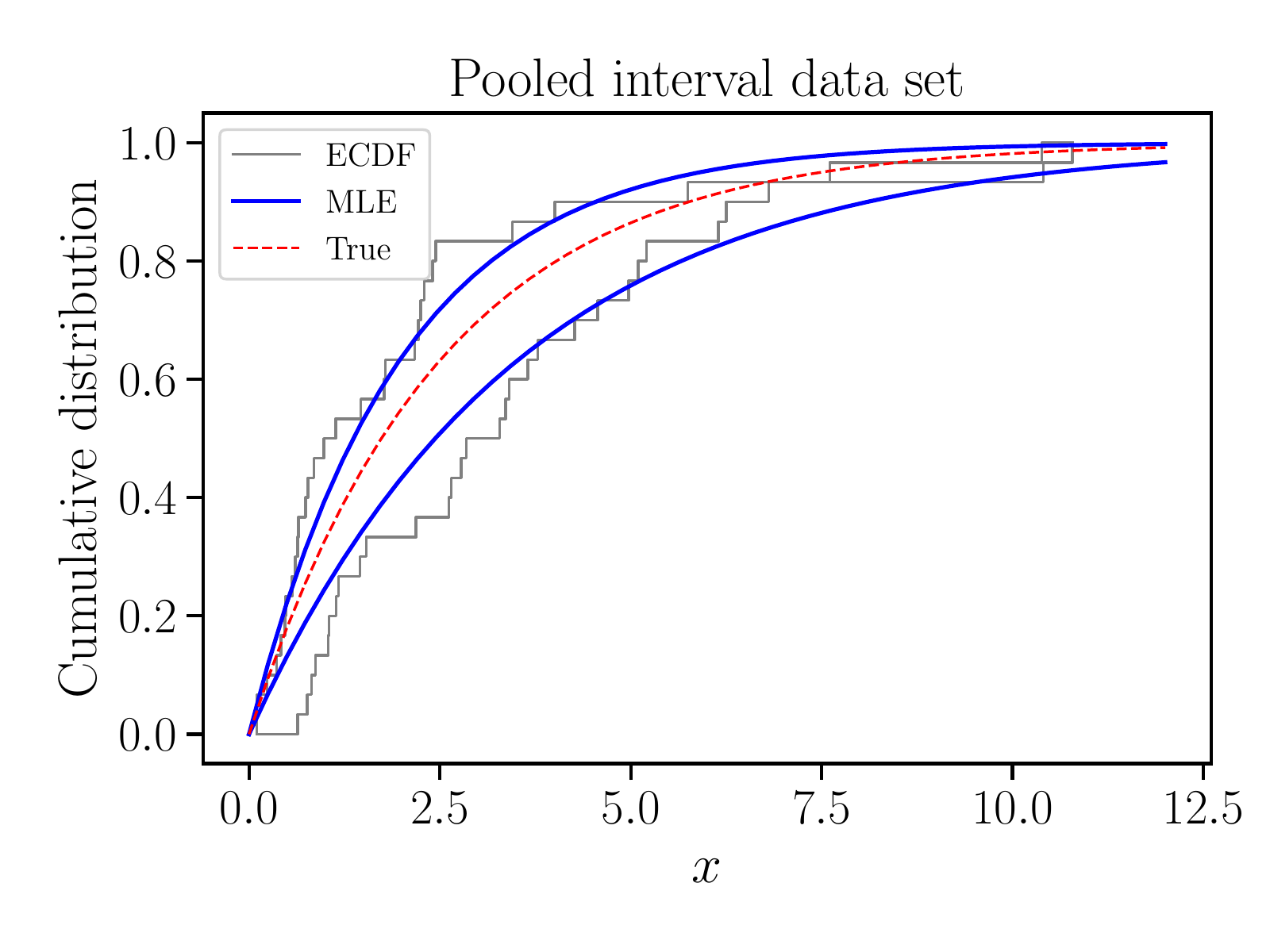}
		\caption{}
	\end{subfigure}
	\caption{Empirical cumulative distribution function (grey) and the fitted distribution (blue) from the intervalised maximum likelihood approach. The skinny and pooled interval data sets are independent and identically distributed as Exp($0.4$) (red dotted line).}
	\label{fig:MLEinterval_example}
\end{figure}

Fig. \ref{fig:MLEinterval_example} shows an example application of the intervalized maximum likelihood method.
The skinny interval data set has 15 samples that are independent and identically distributed as Exp($0.4$) with $\Delta=0.2$. The pooled data set includes 15 additional puffy samples with $\Delta=1.4$.

\textit{Confidence intervals.} 
To illustrate the estimation of confidence intervals for the intervalized maximum likelihood method, two example distributions with different tail weights are used: an exponential with a long tail, and a uniform with a much more steeply decreasing tail.

    In the case of the exponential distribution, the confidence interval for the scale parameter $\lambda$ can be estimated using (see \cite{Ross2009-rn} for details)
    \begin{equation}
        P_\lambda \left[ \dfrac{\chi^2_{2N}(\alpha/2)}{2 \sum_{i=1}^{N} X_i} \leq \lambda\leq  \dfrac{\chi^2_{2N}(1 - \alpha/2)}{2 \sum_{i=1}^{N} X_i} \right] = 1-\alpha,
    \end{equation}
    where $\chi^2_{2N}(\alpha/2)$ is a percentile of the chi-squared distribution. For an interval data set $\interval{\underline{x}_1}{\overline{x}_1}, \dots, \interval{\underline{x}_N}{\overline{x}_N}$ the estimator becomes
    \begin{equation}
        \hat{\lambda} = \interval{\underline{\lambda}}{\overline{\lambda}} = \dfrac{N}{\sum_{i=1}^N \interval{\underline{x}_i}{\overline{x}_i}},
    \end{equation}
    and the confidence interval is $\left[ \underline{a}, \overline{b}\right]$ where
    \begin{equation}
        \quad [a] = \dfrac{\chi^2_{2N}(\alpha/2)}{2 \sum_{i=1}^{N} \interval{\underline{x}_i}{\overline{x}_i}}, \; [b] = \dfrac{\chi^2_{2N}(1 - \alpha/2)}{2 \sum_{i=1}^{N}  \interval{\underline{x}_i}{\overline{x}_i}}.
    \end{equation}
Note that the two ends are themselves intervals; the confidence distribution is defined by the outer bounds.

For a uniform distribution with one unknown limit, let $X_1, \dots, X_N$ be independent and identically distributed random variables with $X_i \sim \mathcal{U}(0,\theta)$.  For the specific case
\begin{equation*}
    f(X; \theta) = 
    \begin{cases}
    1/\theta, &0 \leq X \leq \theta\\
    0, &\text{otherwise}
    \end{cases},
\end{equation*}
the maximum likelihood estimator is $\hat{\theta} = \max (X_1, \dots, X_N) < \theta$, and the $1-\alpha$ confidence interval for $\theta$ is \cite[][Sec. 12]{Mathai2017} 
\begin{equation}
    P_{\theta}\left[ \dfrac{\hat{\theta}}{(1-\frac{\alpha}{2})^{1/N}} \leq \theta \leq \dfrac{\hat{\theta}}{(\frac{\alpha}{2})^{1/N}} \right] = 1-\alpha.
\end{equation}
For an interval data set $\interval{\underline{x}_1}{\overline{x}_1}, \dots, \interval{\underline{x}_N}{\overline{x}_N}$ the estimator becomes
\begin{equation}
    \hat{\theta} = \max (\interval{\underline{x}_1}{\overline{x}_1}, \dots, \interval{\underline{x}_N}{\overline{x}_N}) = 
    [\max(\underline{x}_1, \dots, \underline{x}_N), \;
        \max(\overline{x}_1, \dots, \overline{x}_N)]        
        = \interval{\underline{\theta}}{\overline{\theta}},
\end{equation}
and the confidence interval $\left[ \underline{a}, \overline{b}\right]$ is defined by the intervals
\begin{equation}
    \quad \text{where} \quad [a] = \dfrac{\interval{\underline{\theta}}{\overline{\theta}}}{(1-\frac{\alpha}{2})^{1/N}}, \; [b] =  \dfrac{\interval{\underline{\theta}}{\overline{\theta}}}{(\frac{\alpha}{2})^{1/N}}.
\end{equation}

\begin{figure}[ht!]
	\centering
	\begin{subfigure}{0.325\textwidth}
		\includegraphics[width=\linewidth]{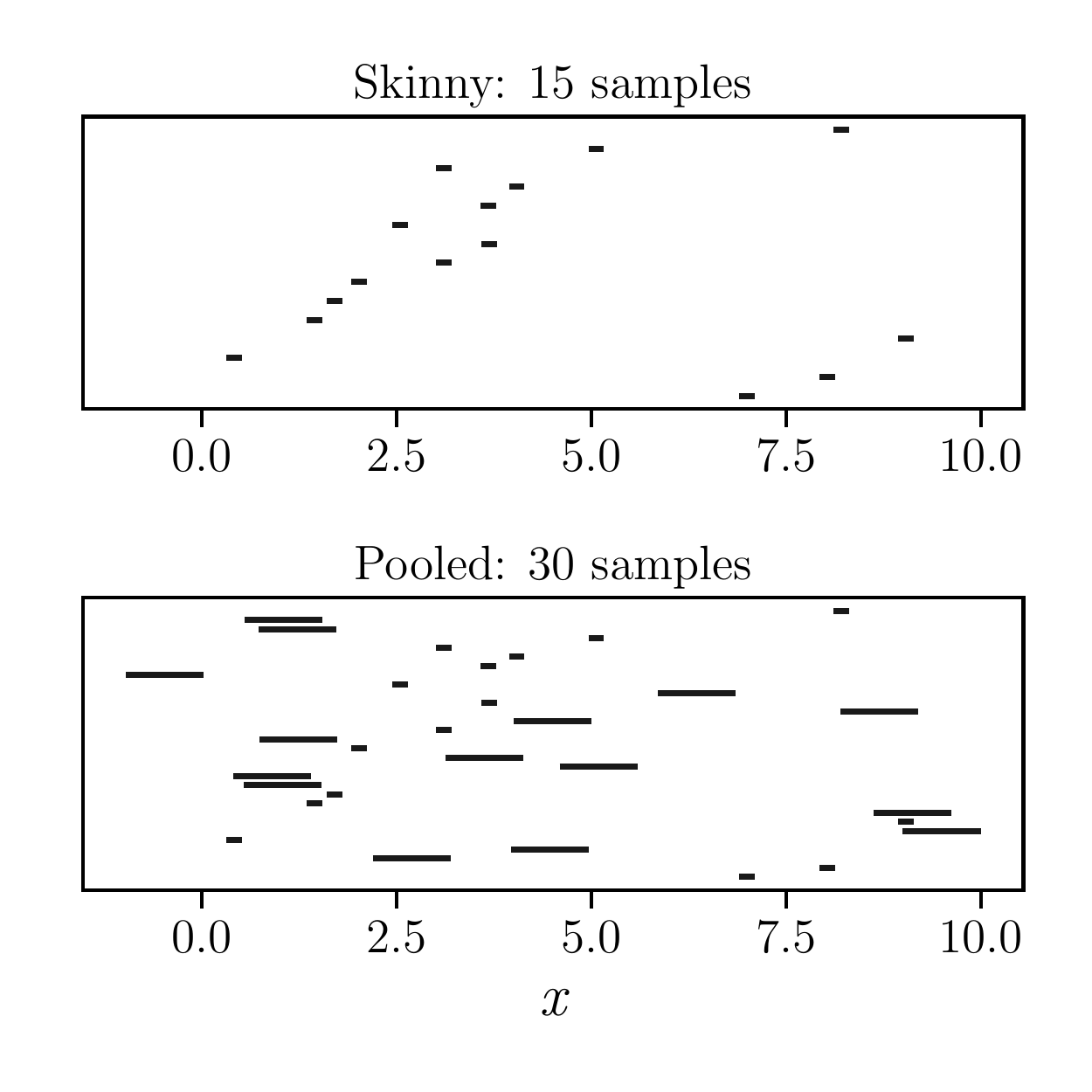}
		\caption{}
	\end{subfigure}
	\begin{subfigure}{0.325\textwidth}
		\includegraphics[width=\linewidth]{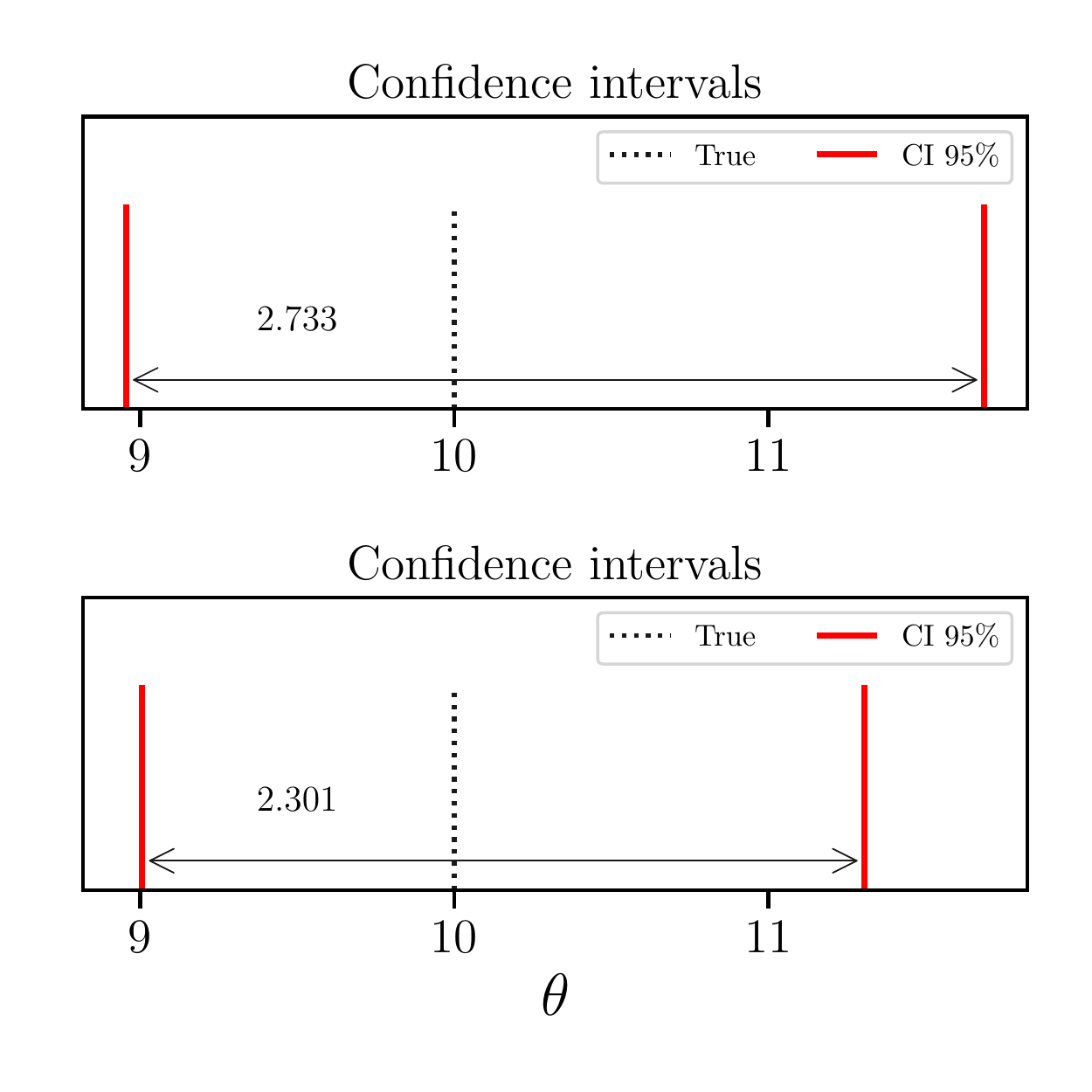}
		\caption{}
	\end{subfigure}
	\begin{subfigure}{0.325\textwidth}
		\includegraphics[width=\linewidth]{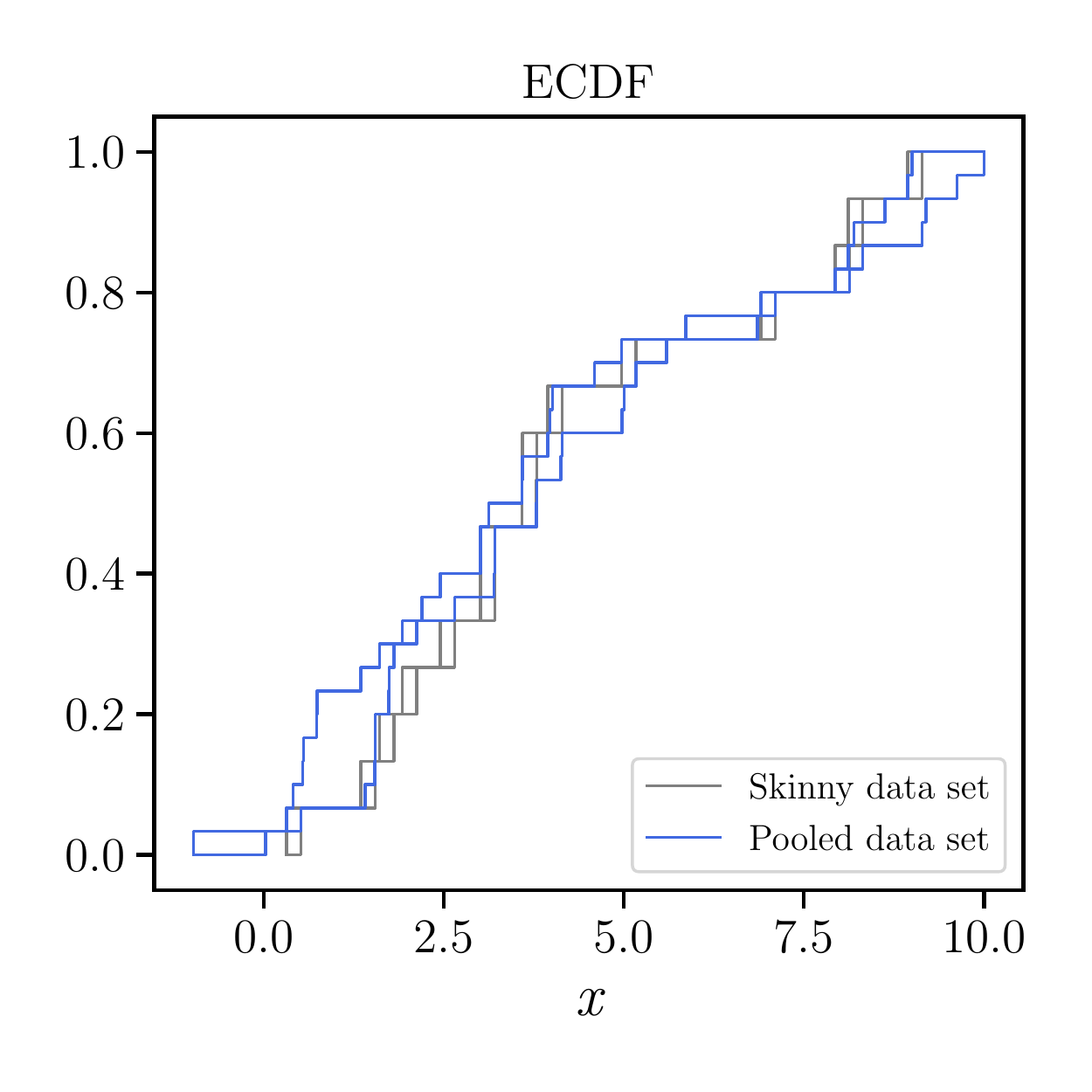}
		\caption{}
	\end{subfigure}
	\begin{subfigure}{0.325\textwidth}
		\includegraphics[width=\linewidth]{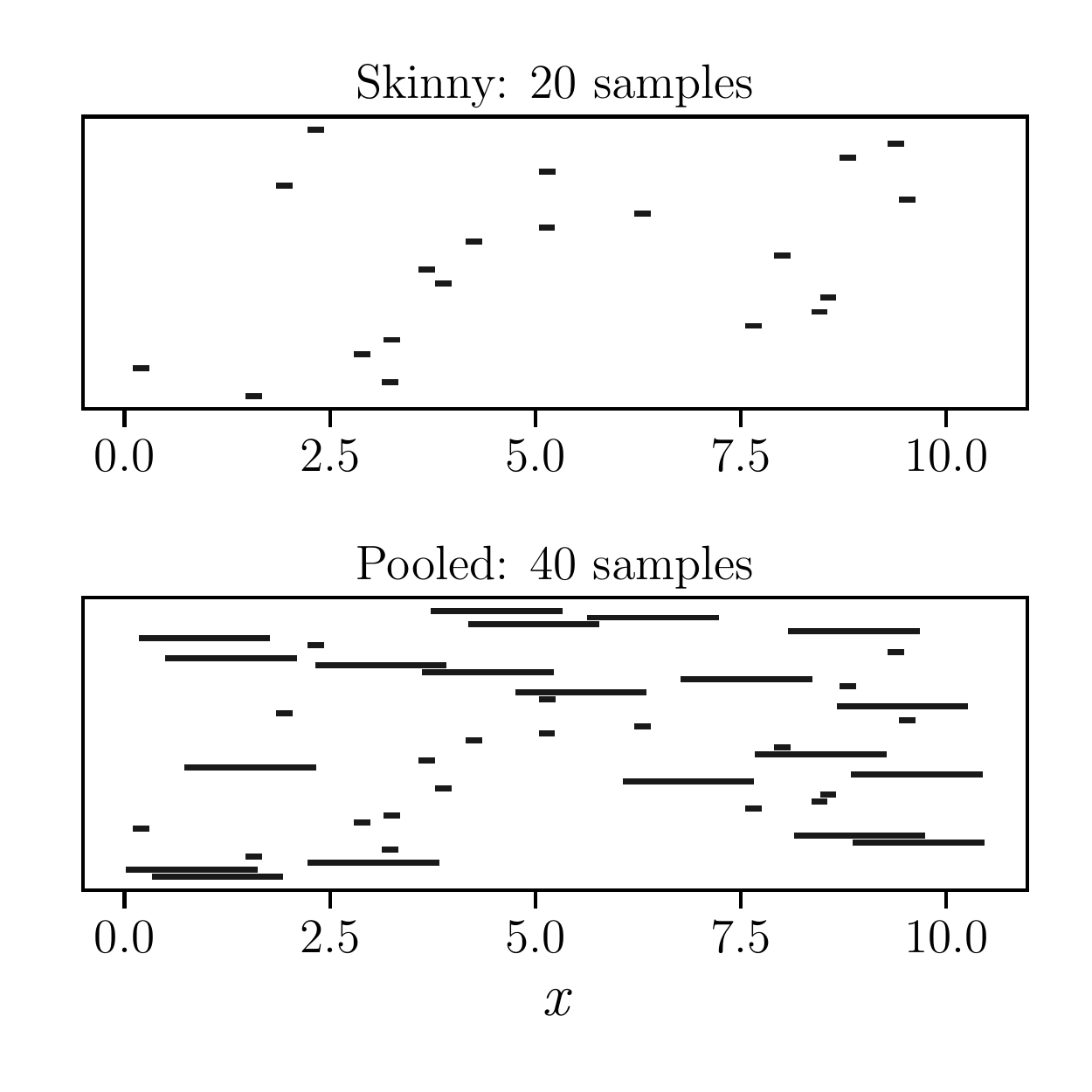}
		\caption{}
	\end{subfigure}
	\begin{subfigure}{0.325\textwidth}
		\includegraphics[width=\linewidth]{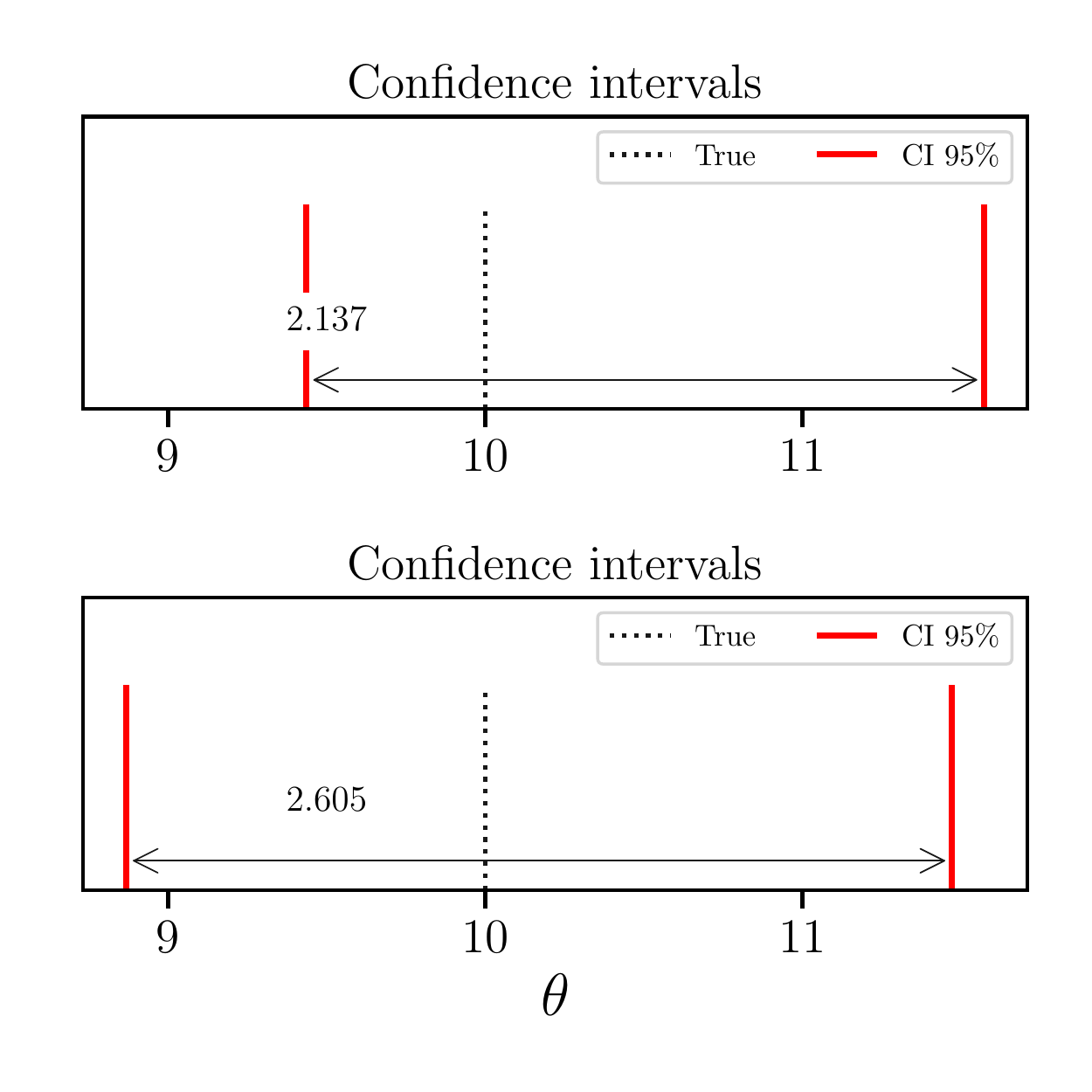}
		\caption{}
	\end{subfigure}
	\begin{subfigure}{0.325\textwidth}
		\includegraphics[width=\linewidth]{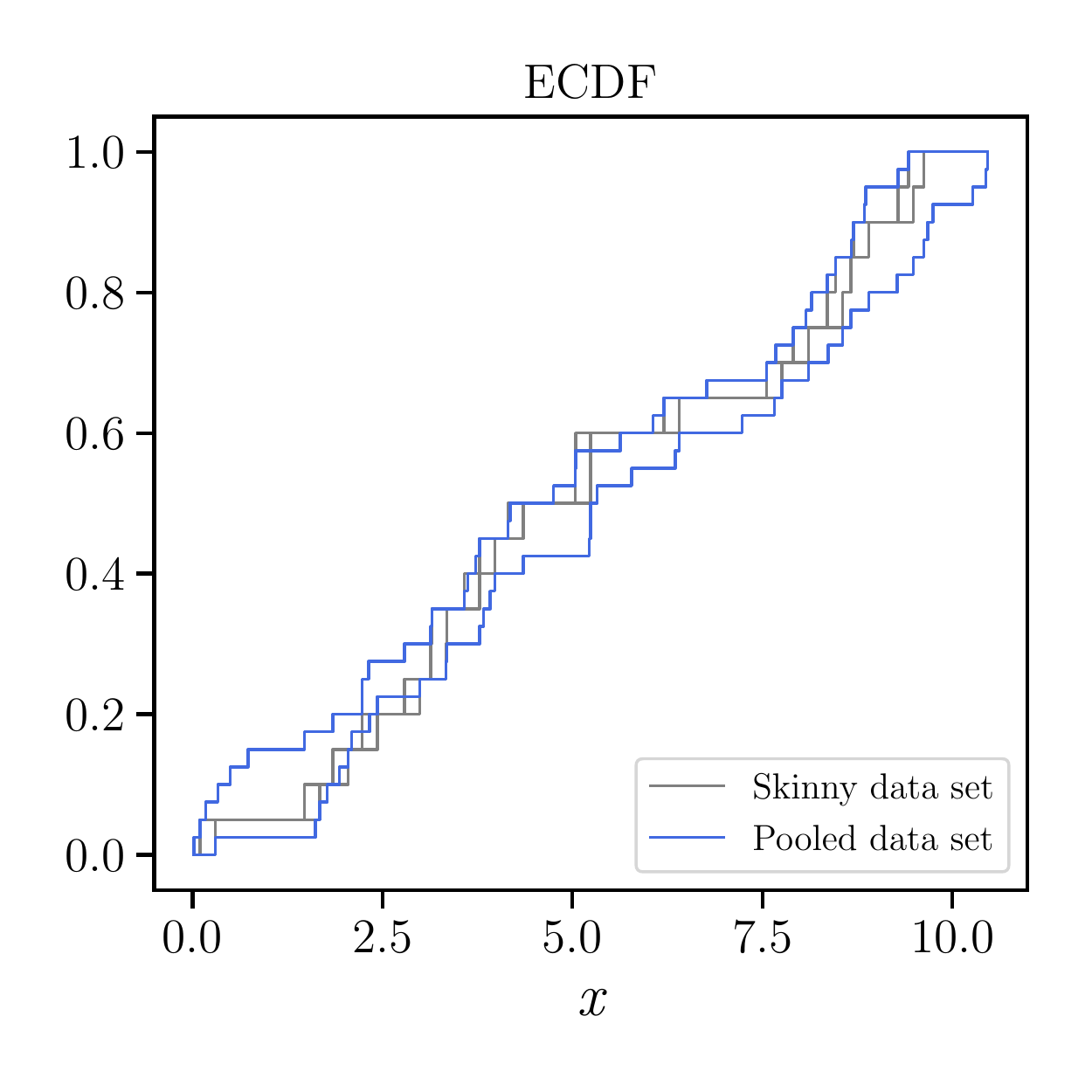}
		\caption{}
	\end{subfigure}
	\caption{Confidence intervals for the maximum likelihood estimator for a uniform distribution with one unknown: (a)~precise and imprecise interval data sets randomly generated as $\interval{\underline{x}_i}{\overline{x}_i} \sim \mathcal{U}(0,10)$, $\Delta_{\text{skinny}} = 0.1$, $f=5$; (d) $\interval{\underline{x}_i}{\overline{x}_i} \sim \mathcal{U}(0,10)$, $\Delta_{\text{skinny}} = 0.1$, $f=8$; (b, e)~estimated  95\% confidence interval for $\theta$; (c, f)~empirical cumulative distribution functions for the skinny and pooled data sets.}
	\label{fig:MLE_uni_CI_examples}
\end{figure}

\begin{figure}[ht!]
	\centering
	\begin{subfigure}{0.325\textwidth}
		\includegraphics[width=\linewidth]{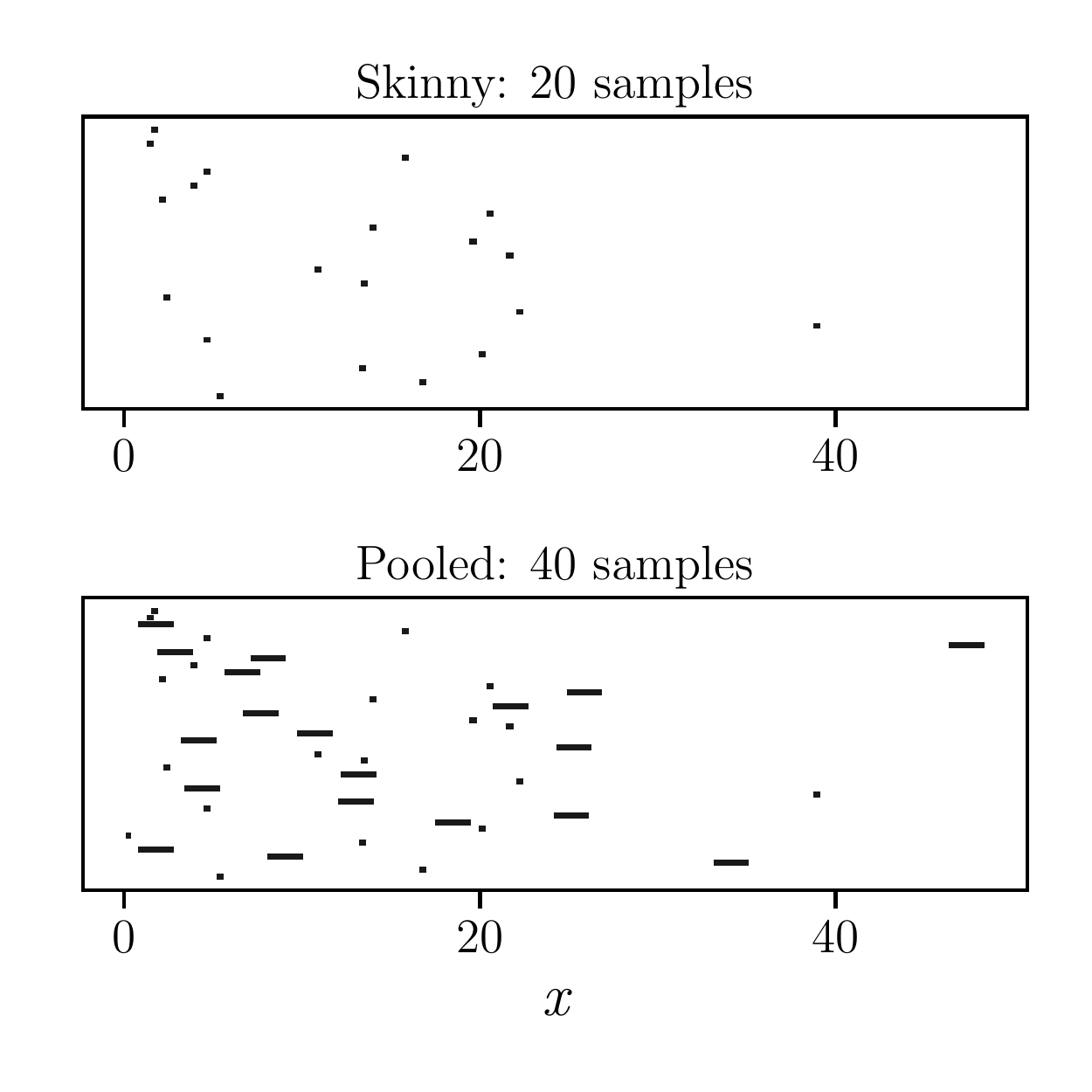}
		\caption{}
	\end{subfigure}
	\begin{subfigure}{0.325\textwidth}
		\includegraphics[width=\linewidth]{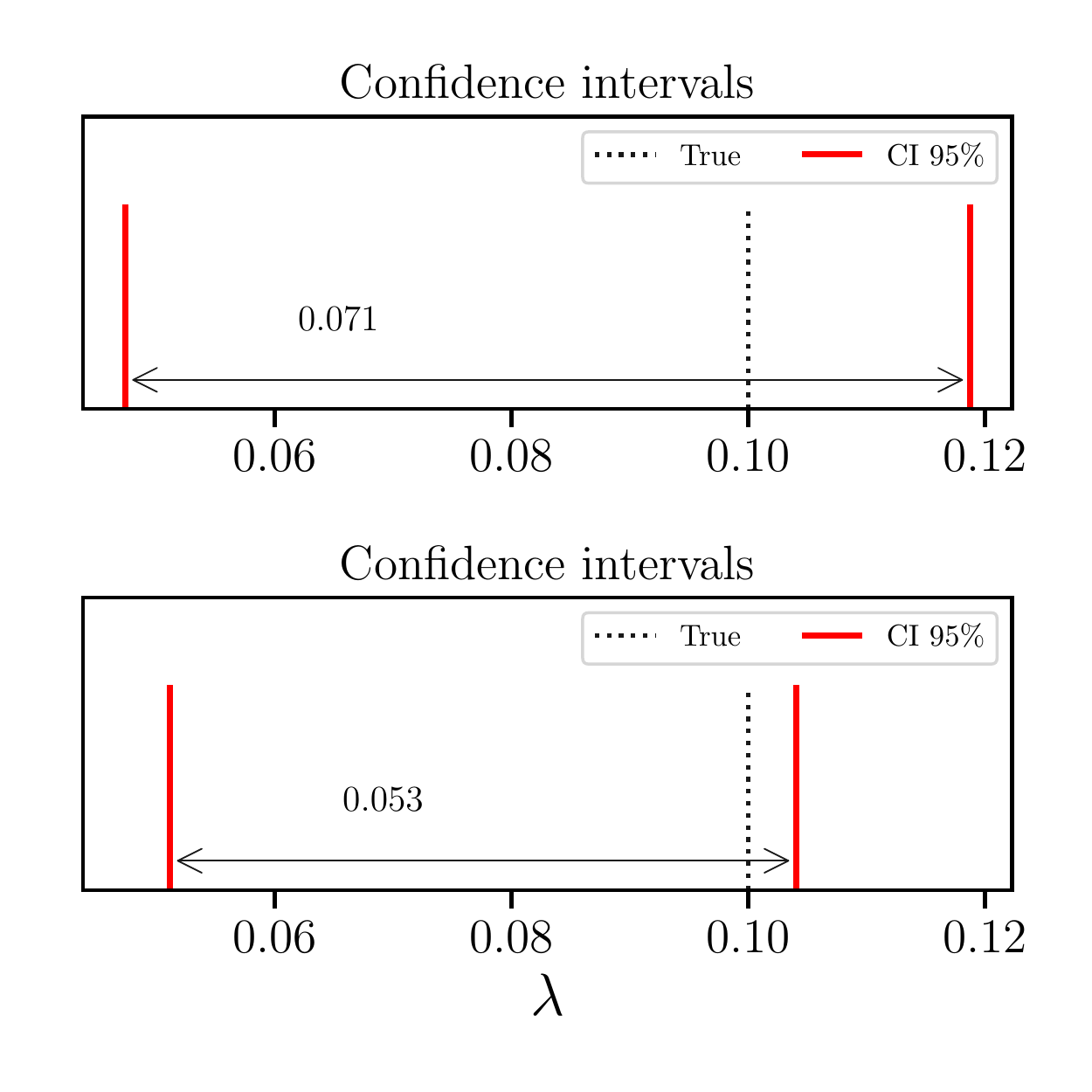}
		\caption{}
	\end{subfigure}
	\begin{subfigure}{0.325\textwidth}
		\includegraphics[width=\linewidth]{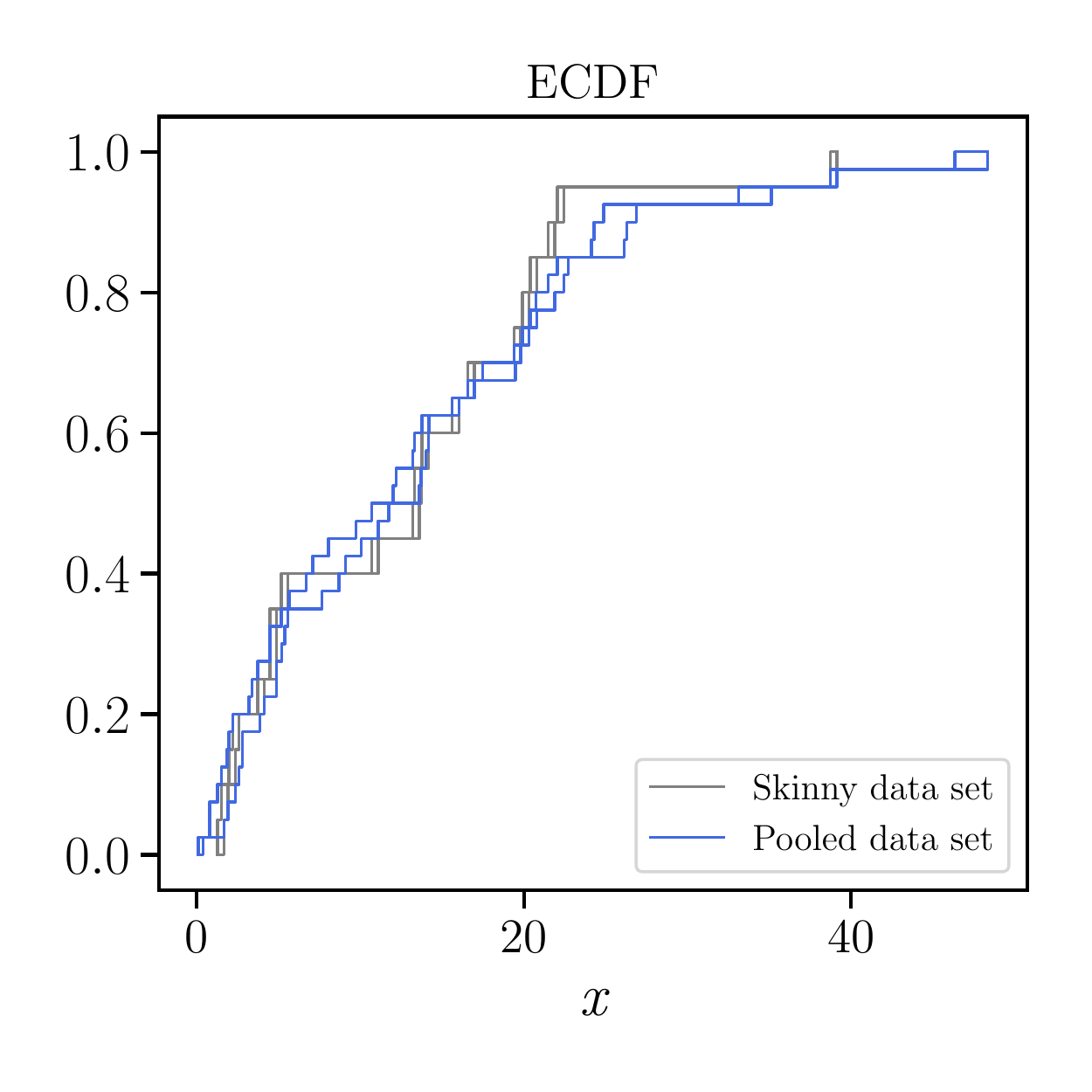}
		\caption{}
	\end{subfigure}
	\begin{subfigure}{0.325\textwidth}
		\includegraphics[width=\linewidth]{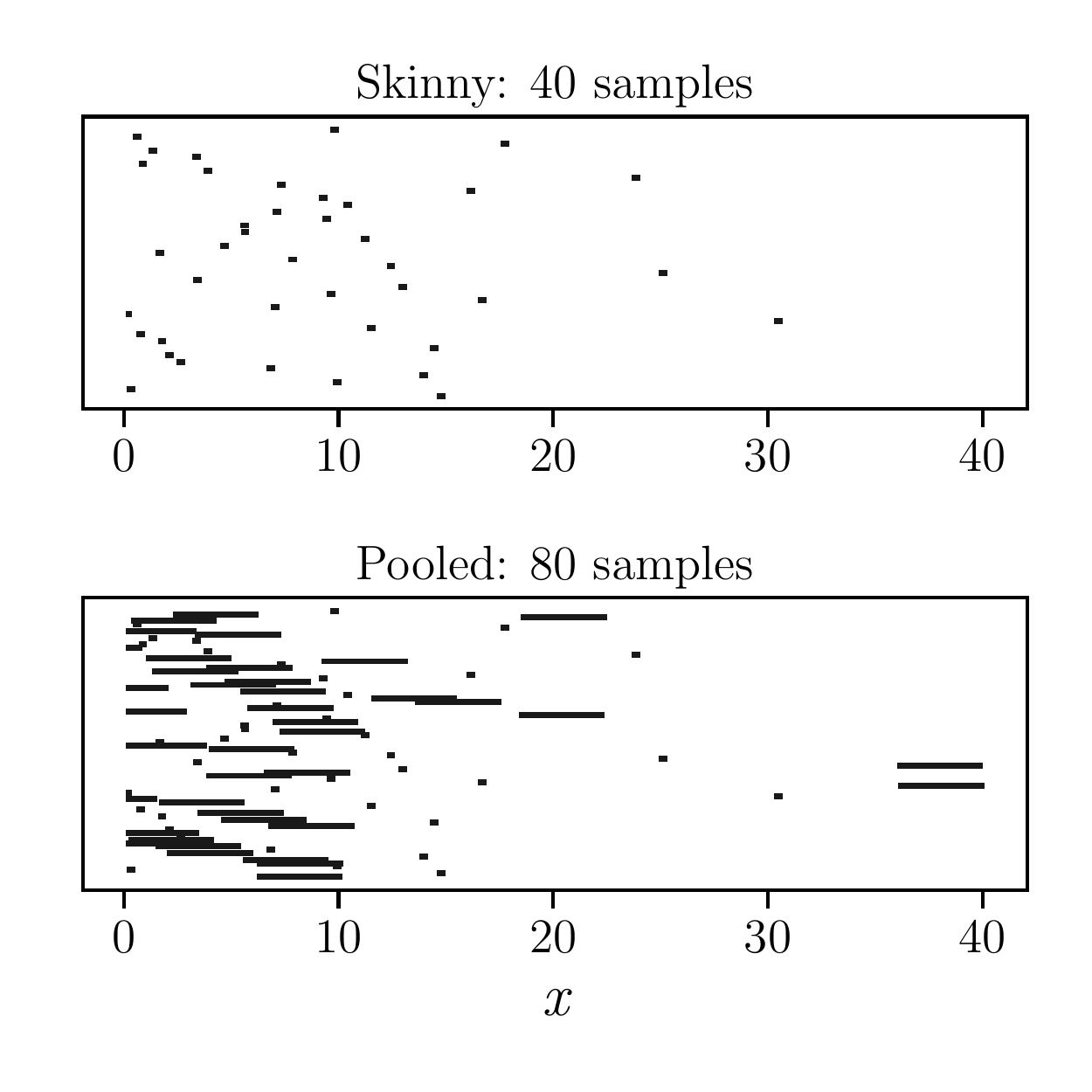}
		\caption{}
	\end{subfigure}
	\begin{subfigure}{0.325\textwidth}
		\includegraphics[width=\linewidth]{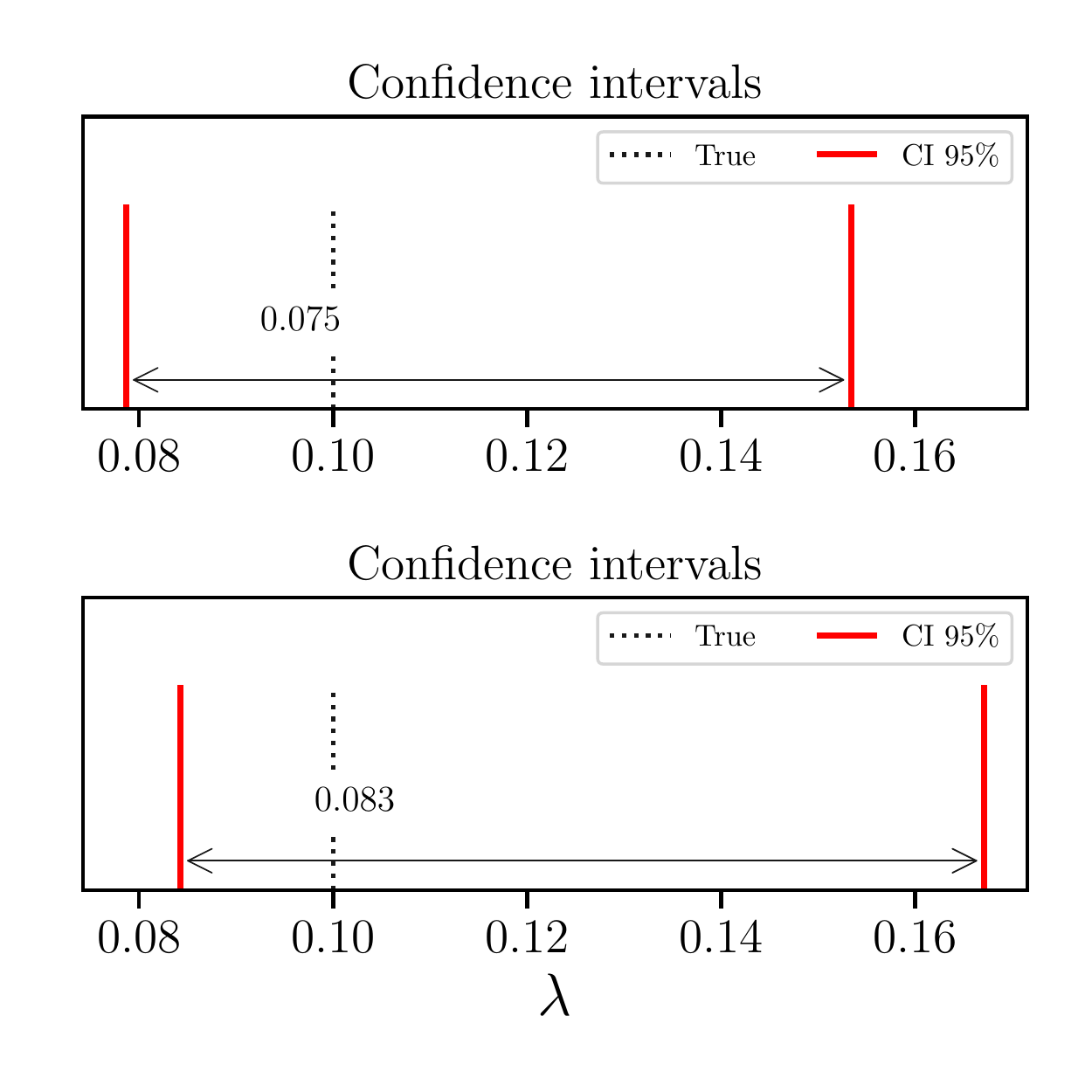}
		\caption{}
	\end{subfigure}
	\begin{subfigure}{0.325\textwidth}
		\includegraphics[width=\linewidth]{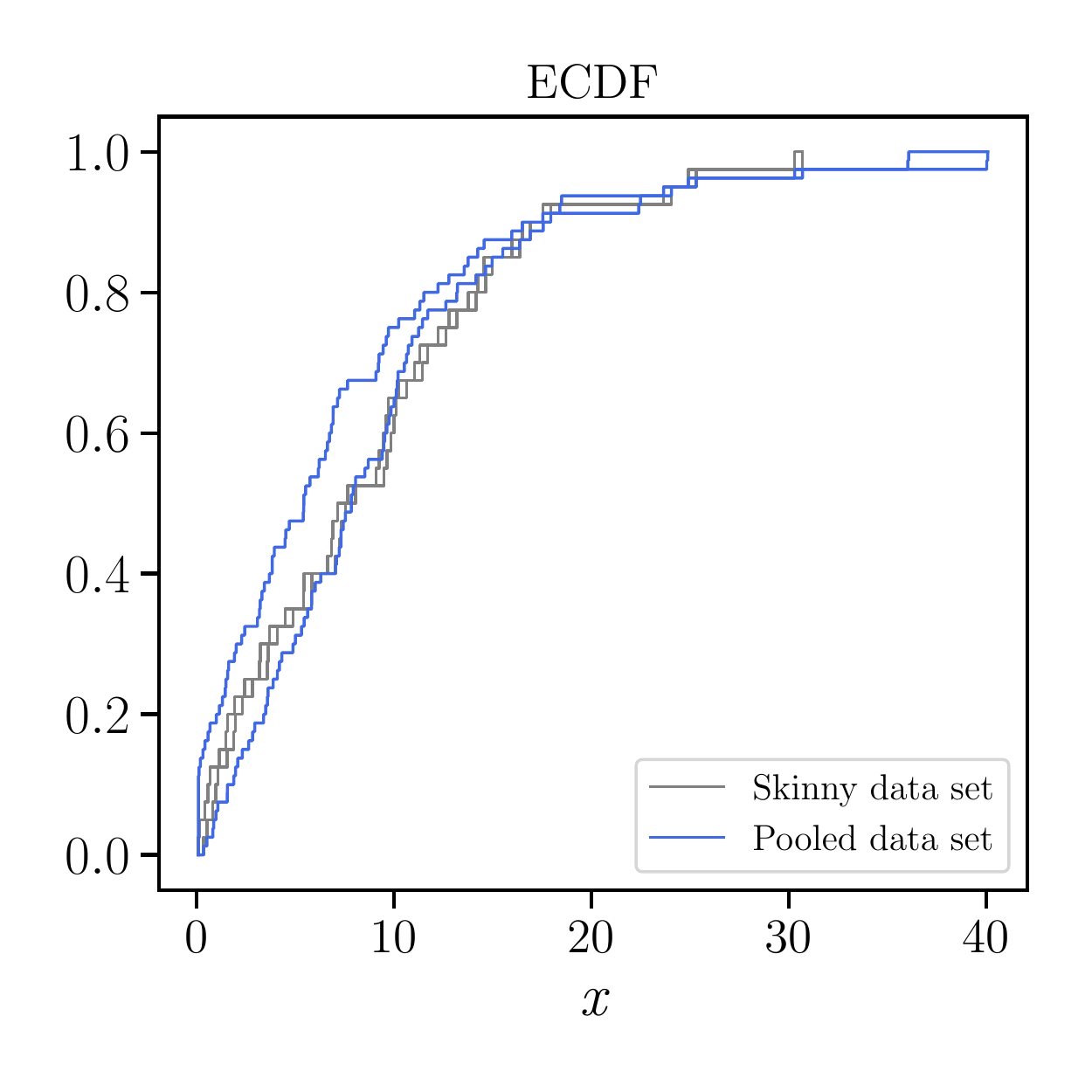}
		\caption{}
	\end{subfigure}
	\caption{Confidence intervals for the maximum likelihood estimator for an exponential distribution: (a)~precise and imprecise interval data sets randomly generated as $\interval{\underline{x}_i}{\overline{x}_i} \sim \text{Exp}(0.1)$, $\Delta_{\text{skinny}} = 0.1$, $f=5$; (d) $\interval{\underline{x}_i}{\overline{x}_i} \sim \text{Exp}(0.1)$, $\Delta_{\text{skinny}} = 0.1$, $f=10$; (b,~e)~estimated  95\% confidence interval for $\lambda$; (c,~f)~empirical cumulative distribution functions for the skinny and pooled data sets.}
	\label{fig:MLE_exp_CI_examples}
\end{figure}

Fig.~\ref{fig:MLE_uni_CI_examples} and \ref{fig:MLE_exp_CI_examples} depict confidence intervals for the parameters of $\mathcal{U}(0,\theta)$ and Exp($\lambda$) distributions, using $\theta=10$ and $\lambda=0.1$ to synthesize example data.
The skinny and pooled data are depicted on the left, and their respective empirical distribution functions are shown on the right. The middle graph shows the resulting confidence intervals for the maximum likelihood estimators as red lines.

\begin{figure}[ht!]
	\centering
	\begin{subfigure}{0.325\textwidth}
		\includegraphics[width=\linewidth]{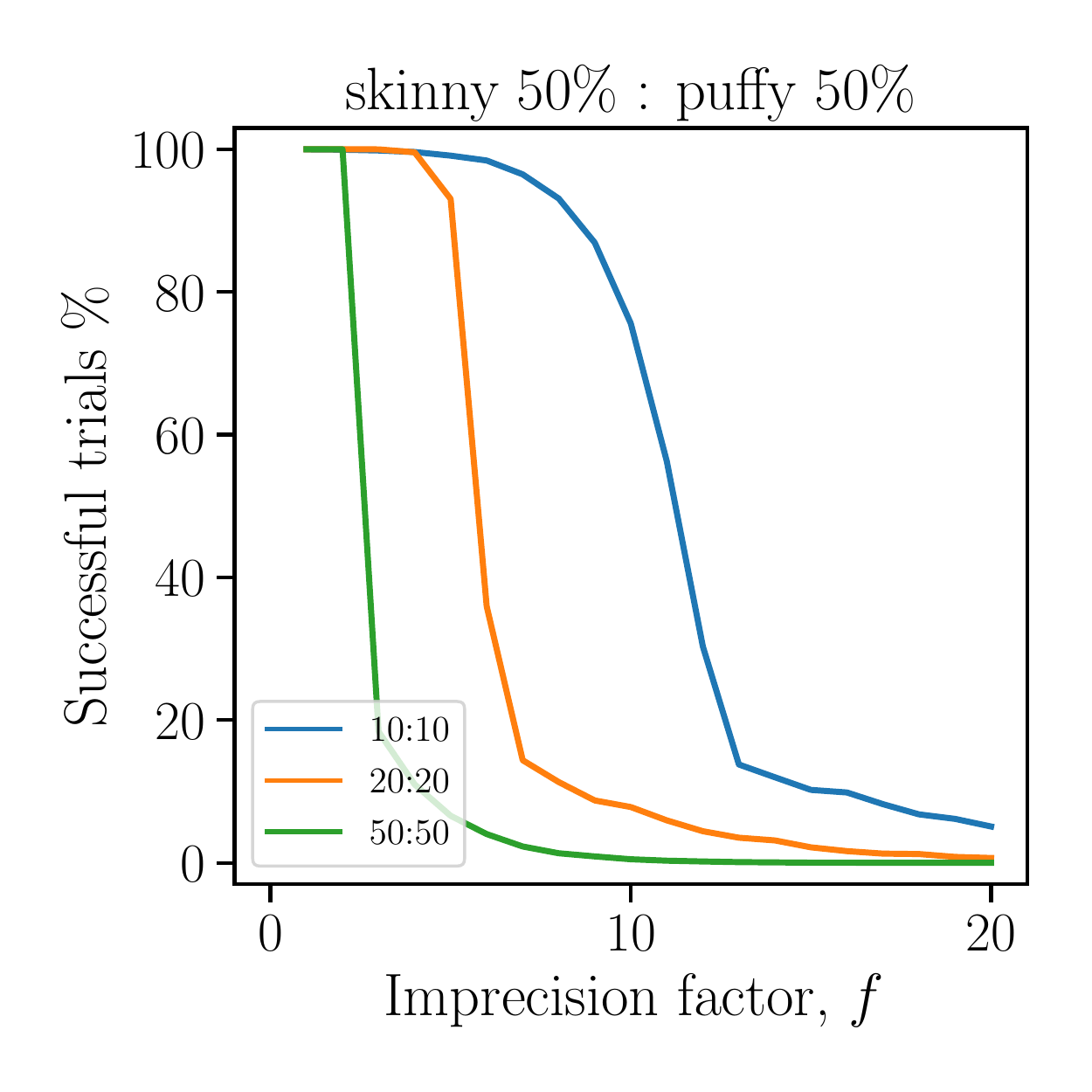}
		\caption{Uniform distribution}
	\end{subfigure}
	\begin{subfigure}{0.325\textwidth}
		\includegraphics[width=\linewidth]{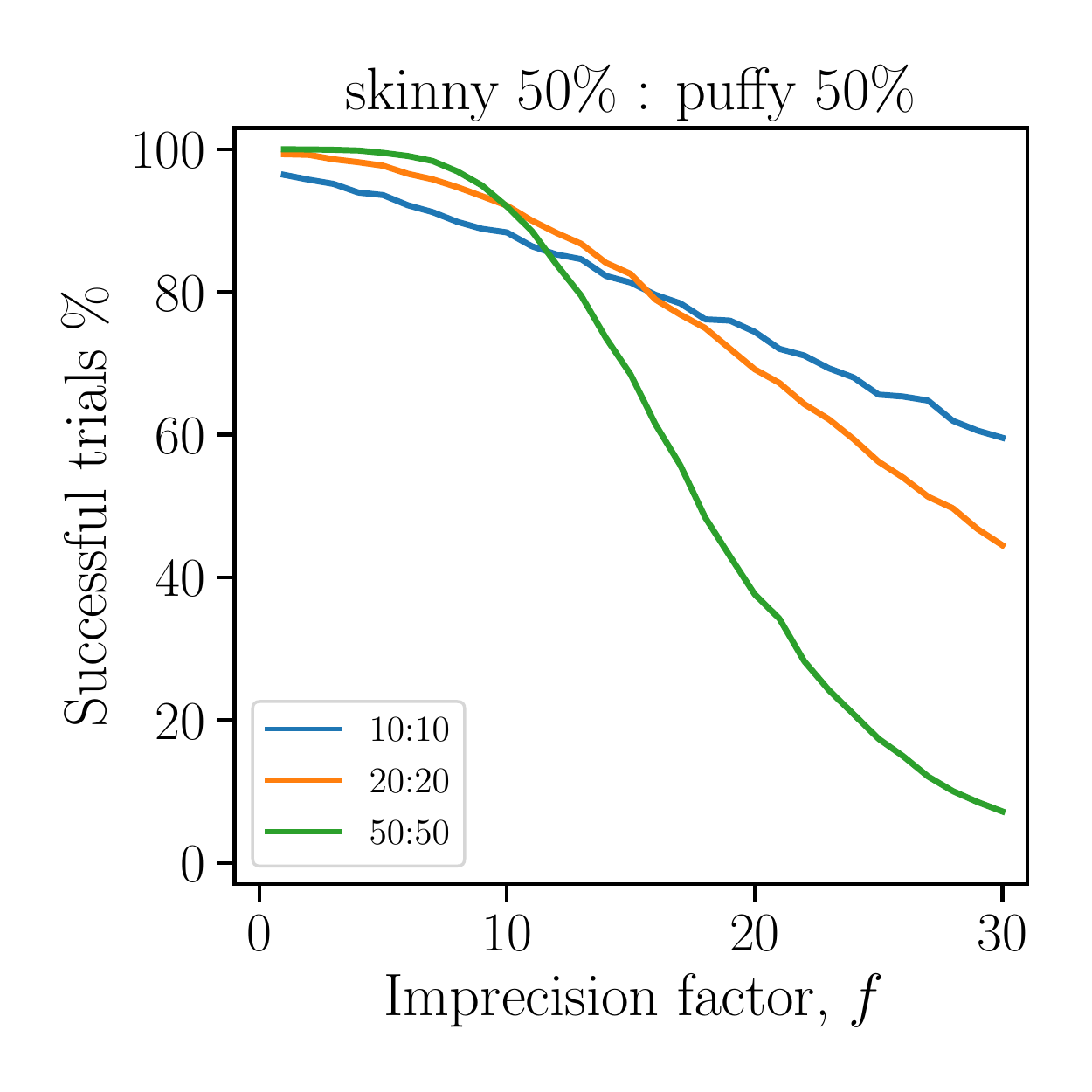}
		\caption{Exponential distribution}
	\end{subfigure}
	\caption{Percent of trials in which the pooled data set has less uncertainty when measuring the width of the confidence intervals for the maximum likelihood estimator: (a) $\interval{\underline{x}_i}{\overline{x}_i} \sim \mathcal{U}(0,10)$ (b) $\interval{\underline{x}_i}{\overline{x}_i} \sim \text{Exp}(0.1)$. Insets denote sample sizes for the skinny and puffy data sets.}
	\label{fig:MLE_CI_simulations}
\end{figure}

We carried out additional numerical simulations (similar to those in Sec.~\ref{Sec_CI} and \ref{KolmogorovSmirnov_bands}) and compared the widths of the confidence interval bounds for the skinny and pooled data sets depending on sample size and imprecision factor. Fig.~\ref{fig:MLE_CI_simulations} shows the results obtained from 
calculating confidence intervals for the estimated parameter by the interval maximum likelihood method.
For each experiment a fixed number of samples were randomly chosen from the corresponding distribution and intervalized using the uniformly biased method with $\Delta_{\text{skinny}} = 0.1$ and $\Delta_{\text{puffy}} = f \Delta_{\text{skinny}}$. The experiment was considered successful if the width of the confidence interval for the pooled data set was less than that for pure skinny (precise) data set $D_\text{pooled} < D_\text{skinny}$. Curves for the uniform and exponential distributions have the same tendency as the Kolmogorov--Smirnov confidence bands and confidence intervals for the mean (see Sec. \ref{Sec_CI}). The chance that pooling is better monotonically decreases for greater imprecision in interval data sets.

\subsection{Summary of MLE methods}
We considered two ways of generalizing the maximum likelihood method for interval data. The traditional and interval approaches for fitting named distributions to imprecisely measured values lead to different results when pooling data sets with disparate precision. This is not surprising, because the traditional approach yields a single distribution which almost neglects the incertitude in imprecise observations.  In contrast, the interval approach  considers a set of distributions and makes weaker assumptions about the underlying sample data. 

As shown in Fig.~\ref{fig:Likelihood_examples}b, the traditional approach yields predicted distributions that are similar despite high discrepancy in the initial data sets. The traditional approach is relatively insensitive to the incertitude in data and essentially masks the difference between the skinny and pooled data sets, despite high disparity of their uncertainties. Using the traditional approach, simulations always suggest comparable or better accuracy for pooled data sets, no matter how imprecise the wide intervals in them may be. 

The interval approach, in contrast, is notably sensitive to the magnitude of imprecision and the effects of differences in imprecision when pooling data.  The results obtained from calculating the confidence intervals for the intervalized maximum likelihood estimator show that, depending on the disparity of the imprecision, pooling may or may not be favoured. This behaviour is the same as shown by the K--S confidence bands and confidence intervals for the mean under a normality assumption in Sec. \ref{Sec_CI}. The chance that pooling is better decreases monotonically under greater disparity in the imprecision of the interval data sets.
The examples of the uniform and exponential distributions suggest that preference for pooling generally depends on $f$, irrespective of tail weight, although the steepness of the relationship between the chance that pooling is preferable and the the imprecision factor $f$ depends on multiple parameters, including sample size, mixture proportion, distribution shape, and tail weight.

The intervalized maximum likelihood approach \cite{Walley1991}, \cite[Sec. 4.9.3]{SAND2007-0939} asymptotically converges to \emph{enclose} the true value being estimated so long as the maximum likelihood estimator converges to the true value with asymptotically many precise data points. However, the use of the interval approach to calculate the maximum likelihood estimator parameters of some distributions can be difficult, i.e., mathematical expressions cannot always be easily generalized to handle interval data. The problem arises when an interval value occurs several times in a calculation \cite{Moore2009}. This issue with repeated uncertain variables is sometimes called the \emph{dependency problem}. 
Computing standard deviation and other quantities of dispersion in general for interval data may be computationally prohibitive as it is an NP-hard problems \cite{NP-Hard_Ferson}. The general overview of NP-hardness of computational problems in interval context is given in \cite{Kreinovich1998}. However, efficient and practical algorithms exist for such calculations for several special cases, including the most common situations of binned data and constant-width intervals.

Similar difficulties accompany computation of confidence limits in general \cite{SAND2007-0939,Kreinovich1998}. 
Workable formulas are described here for normal, uniform, and exponential distributions, but analogous efficient formulas are needed for computing parametric confidence limits for other distributions from data sets involving intervals.
An approach employing the distribution-free K--S bounds works for any distribution shape.
\section{\reva{Discussion}}\label{Discussion}

Different statistical approaches make different assumptions, but as depicted in Fig.~\ref{fig:Continuum}, assumptions sets can sometimes be ordered. Compared to traditional approaches for missing and censored data, the proposed interval statistics approach makes generally \emph{weaker} assumptions about the available data.  But its conclusions are consequently more reliable. \revb{The only assumption used by the interval statistics approach is that the true values are surely inside the respective intervals that represent measurement uncertainty.}

The assumption of irrelevantly imprecise means that the imprecision has to do only with the mensurational or observation process, rather than the measurand's value itself. 
Our assumptions can be contrasted with the \say{coarsened at random} (CAR) assumption suggested by Heitjan and Rubin \cite{Heitjan1991}
which generalises the idea of missing at random \cite{RUBIN1976}.
CAR depends on knowledge about why and how the imprecision was formed during the measurement process. The notion of coarsening at random was introduced to refer to situations in which the coarsening mechanism can be ignored when making inferences on the distribution using a traditional (precise) likelihood function, which Gill et al.\ \cite{Gill1997} argued can be dangerous under uncritical application.
\revb{The assumption made by the interval statistics approach presumes that these point values are within their respective intervals, which is a coarsening in the language of Heitjan and Rubin, but the assumption of irrelevantly imprecise is not a special case of coarsened at random.  Neither does being coarsened at random imply that data will be irrelevantly imprecise as different groups can be differently biased, and coarsened at random refers only to how intervals are identified with the underlying true point values.} 



All of the various theories 
\cite{Billard_Diday2006, Italians2005, Augustin2021, Heitjan1991, Ferson_Siegrist2011} for handling epistemic uncertainty of measurements in statistical analysis seem to refer to similar mechanisms such as censoring, rounding, and, in the limit, missingness, and they all make recourse to similar mathematical notions such as intervals or sets.
But their perspectives sometimes differ sharply. 
The conception of Augustin \cite{Augustin2021}, and the one we adopt here, is that there is a true, real-valued quantity, and we are using an interval to characterise our empirical uncertainty about that precise value.
Heitjan and Rubin \cite{Heitjan1991} seem to be thinking of the true epistemic uncertainty about a quantity as an interval perhaps, but they have only a real-valued number to characterise that interval, such as its limit or its centroid.
The conception used by Billard and Diday \cite{Billard_Diday2006} seems similar to ours referring to interval-censoring, but it holds that it is reasonable to use a uniform distribution to characterise that interval.
All of these approaches are reasonable in their own contexts, and the variety of approaches is essential for correct analysis of disparate situations where different assumptions should be made.

If we can assume that the data are irrelevantly imprecise, meaning that the width of each interval is not related to the magnitude of the underlying unknown value, then it can sometimes make sense to throw out the imprecise data when it increases overall uncertainty.
Without this assumption, however, pooling is required no matter the level of the imprecision.  Moreover, without the assumption, methods that do not neglect or understate the effect of the imprecision are needed to obtain reliably conservative results.
The classical maximum likelihood approach may be reasonable if its assumptions, including common error structure, are satisfied.  But the robust interval maximum likelihood approach described in Section~\ref{MLE} would be preferable in other cases.


\revb{Different assumptions about imprecision are possible.
The results of this paper are obviously sensitive to our assumptions about the cause and structure of the imprecision.  
We do not know whether conclusions similar to those of this paper about omitting data would be possible for data that are, for instance, coarsened at random but are not irrelevantly imprecise.  This question deserves additional study.}

\revb{This paper assumes that there is an underlying true (point) value for each interval but that it is observed in an imprecise way. However, the methods and conclusions herein may extend to situations in which there is ontic uncertainty 
where there is no point value that could be called `true' because the underlying measurand does not exist as a point value or is imperfectly defined.}

\section{Conclusion}\label{Conclusion}

One of the most important conclusions arising in studies of this kind is that measurement imprecision sometimes trumps sampling uncertainty.  Sufficiently large measurement imprecision can increase uncertainty more than larger sample size decreases uncertainty. This consideration determines whether we want to pool bad data with good data that are drawn from the same distribution. The alternative to pooling is to discard the bad data because of its imprecision, which can be reasonable to do under the assumption that the measurement imprecision is independent of the magnitudes of the measurands. These findings are relevant when analysts can assume that the data are irrelevantly imprecise, \reva{meaning that the true values are inside the intervals and the widths of those intervals are independent of those true values.}

We employed numerical simulations to measure uncertainty when pooling interval synthetic data sets derived from various distribution families.
In these studies we assumed that empiricists generally know the precision/imprecision of their data. We generated a precise data set with a small, fixed level of measurement uncertainty, and an imprecise data set with a larger level of uncertainty prescribed by a relative imprecision factor. The two data sets were combined to obtain a pooled data set.  Comparisons were made between the pooled data set and the smaller but more consistent precise data set.  We compared multiple outcomes, including confidence intervals on the mean, K--S limits around the median, and confidence intervals on maximum likelihood estimators for parameters of multiple distributions with various tail weights. The comparisons assessed the overall uncertainty that combines sampling error and the imprecision about individual measurements. 

We observed functional dependencies for the percent of trials in which the pooled data set has less overall uncertainty compared to data sets with only precise data. These functional dependencies decrease monotonically when the imprecision factor increases. These findings reveal the efficiency of pooling at various sample sizes, balances, imprecision factors, and dispersions. The more uncertainty in the imprecise data set, the more likely that pooling will damage the high-quality data yielding an increase in the overall level of uncertainty.

The simulation results suggest that pooling is preferred when the low-quality data set does not exceed a particular level of imprecision. We found that all four parameters (sample size, balance, dispersion, and imprecision factor) play a role in determining whether the data sets should be pooled.
If the level of uncertainty in the low-quality data set is only several (up to 5) times higher than the precise data, then pooling is desirable, and numerical simulations suggest that the chance of reducing the overall uncertainty is over $80\%$ when using confidence intervals on either the mean or K--S limits.
When the imprecision in the low-quality data exceeds ten times that of the precise data, then the chance that pooling is beneficial falls below $50\%$ depending on sample size. 
Data sets with small sample size less than $N \approx 20$ with high dispersion should be pooled even if the level of uncertainty in the imprecise data set is up to 15 times larger than the high-quality data. 

For a very unbalanced case such as when the pooled data set contains mostly precise data and a small amount of low-quality data, pooling may be beneficial only if the uncertainty in the low-quality data set is no more than several times higher than the precise data. For example, simulations over multiple distribution families suggest that adding a single low-quality measurement to a data set with $20$ samples will yield a statistical improvement in about 70\% of cases when the imprecision factor equals two.  
For large precise data sets ($N > 80$) pooling will improve statistics only if the uncertainty in the imprecise data set does not exceed twofold, otherwise it may be acceptable to ignore the imprecise data entirely no matter how many samples it contains. 
\reva{These quantitative trends suggest the following generic implications:
\begin{enumerate}
\item large imprecision favors omitting imprecise data,
\item high dispersion favors pooling imprecise data,
\item small sample size favors pooling imprecise data,
\item imbalance toward precision favors omitting imprecise data.
\end{enumerate}}

This paper has characterised the general trade-off between more data versus better data. It considers a question in the use of data quality rules that govern whether relatively poor data should be collated with better data. Usually, standards-based data quality objectives are used to determine whether data are allowed to be used in an analysis, based on, for example, whether the number of missing values is low and how large the empirical precision happens to be.  
We suggest that a \emph{performance-based} determination makes more sense. Even very imprecise data may be useful if including it reduces sampling uncertainty more than its imprecision increases measurement uncertainty. This consideration of the trade-off between sample size and precision on the overall uncertainty obviously should also be helpful in planning future empirical effort.

\section{Acknowledgements}\label{acks}
We thank Jack Siegrist, William Huber, Vladik Kreinovich, Marco De Angelis, Nick Gray, George Dale, and the anonymous referees for their very helpful comments. This work was supported by the UK Engineering and Physical Science Research Council through grant number EP/R006768/1.

\bibliographystyle{ieeetr}
\bibliography{lib.bib}

\end{document}